\documentclass[aps,amsmath,amssymb,showpacs,showkeys]{revtex4}
\usepackage[dvips]{graphicx,color}
\usepackage{times}
\usepackage{xcolor}
\usepackage[%
  colorlinks=true,
  urlcolor=blue,
  linkcolor=red,
  citecolor=blue
]{hyperref}

\begin{document}
\title{A new $f(R)$ Gravity Model and properties of Gravitational Waves in it}

\author{Dhruba Jyoti Gogoi}
\email[Email: ]{moloydhruba@yahoo.in}

\affiliation{Department of Physics, Dibrugarh University,
Dibrugarh 786004, Assam, India}

\author{Umananda Dev Goswami}
\email[Email: ]{umananda2@gmail.com}
\affiliation{Department of Physics, Dibrugarh University,
Dibrugarh 786004, Assam, India}

\begin{abstract}
In this paper, we have introduced a new $f(R)$ gravity model as an attempt to 
have a model with more parametric control, so that the model can be used to 
explain the existing problems as well as to explore new directions in physics
of gravity, by properly constraining it with recent observational data. Here 
basic aim is to study the properties of Gravitational Waves (GWs) in this new
model. In $f(R)$ gravity metric formalism, the model shows the existence of 
scalar degree of freedom as like other $f(R)$ gravity models. Due to this 
reason, there is a scalar mode of polarization of GWs present in 
the theory. This polarization mode exists in a mixed state, of which one is 
transverse massless breathing mode with non-vanishing trace and the other is 
massive longitudinal mode. The longitudinal mode being 
massive, travels at speed less than the usual tensor modes found in General 
Relativity (GR). Moreover, for a better understanding of the model, we have 
studied the potential and mass of scalar graviton in both Jordan frame and 
Einstein frame. This model can pass the solar system tests and can explain 
primordial and present dark energy. Also, we have put constraints on the 
model. It is found that the correlation function for the third
mode of polarization under certain mass scale predicted by the model agrees 
well with the recent data of Pulsar Timing Arrays. It seems that this new 
model would be useful in dealing with different existing issues in the areas of 
astrophysics and cosmology.  
\end{abstract}

\pacs{04.30.Tv, 04.50.Kd}
\keywords{Modified Gravity; Gravitational Waves; Constraints; Pulsar Timing Arrays}

\maketitle
\section{Introduction}
Einstein's General Relativity (GR) has been the most widely accepted theory 
capable of explaining a number of phenomena and the geometry of spacetime in 
general. However, recent experimental observations showed the existence of
phenomena with deviations from GR predictions. Among them the present 
acceleration of the universe has been a big problem in cosmology lacking of a 
proper or satisfactory explanation till now. This problem was 
discovered around 22 years ago with the help of type Ia supernovae observations
\cite{Riess_1998, Perlmutter1999, Bahcall1999, Kirshner_1999}. It put a big 
question on the viability of GR, specially at cosmological scale. However, if 
one still wishes to stick with GR then he has to bear with the problem of 
mysterious invisible and exotic dark energy, which is responsible for around 
76\% of total energy content of the universe. Moreover, a theoretical problem 
associated with GR is that it is not renormalizable based on the conventional 
methods. In order to overcome these drawbacks of GR different modifications 
have been proposed. In most of the cases the modifications were introduced to 
solve some specific problems and as expected/ eventually they redefined 
the spacetime geometry and imprinted some changes in other sectors/ranges also. These changes can be a measure of how much the new theory is deviating from GR.
A very important result from GR is the Gravitational Waves (GWs). In modified 
theories of gravity, the properties of GWs also change and can result massive 
modes of polarization apart from the usual GR modes or tensor modes of 
polarization \cite{corda01}. These massive modes travel with less speed than 
the tensor modes. Besides the presence of massive modes of polarizations, the 
generation and propagation of GWs are also affected in different modified 
theories of gravity. The study of these variations can be a good tool to test 
the modified theories of gravity. With the first detection of GWs in 2015 by 
the LIGO and Virgo collaborations \cite{Abbott_2016} followed by many other 
detections till now, a new and promising direction of studying gravitational 
theories has begun. These experimental evidences of GWs put a new set of 
constraints on GR as well as on modified theories of gravity. 

A straight forward and simple modification to GR is the $f(R)$ theory of 
gravity. In this theory, the Ricci scalar $R$ of Einstein - Hilbert action is 
replaced by a function of $R$. Till now many models of $f(R)$ theory have been 
proposed. Some successful models capable of explaining the drawbacks of GR 
upto a significant range are the Starobinsky model \cite{Starobinsky_2007}, 
Hu-Sawicki model \cite{Hu_2007}, and Tsujikawa model \cite{Tsujikawa_2008, 
Zhang_2006, Cognola_2008, Linder_2009} or exponential gravity. The properties 
of GWs in these models have been studied earlier. These studies show that GWs 
in $f(R)$ gravity vary significantly from those in GR. As mentioned above, in 
GR, there are two polarization modes of GWs, viz., the tensor plus mode and 
tensor cross mode. These modes are massless in nature and they propagate with 
the speed of light in spacetime. In metric formalism $f(R)$ gravity, there 
exists a scalar degree of freedom in the theory and hence the total degrees of 
freedom in the theory increases, which leads to increase the polarization 
modes of GWs in such theories \cite{corda01, corda02}. It is found that total 
number of polarization modes of GWs that can exist in $f(R)$ theories of 
gravity is 3 \cite{Liang_2017}. Recent studies show that there exist a 
massless breathing mode and a massive longitudinal mode in a mixed state as a single polarization mode of 
GWs in $f(R)$ theories of gravity. The existence of this extra polarization mode can be checked with the help of modified Newman-Penrose scalars and also by geodesic deviation equations in the theory 
\cite{Liang_2017}. However, another study shows that the existence of third mixed polarization mode of GWs are model dependent and there exists a model in $f(R)$ 
gravity where the massive longitudinal mode vanishes and the third polarization mode becomes a pure massless breathing mode 
\cite{Gogoi_2019}.

In this work, we have used a new $f(R)$ gravity model as a toy model and 
studied the behaviour of the potential and scalar field mass both in Jordan 
frame and Einstein frame. We have also checked the polarization modes of GWs 
present in this model and tried to constrain the model.

The paper is organized as follows. In the next section, we have introduced our
new $f(R)$ model along with motivations. The characteristics of this 
model and the behaviour of the associated scalar field both in Jordan frame and Einstein frame have been studied in this section. Also, in this section, we 
have checked the viability of the model in terms of solar system tests. In the 
third section, we have compared the model with two other most viable models viz., the Starobinsky model and the Hu-Sawicki model. In the fourth section, the model has been constrained. In section five, we have studied
the polarization modes of GWs present in the model by using the perturbed field equation and the modified Newman-Penrose formalism. In the sixth section, we have 
reviewed a way to detect the polarization modes of GWs experimentally and 
discussed the possibilities of experimental validation of the model. In the 
last section, we conclude the paper with a very brief discussion of the 
results and the future aspects of the model in such type of studies.
\section{A New Model of f(R) Gravity}\label{sec2}
Although we have a pretty good number of $f(R)$ gravity models, no $f(R)$ 
gravity model can explain all cosmological and astrophysical aspects of the 
present universe completely. Moreover, as like GR, different $f(R)$ gravity 
models come with different types of drawbacks. However, we should mention that 
including the Starobinsky and Hu-Sawicki models there are several viable 
models in $f(R)$ gravity, which were proposed in order to overcome the 
drawbacks of GR. 
Although the asymptotic behaviour of such viable models are almost same, the functional forms are completely different. In such $f(R)$ gravity models the modifications of geometry is done in a unique manner by different $f(R)$ functions, which might result 
different and unique cosmological and astrophysical features in the intermediate curvature regime. Thus a new $f(R)$ gravity model capable of explaining the drawbacks of GR might have some different implications in different perspectives and also can have drawbacks or anomalies in different 
realms. Furthermore, a $f(R)$ gravity model with more parametric control is 
more suitable in this respect in the sense that such a model can be 
constrained properly with the observational data by controlling its 
parameters and hence can be used easily to overcome the drawbacks of GR. With 
these motivations, here we introduce another $f(R)$ gravity model containing 
three parameters, over and above to the existing ones, as given by:
\begin{equation}\label{model}
f(R) = R-\frac{\alpha}{\pi }\, R_c \cot ^{-1}\!\left(\tfrac{R_c^2}{R^2}\right)-\beta\,  R_c\! \left[1-\exp\left({-\,\tfrac{R}{R_c}}\right)\right],
\end{equation}
where $\alpha$ and $\beta$ are two dimensionless positive constants and $R_c$ 
is a characteristic curvature constant having dimensions same as curvature 
scalar $R$. This model has two correction terms:
 $$ \frac{\alpha}{\pi }\,  R_c \cot ^{-1}\!\left(\tfrac{R_c^2}{R^2}\right)\;\; 
\text{and}\;\; \beta\,  R_c\! \left[1-\exp\left({-\,\tfrac{R}{R_c}}\right)\right].$$ 
The first correction factor is estimated by two parameters $\alpha$ and 
$R_c$. Similarly, the second correction factor has also two parameters 
$\beta$ and $R_c$ and it mimics the exponential $f(R)$ gravity model. We'll 
show that this model passes the basic requirements of a viable $f(R)$ gravity 
model including the solar system tests.

The requirements for any $f(R)$ gravity model to describe the late-time dark 
energy problem are \cite{Chen_2019,Amendola_2015, Amendola_2007, Bamba_2010}:

(1) A sufficient and suitable chameleon mechanism which allows $f(R)$ gravity 
to pass the constraints of local systems. In case of our model, we'll show that
it can pass the solar system tests. A detailed study of chameleon mechanism in 
this model is beyond the scope of this paper.
 
(2) A late-time stable de-Sitter solution. The condition for the existence 
of de-Sitter solution for a model is 
\begin{equation} \label{de_sitter_eq}
2 f(R_0) - R_0 f'(R_0)=0,
\end{equation}
 where
$R_0$ is the de Sitter curvature. To ensure the stability of the de Sitter solution, the model needs to satisfy the following condition \cite{motohashi}:
\begin{equation} \label{stability_primary_eq}
\frac{f'(R_0)}{f''(R_0)}>R_0.
\end{equation}
For simplicity, considering $R_0= R_c$ in the Eq.\ \eqref{de_sitter_eq} and solving for $\beta$ we find,
\begin{equation}\label{eq_four}
\beta = -\frac{e (2 \alpha -\pi  (\alpha -2))}{(6-4 e) \pi },
\end{equation}
where $e$ is the natural exponential. This is the de Sitter solution for the 
case $R_0= R_c$. Now the stability condition \eqref{stability_primary_eq} gives,
\begin{equation}
(e\, \alpha +\pi\,  \beta ) (2\, e\, \alpha +2\, \pi\,  \beta -e\, \pi )<0.
\end{equation}
Using the expression for $\beta$ from Eq.\ \eqref{eq_four} in the above 
expression, we find the allowed range of $\alpha$ as 
\begin{equation}
-1.68381<\alpha <0.367545.
\end{equation}
Thus for $R_0= R_c$, the model can have stable de Sitter solutions if $\alpha$ 
lies in the above range. 
In section \ref{sec3_comparison}, we explicitly showed that the model has a stable de Sitter solution and also compared it with two other viable models.
 
(3) A positive effective gravitational coupling, leading to $f'(R) > 0$. By 
putting our model in this condition gives, $$1-\frac{2\, \alpha\,  R\, R_c^3}{\pi  R^4+\pi  R_c^4}-\beta  \exp\left(-\,\tfrac{R}{R_c}\right) > 0.$$
 
(4) A stable cosmological perturbation and a positivity of the GWs for the 
scalar mode, causing to $f''(R) > 0$. Using our model in this condition we 
find, $$\frac{\beta  \exp\left(-\,\tfrac{R}{R_c}\right)}{R_c}-\frac{2\, \alpha\,  R_c^3 \left(R_c^4-3 R^4\right)}{\pi  \left(R^4+R_c^4\right)^2} > 0.$$ This 
condition ensures the absence of tachyonic instabilities in the model, i.e.\ 
this condition results $m^2_\phi > 0$.

(5) An asymptotic behaviour to the $\Lambda CDM$ model in the large curvature 
region. In case of our model, at large curvature region i.e.\ at $R \rightarrow 
\infty$, we have $f(R)-R \rightarrow -\,\frac{1}{2} R_c (\alpha +2 \beta ) = \text{constant}$, 
which mimicks the $\Lambda CDM$ model in the large curvature region. Again, 
at $R \rightarrow 0$, we have $f(R)-R \rightarrow 0$.

Thus it is seen that this model is capable of explaining the late-time dark 
energy problem. In the following part of this paper, we'll study the behaviour 
of scalar degrees of freedom of the model as well as the nature of GWs in it.

\subsection{Scalar Degrees of Freedom in Jordan Frame} 
\begin{figure}[htb]
\centerline{
   \includegraphics[scale = 0.3]{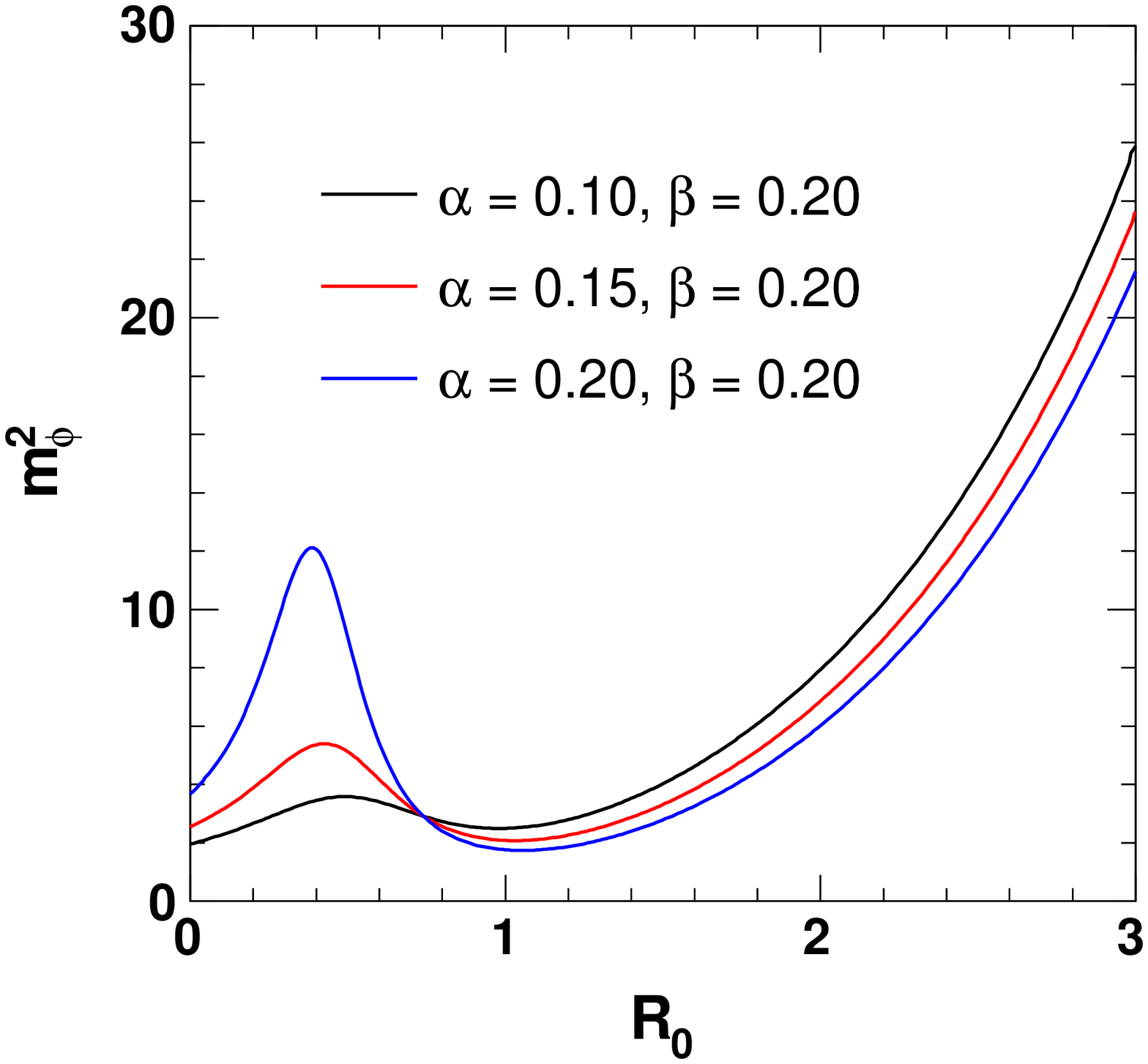}\hspace{1cm}
   \includegraphics[scale = 0.3]{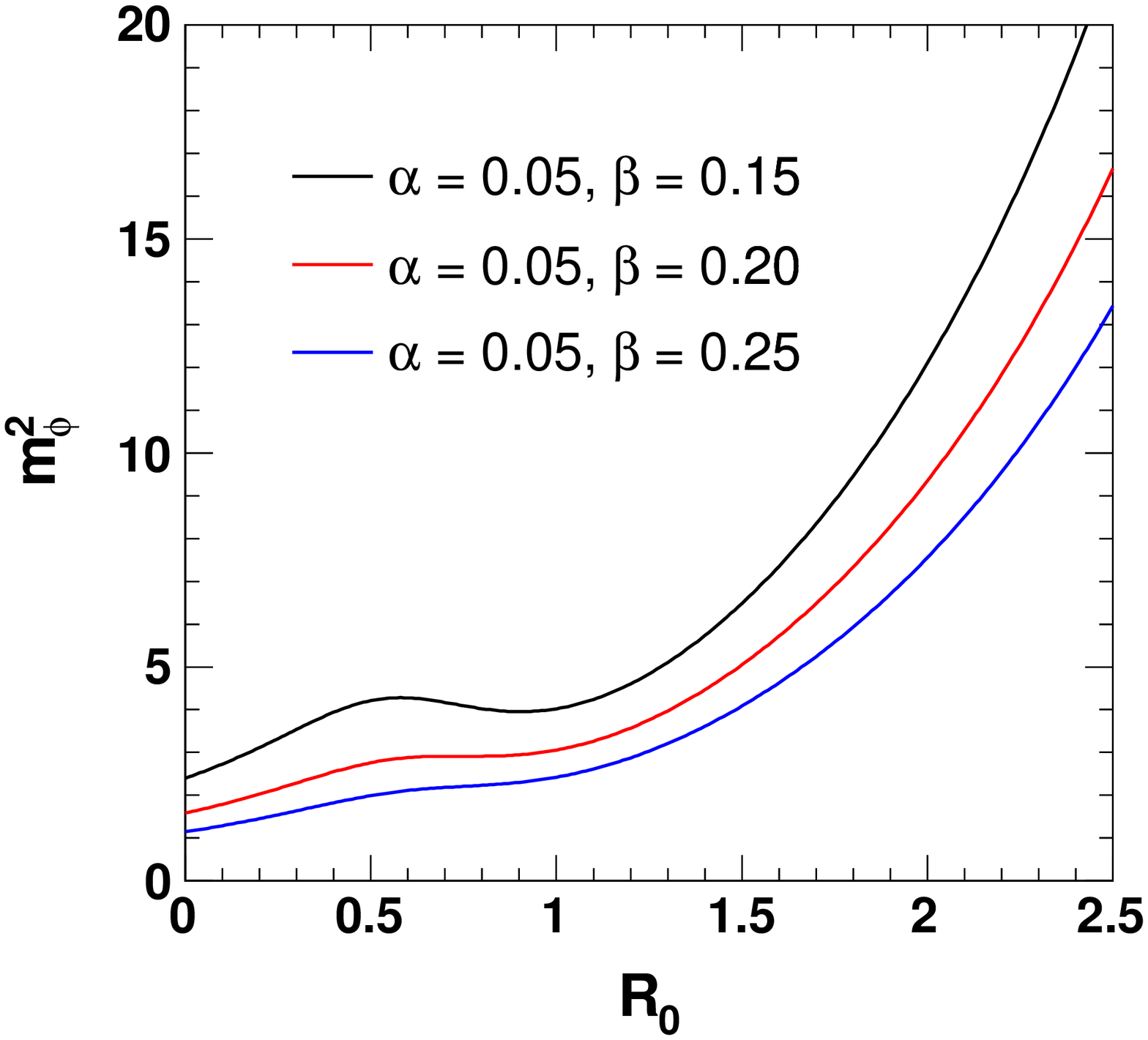}}
\vspace{-0.2cm} 
\caption{Mass square of the scalar field as a function of the background 
curvature for different sets of values of $\alpha$ and $\beta$ with 
characteristic curvature constant $R_c = 1$ in arbitrary units.}
\label{fig01}
\end{figure}

The action of a generic $f(R)$ gravity model is given as
\begin{equation}\label{action}
S = \dfrac{1}{2\kappa} \int d^4 x \sqrt{-g} f(R) + \int d^4 x \sqrt{-g}\, \mathcal{L}_m\! \left[ g^{\mu\nu}, \bar{\psi} \, \right]\!. 
\end{equation}
In the above equation, the function $f(R)$ stands for any arbitrary function 
of Ricci curvature scalar R, $g_{\mu\nu}$ is the metric, $\kappa^2 = 8 \pi G =
M_{pl}^{-2}$ and $\hbar = c=1$. Here $M_{pl} \approx 2 \times 10^{18}\, 
\text{GeV}$ is the reduced Planck's mass. $\mathcal{L}_m\! \left[ g^{\mu\nu}, 
\bar{\psi} \, \right]$ is the Lagrangian for a matter 
field $\bar{\psi}$. Variation of the action (\ref{action}) with respect to the 
metric gives the following field equation:
\begin{equation}\label{field_equation}
 f'(R)R_{\mu\nu} -\dfrac{1}{2} f(R)g_{\mu\nu} - \nabla_\mu \nabla_\nu f'(R) 
 + g_{\mu\nu}\, \square f'(R) = \kappa^2\, T_{\mu\nu}(g^{\mu\nu}, \bar{\psi}), 
\end{equation}
where $\square \equiv \nabla^\mu \nabla_\mu$,\; $T_{\mu\nu}(g^{\mu\nu}, \bar{\psi}) = \dfrac{-\,2}{\sqrt{-g}}\dfrac{\delta \left(\sqrt{-g}\,\mathcal{L}_m\!\left[ g^{\mu\nu}, \bar{\psi} \, \right]\right)}{\delta g^{\mu\nu}}$ is the matter
energy-momentum tensor and $f'(R) = \partial_R f(R)$.
Trace of Eq.\ (\ref{field_equation}) is
\begin{equation} \label{trace_field_eq}
f'(R) R +3\, \square f'(R) - 2 f(R) = \kappa^2\, T,
\end{equation}
where $T = g^{\mu \nu} T_{\mu \nu}$ is the trace of $T_{\mu \nu}$.
It is seen that the trace of the field equation is dynamical. This equation 
also indicates the existence of an extra scalar degree of freedom in the 
theory. For a detailed study about this degree of freedom we would like to use
our model given in the Eq.~(\ref{model}). Now, if we define a scalar field as
\begin{equation}\label{field}
\phi = f'(R),
\end{equation}
then for our model the field $\phi$ becomes,
\begin{equation}\label{fieldm}
\phi = 1-\frac{2 \alpha  R_c^3}{\pi  R^3\! \left(\frac{R_c^4}{R^4}+1\right)}-\beta\,\exp\left(-\,\tfrac{R}{R_c}\right).
\end{equation}
This shows that the scalar curvature $R$ can be expressed as a function of 
 the field $\phi$. From this definition of scalar field $\phi$, we may 
view the trace Eq.\ (\ref{trace_field_eq}) as an effective scalar field 
equation of Klein-Gordon with the following identification \cite{Capo_2007}:
\begin{align}\notag
\dfrac{dV}{d\phi} &\equiv \dfrac{1}{3} \Big[ 2f(R(\phi)) - R(\phi) f'(R(\phi)) \Big] \\[5pt]
&\equiv \dfrac{1}{3} \Big[ 2f(R(\phi)) - R(\phi) \phi \Big], 
\end{align}
where $V$ is the potential of the scalar field $\phi$. Thus, the trace 
Eq.\ (\ref{trace_field_eq}) can be written as a Klein-Gordon type equation for 
the scalar field $\phi$ as given by,
\begin{equation}
\square \phi = \dfrac{dV}{d \phi} + \dfrac{1}{3}\,\kappa^2\, T = 
\dfrac{dV_{eff}}{d\phi}, 
\end{equation}
where $V_{eff}$ is effective potential of the field and is define as 
\begin{equation}\label{adeq}
\dfrac{d V_{eff}}{d \phi} = \dfrac{1}{3}\Big[ 2f(R(\phi)) - R(\phi)\phi +\kappa^2\, T \Big].
\end{equation}
At far away from the source or in absence of any matter source 
$V_{eff} \equiv V$. Again, from the stationary condition:
\begin{equation} \label{stationary_condition}
\dfrac{dV_{eff}}{d\phi} = 0,
\end{equation}
we can have $\phi =\phi_0$ satisfying $\phi_0 = f'(R_0)$. From this condition, 
the mass of the scalar field (or the scalaron mass) can be obtained by 
differentiating the Eq.\ \eqref{adeq} with respect to $\phi$ as
\begin{equation} \label{mass_scalar_field}
m^2_{\phi} \equiv \dfrac{dV^2_{eff}}{d\phi^2}\Big\vert_{\phi = \phi_0} 
= \dfrac{1}{3}\!\left[ \dfrac{f'(R_0)}{f''(R_0)} - R_0 \right]\!.
\end{equation}
Here $R_0$ is the 
background curvature corresponding to $\phi_0$. From the above equation, we 
can see that avoiding of tachyonic instabilities demands $\frac{f'(R_0)}{f''(R_0)} - R_0 \geq 0$ and to keep the mass term finite we need $f''(R_0) \neq 0$. 
For our model the mass term is found as
\begin{equation}\label{mass_model_01}
m^2_{\phi} =\left[\frac{R_c \exp\!\left(R/R_c\right) \left(\pi  \left(R^4+R_c^4\right)^2-8 \alpha  R^5 R_c^3\right)-\pi  \beta  (R+R_c) \left(R^4+R_c^4\right)^2}{3 \pi  \beta  \left(R^4+R_c^4\right)^2-6 \alpha  R_c^4 \exp\!\left(R/R_c\right) \left(R_c^4-3 R^4\right)}\right]_{R\, =\, R_0}\!\!\!\!\!\!\!\!\!\!\!\!\!\!\!\!.
\end{equation}
For $R_0 = 0$, this equation gives
\begin{equation}\label{mass_model_02}
m^2_{\phi}\big|_{R_0\, =\, 0} = \frac{\pi  (\beta -1) R_c}{6 \alpha -3 \pi  \beta }.
\end{equation}
This shows that the mass of the scalar field is non-vanishing even at a large 
distance away from the source or in the Minkowski space. The mass $m_{\phi}$ 
of the scalar field depends on the model parameters $\alpha$, $\beta$ and 
$R_c$. Fig.\ \ref{fig01} shows the variation of $m^2_{\phi}$ with respect
to $R_0$ for different sets of model parameters. From the figure we see that,
the mass of the scalar field increases rapidly with the increasing value of the 
background curvature after a hump in the curve for $0<R_0<1$ region, which 
increases when the parameter $\alpha$ takes value closer to parameter $\beta$. 
By increasing the difference between $\beta$ and $\alpha$ $($i.e.\ for 
$\beta \gg \alpha)$ the hump can be minimized. An increase of $\alpha$ 
increases the hump which occurs near the small curvature region as mentioned 
above and comparatively decreases the mass of the scalar field at higher 
curvature region. 

\subsubsection{Scalar Tensor Equivalence of the Model}
To see the origin of the scalar field in the theory, we would like to rewrite 
the action (\ref{action}) by introducing a new auxiliary scalar field 
$\psi$ as \cite{Liang_2017, Teyssandier_1983}
\begin{equation} \label{scalar_tensor_action01}
S = \dfrac{1}{2\kappa^2}\int d^4 x \sqrt{-g}\, \Big[ f_\psi(\psi)R - \big\lbrace f_\psi(\psi)\psi - f(\psi) \big\rbrace\Big]+ \int d^4 x \sqrt{-g}\, \mathcal{L}_m\!\left[ g^{\mu\nu}, \bar{\psi} \; \right],
\end{equation}
where $f_\psi = \dfrac{\partial f\! \left(\psi \right)}{\partial \psi}$.
Now, varying this equation with respect to the new auxiliary scalar field 
$\psi$ we get,
\begin{equation}
f_{\psi \psi}\! \left( \psi \right) \left( R - \psi \right) = 0.
\end{equation}
For finiteness of the previously defined mass square term of the scalar field 
(see Eq.\ (\ref{mass_scalar_field})), we have $f''\!\left( R \right) \neq 0$, 
which is in terms of $\psi$, $f_{\psi \psi}\! \left( \psi \right) \neq 0$. 
With this condition the above equation gives, $R=\psi$. Substituting of 
this result in the action (\ref{scalar_tensor_action01}) we can recover the 
original action (\ref{action}). Moreover, the quantum stability condition 
demands that $f''\! \left( R \right) \geq 0$. This along with the finiteness 
condition of the mass of the scalar field demands that $f''\! \left( R \right) > 0$. Thus, it is always possible to have a scalar tensor representation of 
$f(R)$ theory of gravity. 
Redefining the previously defined scalar field $\phi$ in terms of the new 
auxiliary field $\psi$ as
\begin{equation}
\phi = f_\psi\! \left( \psi \right),
\end{equation}
the action (\ref{scalar_tensor_action01}) can be rewritten as
\begin{equation} \label{scalar_tensor_action02}
S = \dfrac{1}{2\kappa^2} \int d^4 x \sqrt{-g}\, \Big[ \phi R - U\left(\phi \right) \Big] + \int d^4 x \sqrt{-g}\, \mathcal{L}_m\! \left[g_{\mu\nu}, \bar{\psi} \; \right],
\end{equation}
where $U\left( \phi \right) = f_\psi\! \left( \psi \right) \psi - f\! \left( \psi \right) = \phi\, \psi\! \left( \phi \right) - f\! \left(\psi \left( \phi \right) \right)$ is the potential of the scalar field. To be precise, this term $f_\psi\! \left( \psi \right) \psi - f\! \left( \psi \right)$ originates the scalar 
field. Unless and otherwise this term equals to zero, there exists a scalar 
field in the theory. In terms of the Ricci scalar, this term reads, 
\begin{equation} \label{potential_jordan}
V = U\! \left( \phi \right) = f'\!\left(R \right) R - f\! \left( R \right) \big|_{R\, =\, R_0}.
\end{equation}
For $f\!\left(R \right) = R$ and hence for $f'\!\left(R \right) = 1$, the 
action (\ref{scalar_tensor_action02}) recovers GR giving the potential term 
$V = 0$. Using Eq.\ (\ref{model}) in this Eq.\ (\ref{potential_jordan}), 
the scalar field potential for our model can be obtained as
\begin{equation} \label{model_V_jordan}
V = \frac{\alpha }{\pi }\, R_c \cot ^{-1}\!\left(\tfrac{R_c^2}{R^2}\right)+R \left[-\,\frac{2 \alpha  R_c^3}{\pi  R^3 \left(\frac{R_c^4}{R^4}+1\right)}-\beta  \exp\left(-\,\tfrac{R}{R_c}\right)+1\right]+\beta  R_c \left[1-\exp\left(-\tfrac{R}{R_c}\right)\right]-R \; \Big| _{R \,=\, R_0}.
\end{equation}
\begin{figure}[htb]
\centerline{
   \includegraphics[scale = 0.3]{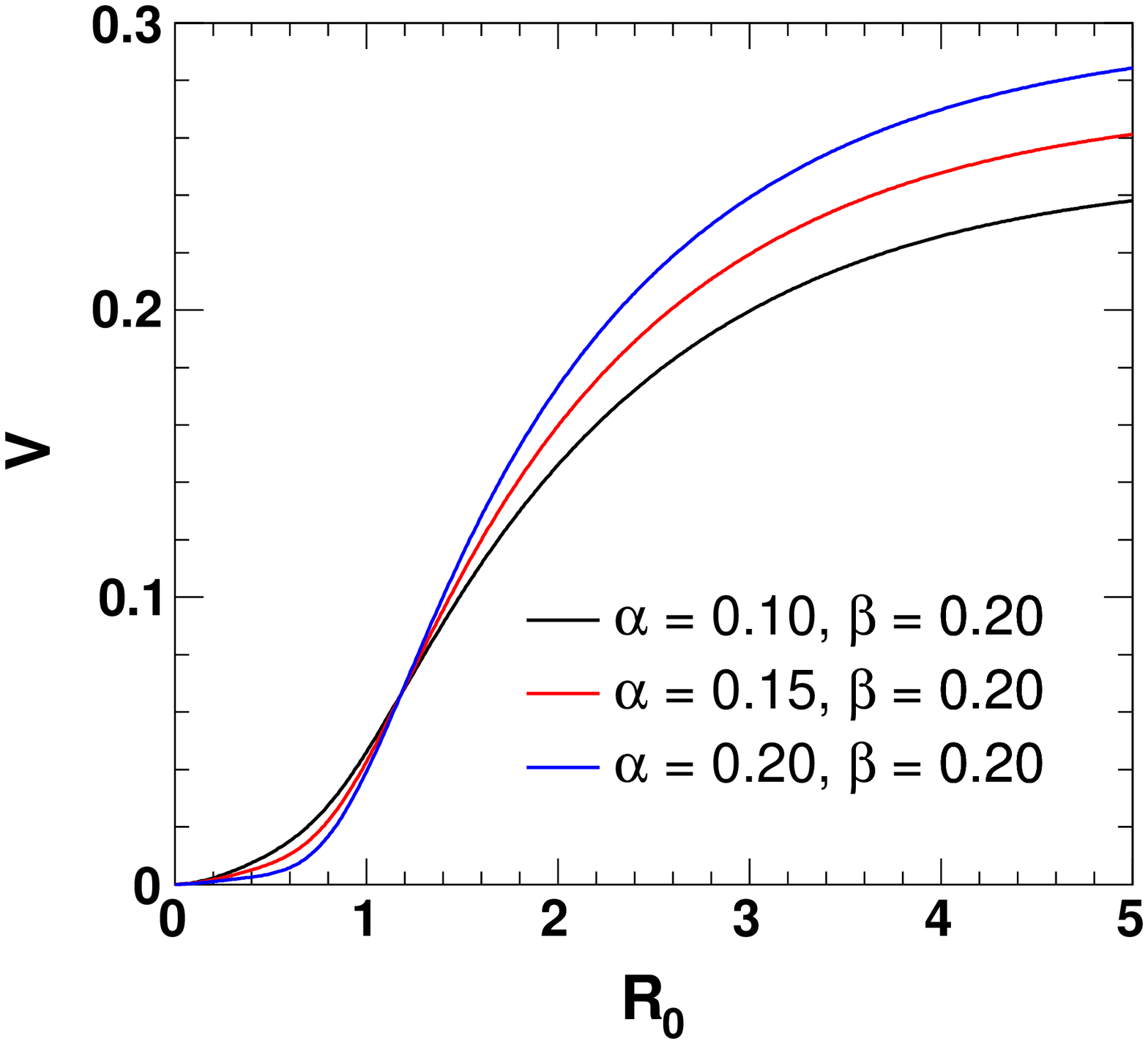}\hspace{1cm}
   \includegraphics[scale = 0.3]{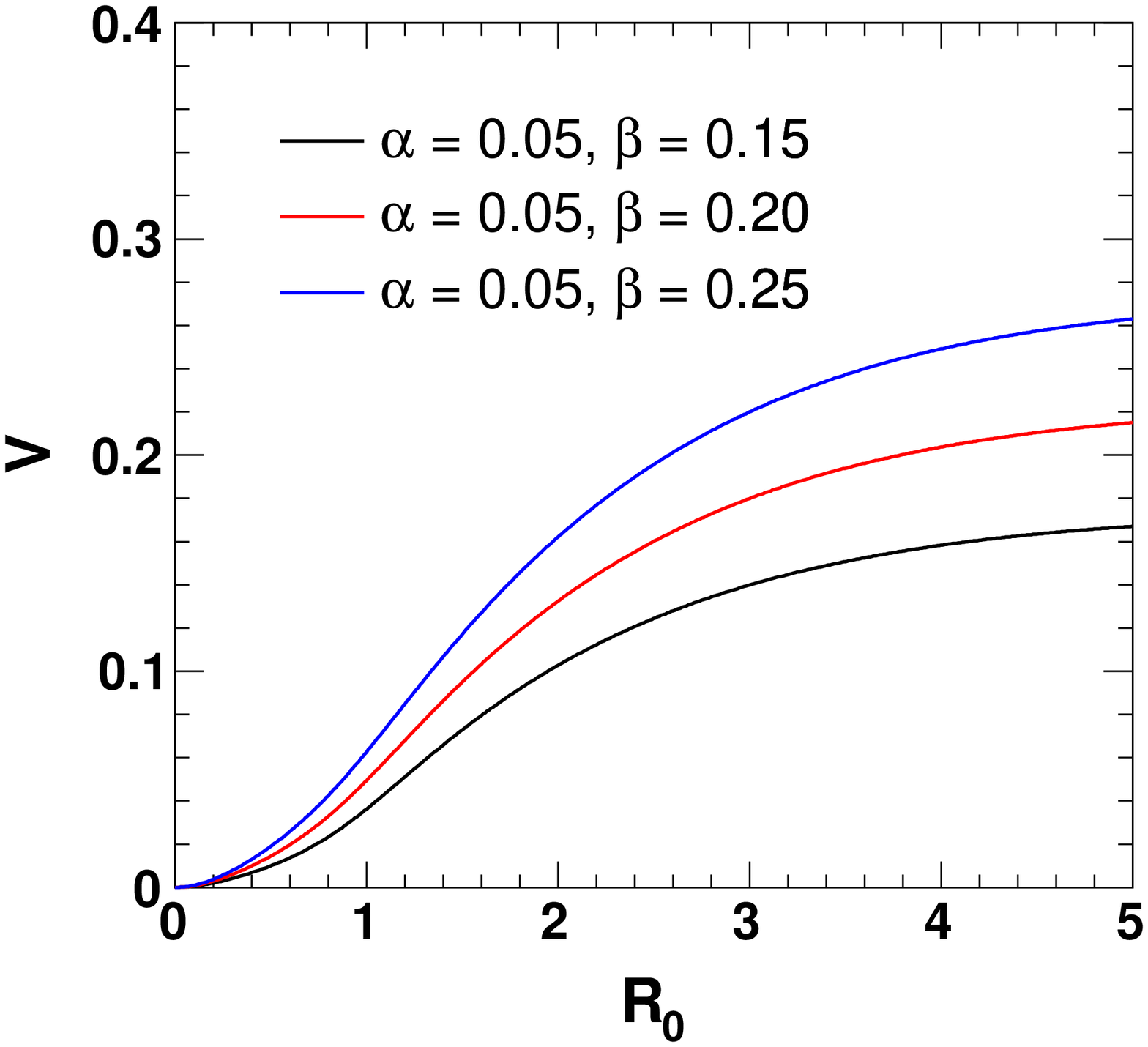}}
\vspace{-0.2cm}
\caption{Variation of the potential (\ref{model_V_jordan}) as a function of 
the background curvature $R_0$ for different sets of values of $\alpha$
and $\beta$ with $R_c = 1$ in arbitrary units.}
\label{fig02}
\end{figure}
The variation of the potential $V$ with respect to $R_0$ for different values 
of $\alpha$ and $\beta$ is shown in Fig.\ (\ref{fig02}) in arbitrary units 
considering $R_c = 1$. The figure shows that the potential in Jordan frame 
increases gradually with respect to the background curvature with some initial
deviations depending upon the values of $\alpha$ and $\beta$. In contrast
to the case of mass square of the scalar field, the potential shows some 
slight amount of dip, but near to the same small curvature region, which 
increases slowly when the value of $\alpha$ moves closer to the value of 
$\beta$. In fact, this dip in the potential is responsible for the hump in
the mass square curve for the corresponding values of $\alpha$ and $\beta$. 
This dip region of the potential curve almost eliminates in the case 
when $\alpha \ll \beta$. Moreover, with the increasing values of both $\alpha$ 
and $\beta$, the potential comparatively increases after the dip region or 
without the dip region. 
\subsection{Model in Einstein Frame}
In order see the behaviour of our model (\ref{model}) in Einstein frame, 
which is usually used to avoid the non - minimal coupling of gravity with the 
scalar field, we would like to study the model in this frame also. In the 
Einstein frame, the following conformal transformation of the metric is
performed \cite{Goswami2013, Chakraborty2019}:
$$\tilde{g}_{\mu\nu} = f'(R) g_{\mu\nu},$$
which for our model takes the form:
\begin{equation}
\tilde{g}_{\mu\nu} =  \left[ 1-\frac{2 \alpha  R_c^3}{\pi  R^3 \left(\frac{R_c^4}{R^4}+1\right)}-\beta  \exp\left(-\,\tfrac{R}{R_c}\right)\right]\! g_{\mu\nu}.
\end{equation}
Consequently, in this frame with $\mathcal{L}_m = 0$, the action changes to 
\cite{Goswami2013, Chakraborty2019} 
\begin{equation} \label{action_einstein}
S = \int d^4 x \sqrt{-g} \left[ \dfrac{1}{2\kappa^2} \tilde{R} - \dfrac{1}{2}\, \tilde{g}^{\mu\nu}\, \nabla_\mu \phi_E \nabla_\nu \phi_E - V(\phi_E) \right]\! ,
\end{equation}
where the scalar field
\begin{equation} \label{phi_R}
\phi_E = -\, \sqrt{\dfrac{3}{2}} \dfrac{1}{\kappa} \ln f'(R)
=  -\, \sqrt{\dfrac{3}{2}} \dfrac{1}{\kappa} \ln\! \left[1-\frac{2 \alpha  R_c^3}{\pi  R^3 \left(\frac{R_c^4}{R^4}+1\right)}-\beta \exp\left(-\,\tfrac{R}{R_c}\right) \right]
\end{equation}
and $V(\phi_E)$ is potential of the scalar field in this frame, and is given by,
\begin{equation}\label{pot_ef}
V(\phi_E) = \dfrac{1}{2\kappa^2} \dfrac{U}{f'(R)^2}
= \dfrac{1}{2\kappa^2} \dfrac{f'(R) R - f(R)}{f'(R)^2} \Big|_{R = R_0}.
\end{equation} 
The Eq.\ (\ref{phi_R}) shows the dependency of the Einstein frame scalar field 
$\phi_E$ on the scalar curvature $R$. Using our model, this potential 
(\ref{pot_ef}) can be expressed as
\begin{equation} \label{potential_einstein}
V(\phi_E) = \dfrac{1}{2\kappa^2}\, \frac{\pi\,  R_c\,\chi\, e^x \Big[-\pi\,  \beta\, \chi\left(x+1\right)+e^x\! \left(\pi  \beta +\pi  \beta\,  x^4-2\, \alpha\,  x^2\right)+\alpha\,\chi\,  e^x \cot ^{-1}\!\left(x^{-\,2}\right)\Big]}{\Big[e^x \left(\pi  x^4-2\, \alpha\,  x+\pi \right)-\pi  \beta\chi \Big]^2},
\end{equation}
where $x = R_0/R_c$ and $\chi = x^4+1$. Hence, the mass square term of the 
scalar field in the Einstein frame is
\begin{align} \label{mass_einstein}\notag
m^2(\phi_E) &= \dfrac{d^2 V(\phi_E)}{d \phi_E^2} 
= \dfrac{1}{3} \left[ \dfrac{1}{f''(R)} + \dfrac{R}{f'(R)} - \dfrac{4f(R)}{f'(R)^2} \right] _{R\, =\, R_0}\\[8pt]\notag
&= \frac{1}{3} R_c \left[\frac{\pi\chi^2  e^x}{2\, \alpha\,  e^x \left(3 x^4-1\right)+\pi  \beta \chi^2}+\frac{\pi \chi  e^x x}{e^x \left(\pi  x^4-2\, \alpha\,  x+\pi \right)-\pi  \beta \chi} \right]\\[10pt]
& \qquad \qquad \qquad \qquad \qquad \qquad \qquad -\frac{4 R_c \left[-\,\pi^{-1}\alpha  \cot ^{-1}\left(x^{-2}\right) + \beta  \left(e^{-x}-1\right)+x\right]}{3\left[2\, \alpha\,  x\left(\pi  x^4+\pi \right)^{-1}+\beta  e^{-x}-1\right]^2}.
\end{align}
In Minkowski space i.e.\ for $R_0 = 0$, this equation takes the form:
\begin{equation}
m^2(\phi_E) = \frac{\pi R_c}{3 (\pi  \beta -2\,\alpha )}.
\end{equation}

\begin{figure}[!htb]
\centerline{
   \includegraphics[scale = 0.3]{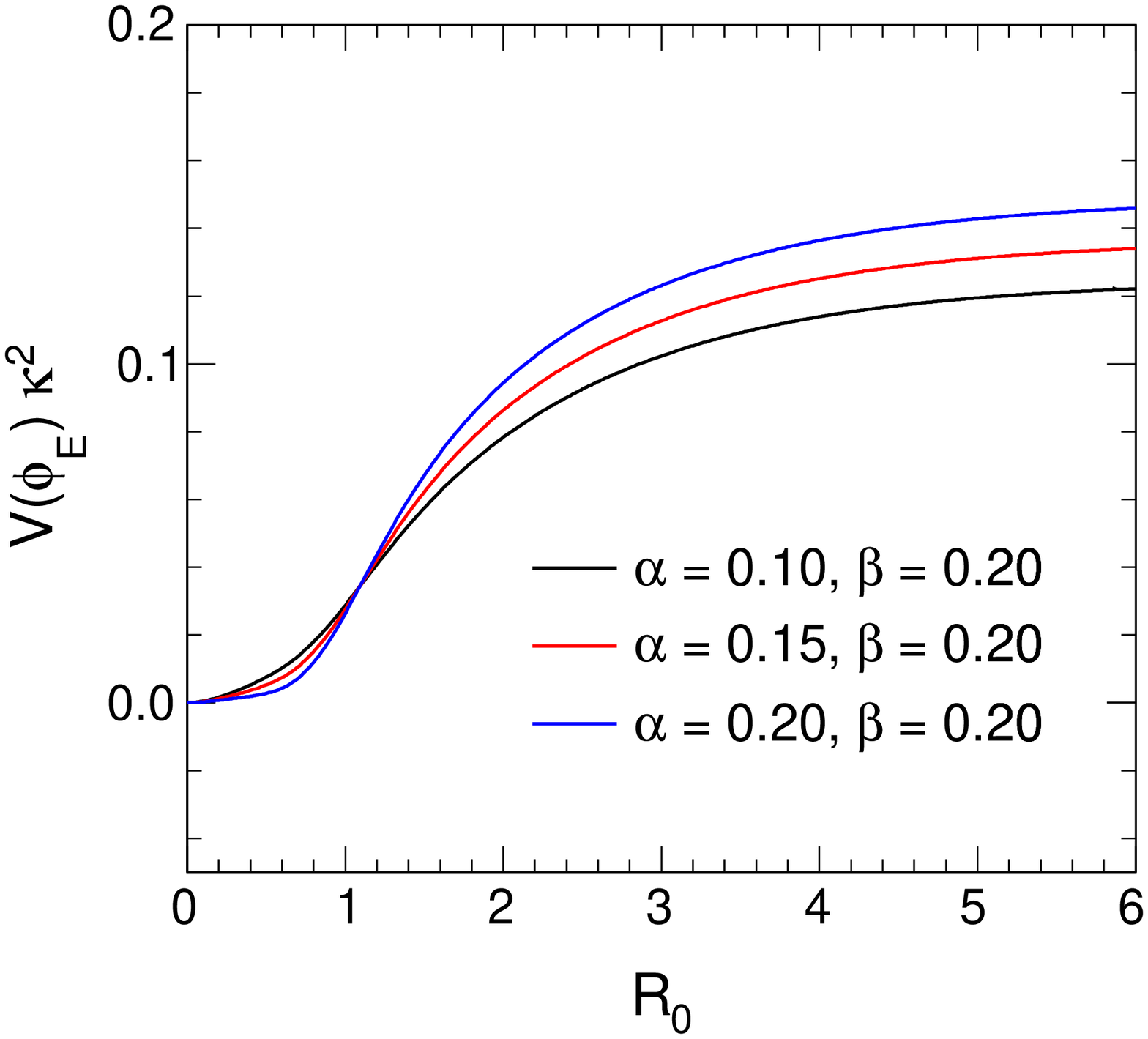}\hspace{1cm} 
   \includegraphics[scale = 0.3]{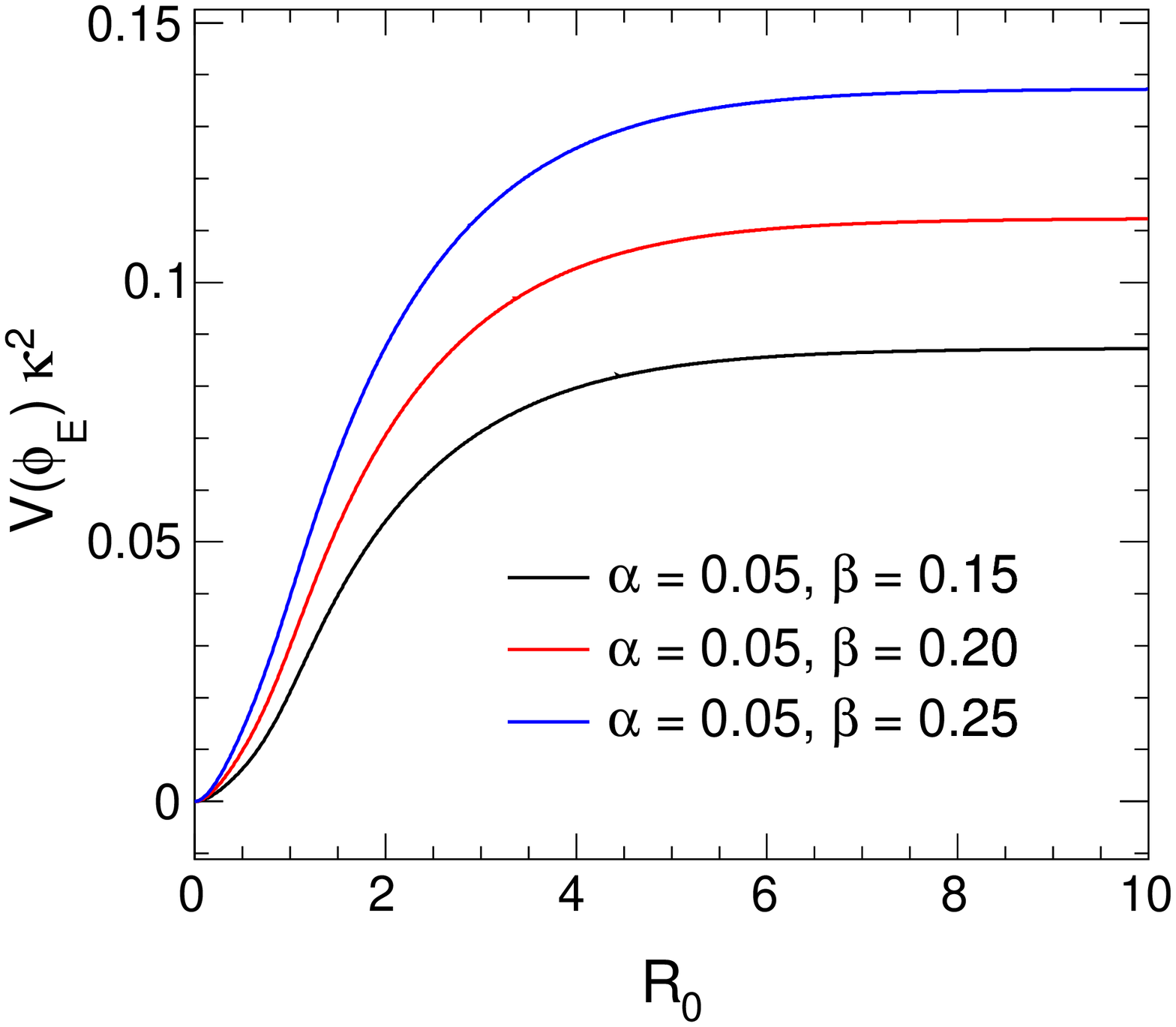}}
\vspace{-0.2cm} 
\caption{Variation of scalar field potential as a function of background 
curvature in Einstein frame for different values of $\alpha$ and $\beta$ 
parameters with $R_c = 1$ in arbitrary units.}
\label{fig03}
\end{figure}
\begin{figure}[!htb]
\centerline{
   \includegraphics[scale = 0.3]{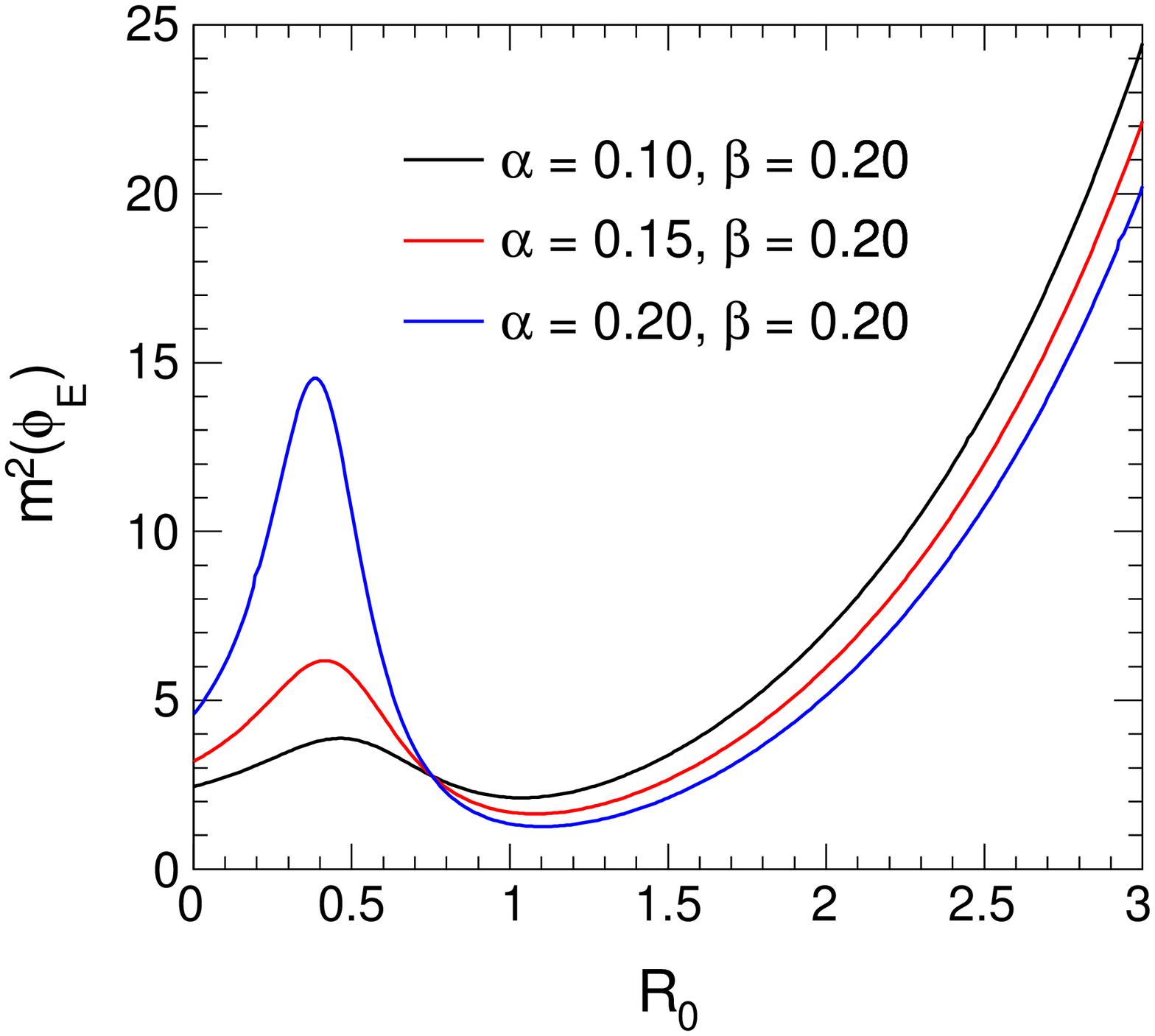}\hspace{1cm}
   \includegraphics[scale = 0.3]{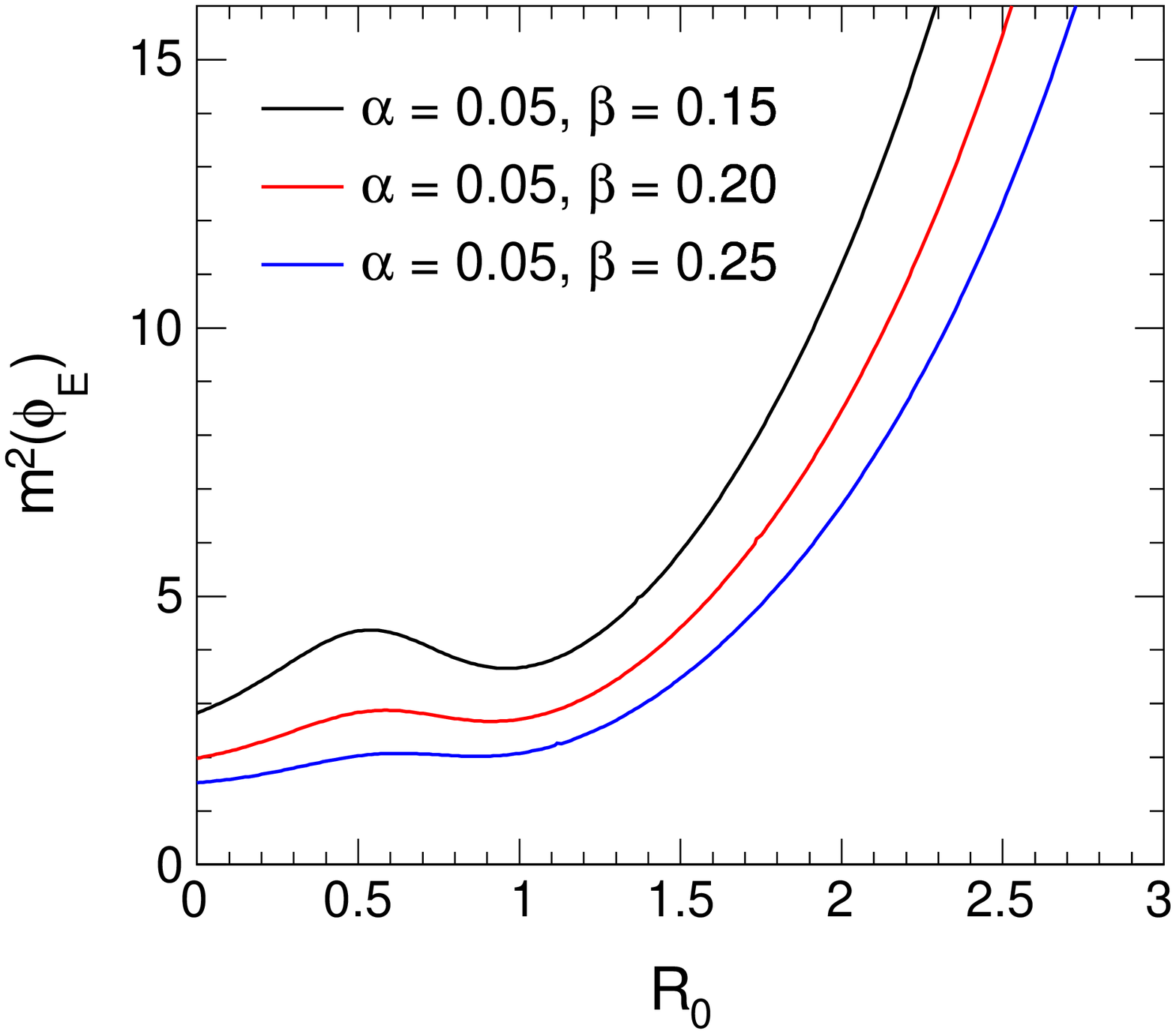}}
\vspace{-0.2cm}
\caption{Variation of scalaron mass square as a function of background 
curvature in Einstein frame for different values of $\alpha$ and $\beta$ 
parameters with  $R_c = 1$ in arbitrary units.}
\label{fig04}
\end{figure}
\begin{figure}[!htb]
\centerline{
   \includegraphics[scale = 0.3]{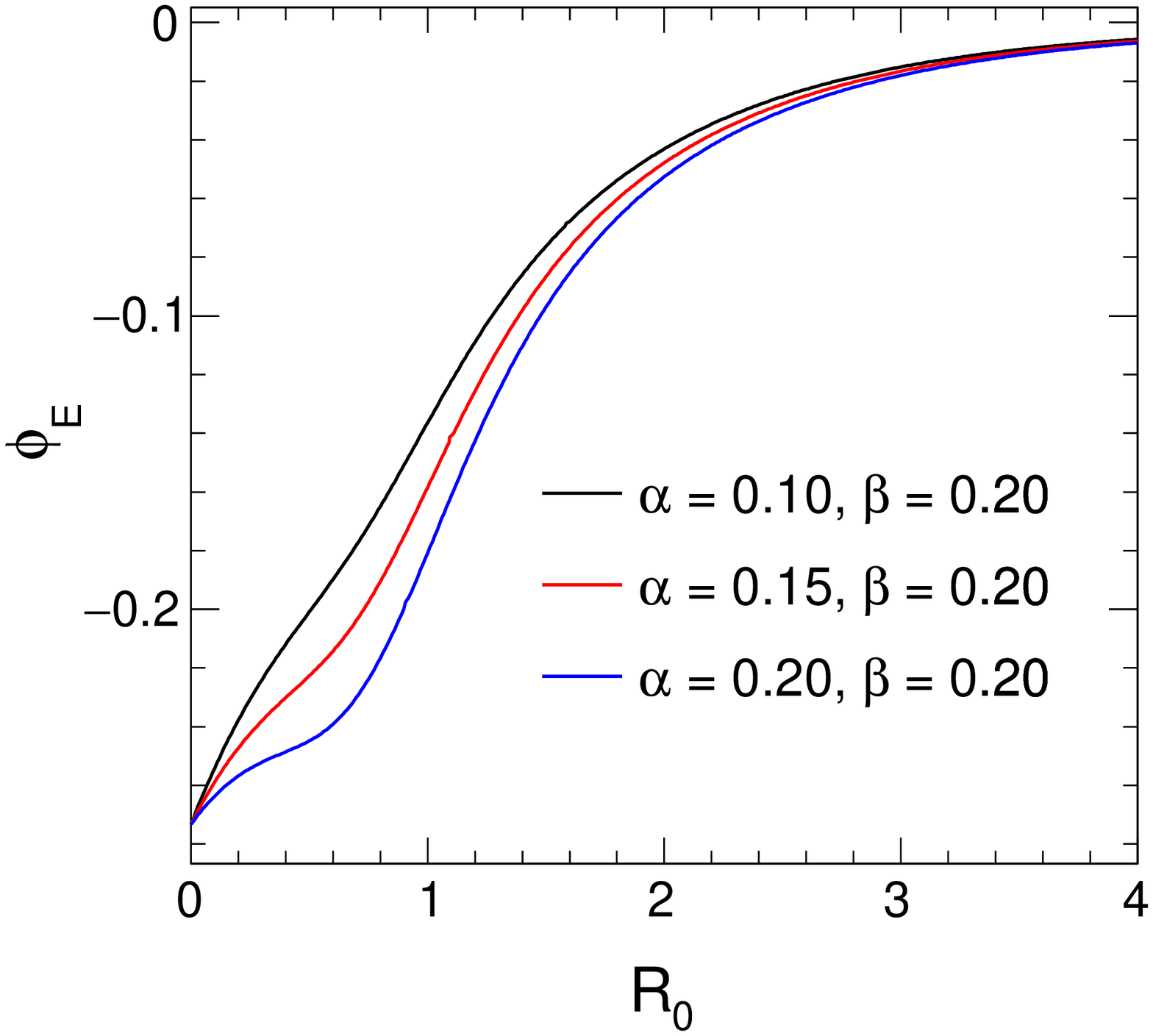}\hspace{1cm} 
   \includegraphics[scale = 0.3]{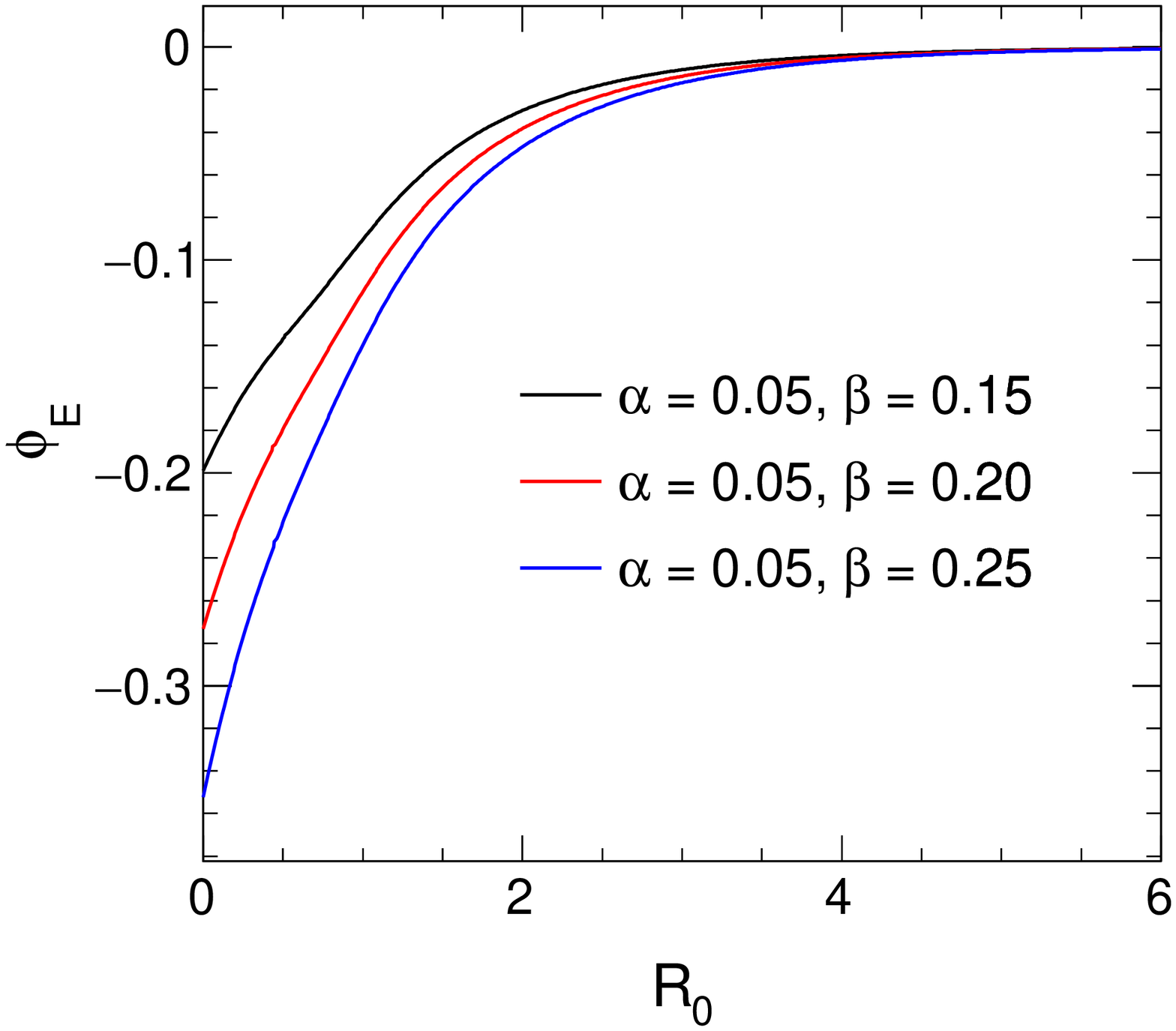}}
\vspace{-0.3cm}
\caption{Variation of scalar field $\phi_E$ as a function of scalar 
curvature in Einstein frame for different values of $\alpha$ and $\beta$ 
parameters with $R_c = 1$ in arbitrary units.}
\label{fig05}
\end{figure}
\begin{figure}[!htb]
\centerline{
   \includegraphics[scale = 0.3]{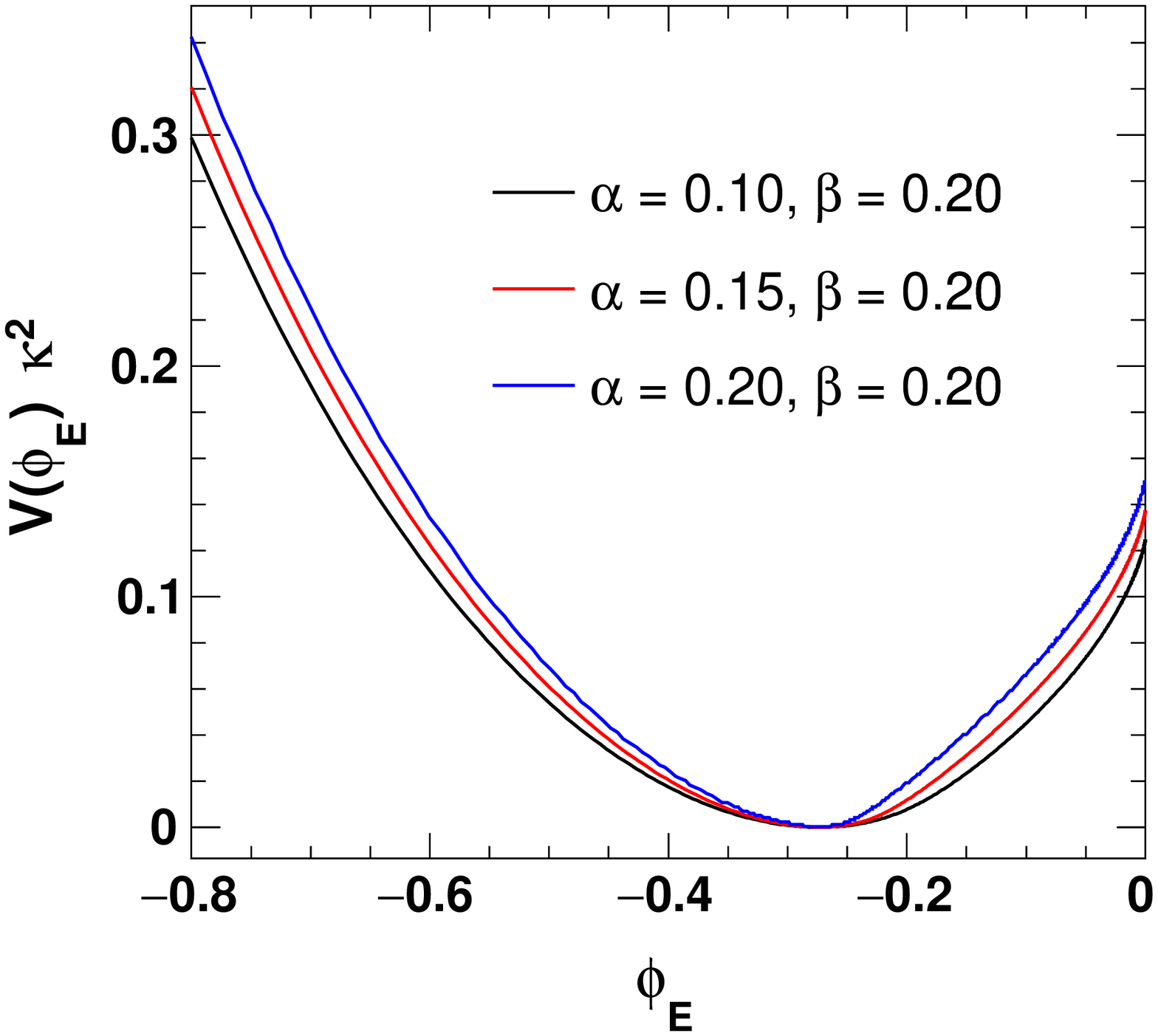}\hspace{1cm} 
   \includegraphics[scale = 0.3]{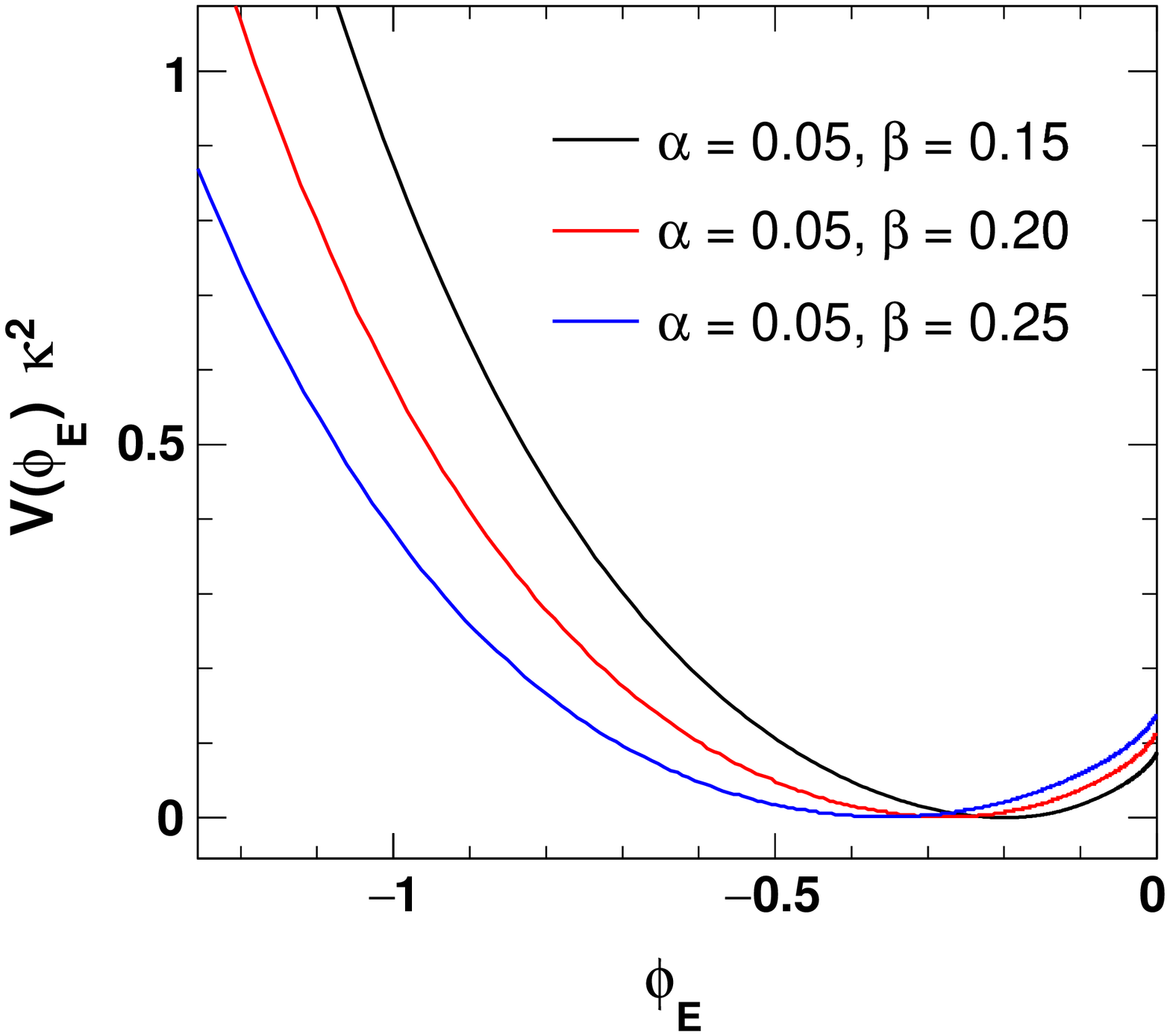}} 
\vspace{-0.2cm}
\caption{Variation of scalar field potential as a function of field $\phi_E$ 
in Einstein frame for different values of $\alpha$ and $\beta$ 
parameters with $R_c = 1$ in arbitrary units.}
\label{fig06}
\end{figure}



From Eq.s (\ref{potential_einstein}) and (\ref{mass_einstein}), we see 
that although the expressions for the scalar field potential and scalaron mass 
square in Einstein frame are little bit complicated in comparison
to that in the Jordan frame, their variations as a function of $R_0$ are found 
to be almost similar as seen from Fig.\ \ref{fig03} and Fig.\ \ref{fig04} 
respectively. That is, the behaviours of potential and the mass term of the 
scalar field are almost identical in both frames. In Fig.~\ref{fig05}, the 
variation of field $\phi_E$ as a function of background curvature is shown. 
The field $\phi_E$ at $R_0=0$ and $R_0 \rightarrow \infty$ is independent of 
$\alpha$, whereas it is independent of $\beta$ only at $R_0 \rightarrow \infty$.
Moreover, $\phi_E$ is non-zero at $R_0=0$ and tends to zero at 
$R_0 \rightarrow \infty$, which is obvious from it's expression. Again from 
Eq.s (\ref{fieldm}) and (\ref{phi_R}) it is clear that all these behaviours of 
$\phi_E$ should be applicable to $\phi$ also, but with positive values of $\phi$
for all values of $R_0$. Variation of the potential (\ref{potential_einstein}) 
as a function of $\phi_E$ is shown in Fig.~\ref{fig06}. It is seen from this 
figure that the minimum of the potential moves towards the higher value of 
$\phi_E$ when $\alpha\ll \beta$ than the case when $\alpha\sim \beta$. 
Obviously, similar behaviour can be attributed to the potential 
(\ref{model_V_jordan}) as a function of the field $\phi$.  Thus, because of 
the similarity of behaviours of the scalar field, it's potential and mass 
square term in both Jordan and Einstein frames, the rest of the study in this 
paper is done in the Jordan frame only.

\subsection{Solar System Tests of the Model}
It is possible to recover GR by introducing the Chameleon mechanism in the 
theory. In this mechanism, the scalar field $\phi = f'(R)$ is coupled with the 
matter density of the environment. Thus, when a model is used inside the solar 
system, due to presence of matter density, the scalar field coupled with it 
gains mass and hence allows the model to pass the solar system tests. 
Clearly, this mechanism implies that the functional form $f(R)$ should 
have a very closer value to the Ricci scalar $R$, for $R$ above or equal to 
the solar system scale. A model is considered viable and consistent if it 
passes the solar system tests. Guo has introduced several methods to test 
whether an $f(R)$ gravity model passes the solar system tests or not in Jordan 
frame \cite{Guo_2014}. According to Guo, a model can pass solar system tests 
if it satisfies the following conditions:

\begin{equation}\label{sstest01}
\left|\, \dfrac{f(R) - R}{R}\, \right| \ll 1,
\end{equation}
\begin{equation}\label{sstest02}
\left|\, f'(R) - 1 \, \right| \ll 1,
\end{equation}
\begin{equation}\label{sstest03}
R f''(R) \ll 1.
\end{equation}

\begin{figure}[!htb]
\centerline{
   \includegraphics[scale = 0.25]{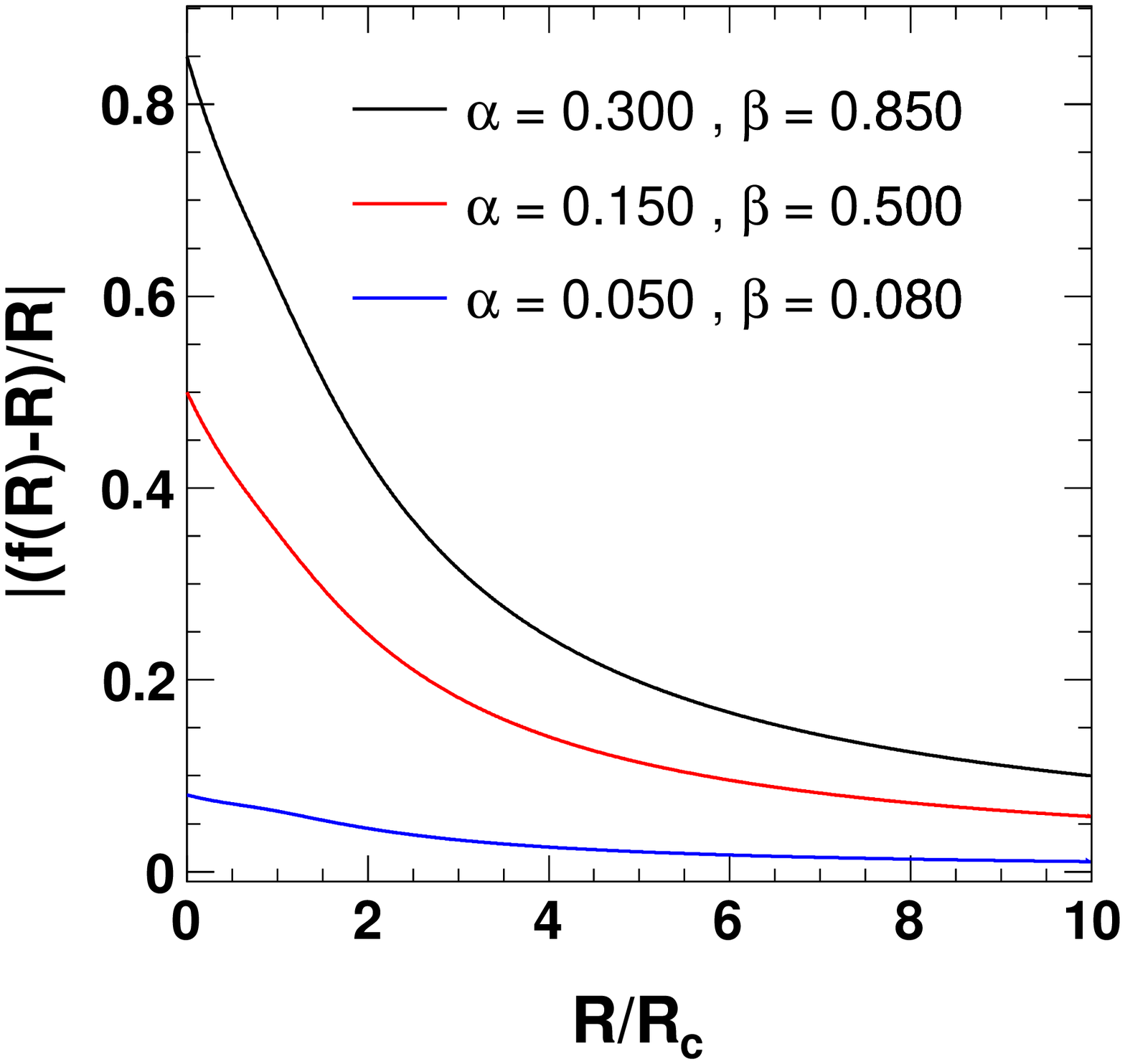}\hspace{0.8cm} 
   \includegraphics[scale = 0.25]{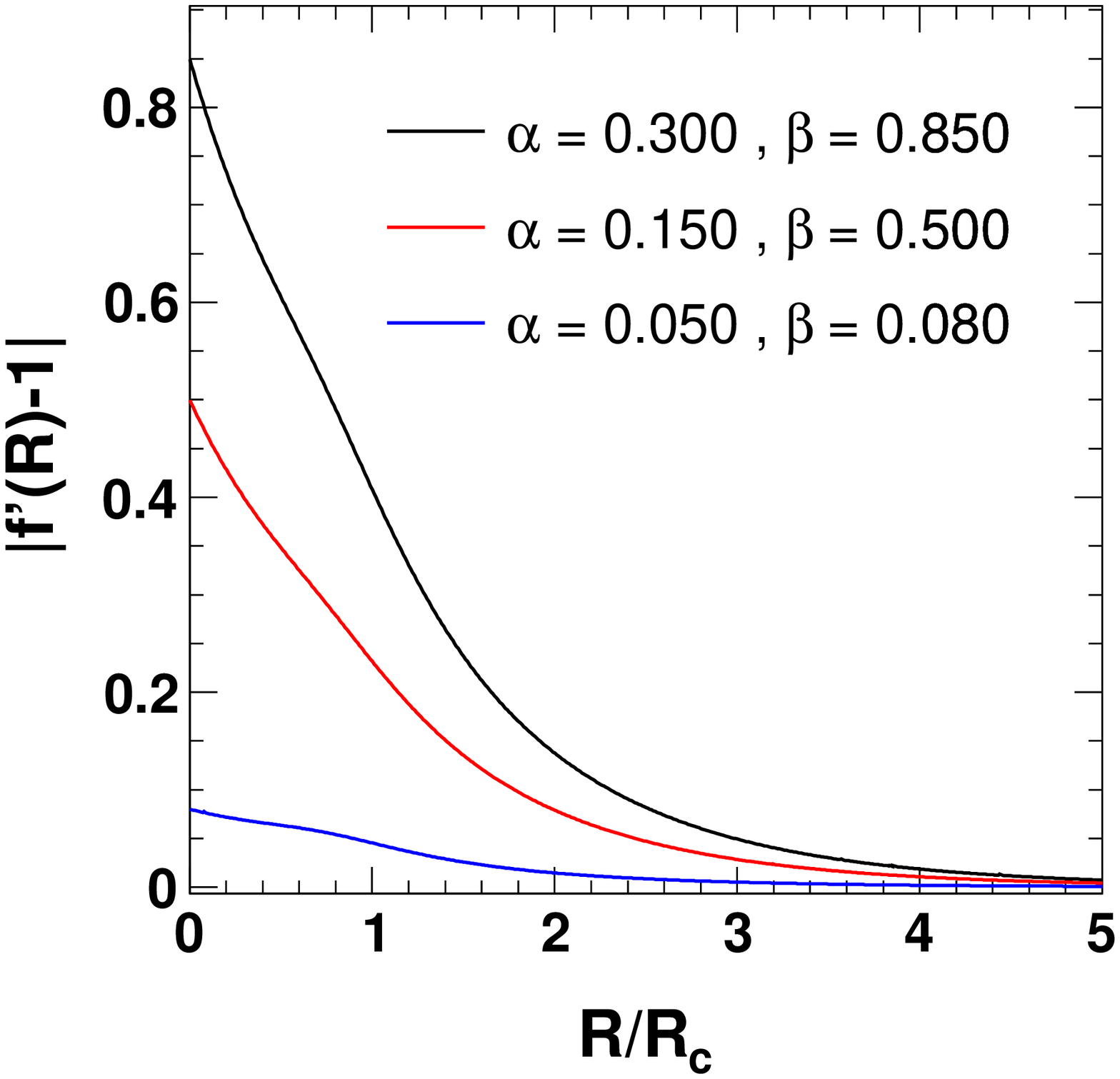}\hspace{0.8cm} 
   \includegraphics[scale = 0.25]{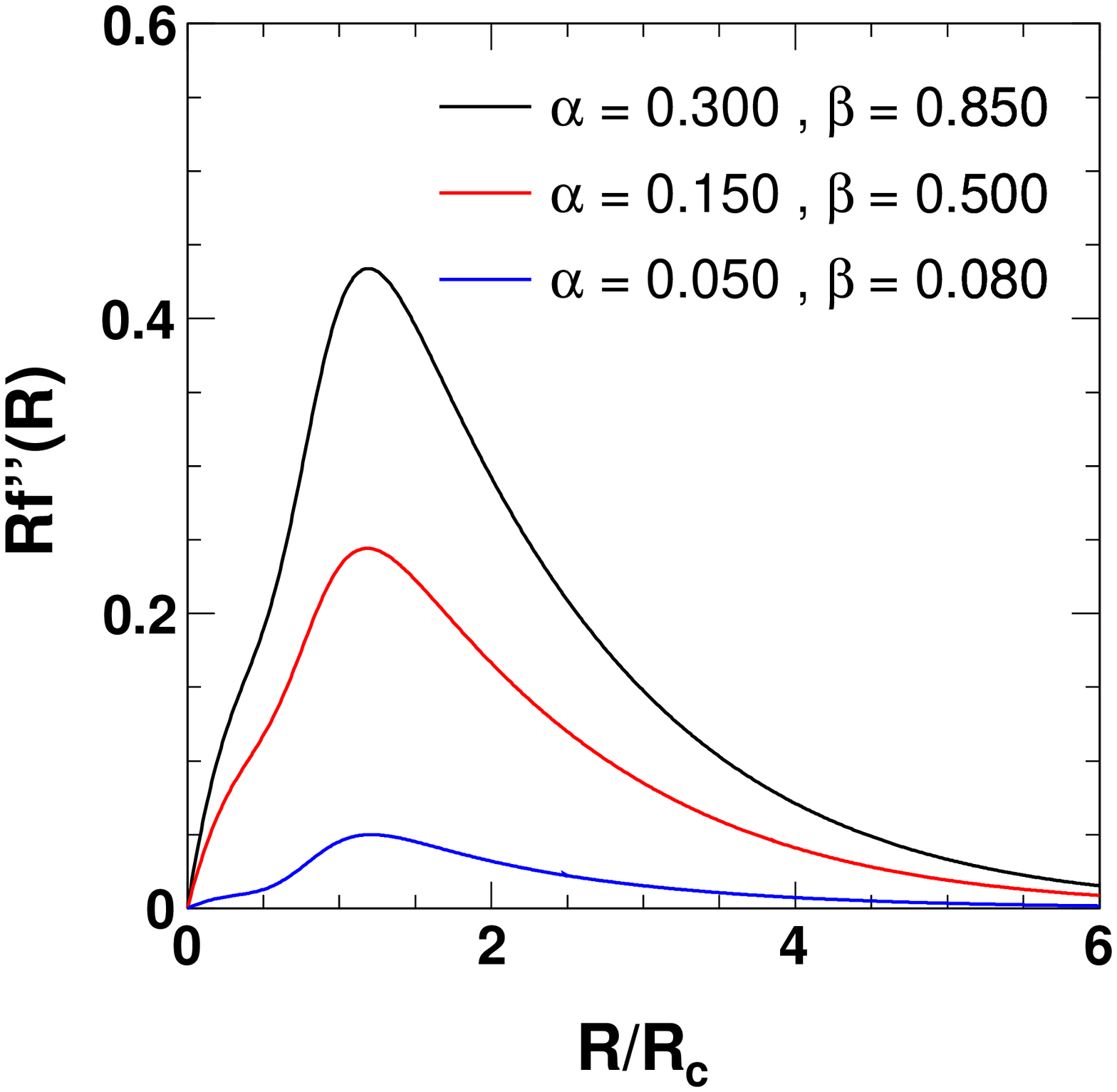}}
\vspace{-0.2cm}
   \caption{Plots of solar system test functions with respect to $R_0/R_c$.}
\label{sstest_fig}
\end{figure}

We've calculated the above functions numerically for our model (see Table
\ref{table01}). These functions are plotted against background curvature in
the units of $R_c$ for different parameters (see Fig.\ \ref{sstest_fig}). These indicate that the model can be made to pass the solar system tests by 
increasing the ratio $R_0/R_c$ or by simply decreasing the parameter 
$R_c$. However, a simple and effective way to make the model solar system 
viable is to decrease all the parameters (i.e. $\alpha, \beta$ and $R_c$) 
sufficiently (see Table \ref{table01}). Thus, within a viable range of 
parameters, the model can easily pass the solar system tests. This is another 
advantage of this model, which allows us to enlist the model as a solar system 
viable model.
\begin{center}

\begin{table}[h]
\caption{Numerical values of solar system test functions for specific values of $R_0/R_c, \alpha$ and $\beta$.}\vspace{0.3cm}
\begin{tabular}{|c|c|c|c|}
\hline 
\rule[-3ex]{0pt}{7ex} Values of Model parameters  & \hspace{0.5cm} $\left | \dfrac{f(R) - R}{R} \right| $ \hspace{0.5cm} & \hspace{0.5cm} $\left | f'(R) - 1 \right| $ \hspace{0.5cm} & \hspace{0.5cm} $R f''(R)$ \hspace{0.5cm} \\ 
\hline 
\rule[-1.3ex]{0pt}{3.5ex} $R_0/R_c = 0.00148 , \alpha = 0.1500 , \beta = 0.500 $  & 0.49970 & 0.49940 & 0.00060 \\ 
\hline 
\rule[-1.3ex]{0pt}{3.5ex} $R_0/R_c = 0.05000 , \alpha = 0.0450 , \beta = 0.500 $  & 0.48842 & 0.47705 & 0.02235 \\ 
\hline 
\rule[-1.3ex]{0pt}{3.5ex} $R_0/R_c = 0.50000 , \alpha = 0.1500 , \beta = 0.500 $  & 0.41686 & 0.34820 & 0.11727 \\ 
\hline
\rule[-1.3ex]{0pt}{3.5ex} $R_0/R_c = 1.00000 , \alpha = 0.0500 , \beta = 0.080 $  & 0.06307 & 0.04534 & 0.04534 \\ 
\hline
\rule[-1.3ex]{0pt}{3.5ex} $R_0/R_c = 1.50000 , \alpha = 0.0050 , \beta = 0.044 $  & 0.02401 & 0.01061 & 0.01657 \\ 
\hline  
\rule[-1.3ex]{0pt}{3.5ex} $R_0/R_c = 2.00000 , \alpha = 0.0005 , \beta = 0.008 $  & 0.00356 & 0.00112 & 0.00226 \\ 
\hline 
\end{tabular}
\label{table01}
\end{table} 
\end{center}

\section{Comparison of the Model with other viable Models} \label{sec3_comparison}
\begin{figure}[htb]
\centerline{
   \includegraphics[scale = 0.8]{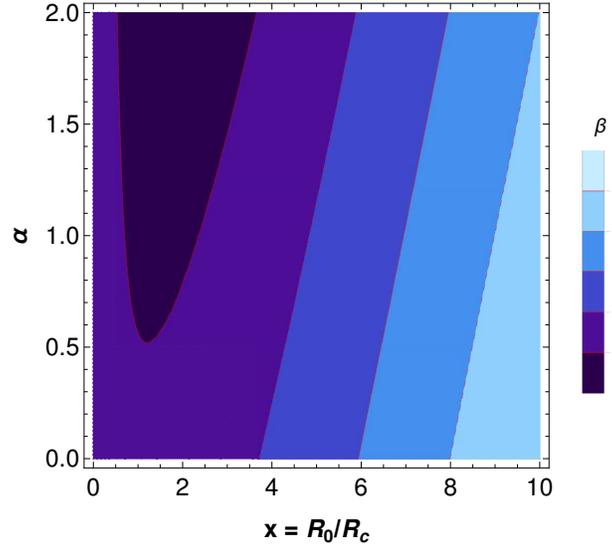}}
\vspace{-0.2cm} 
\caption{$\beta$ as a function of $x$ and $\alpha$. The contour shows the de Sitter solutions of the model.}
\label{fig_contour01}
\end{figure}

\begin{figure}[htb]
\centerline{
   \includegraphics[scale = 0.7]{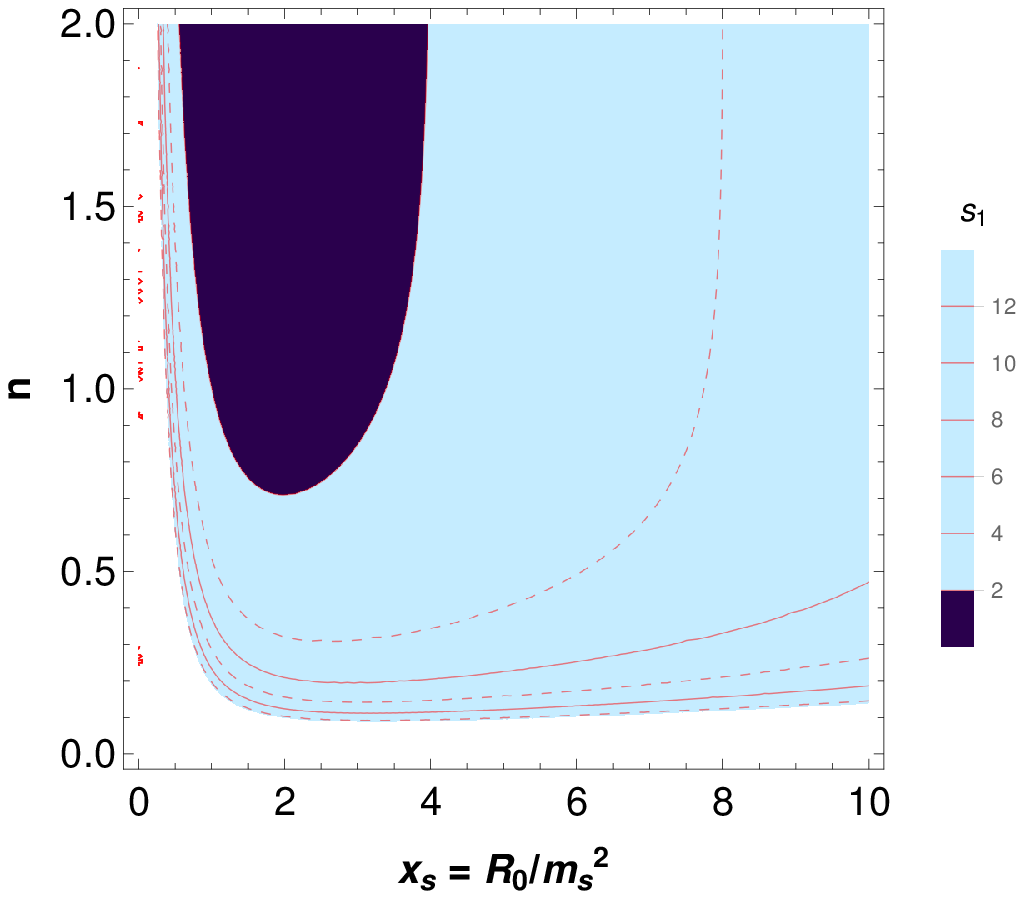}\hspace{1cm}
   \includegraphics[scale = 0.7]{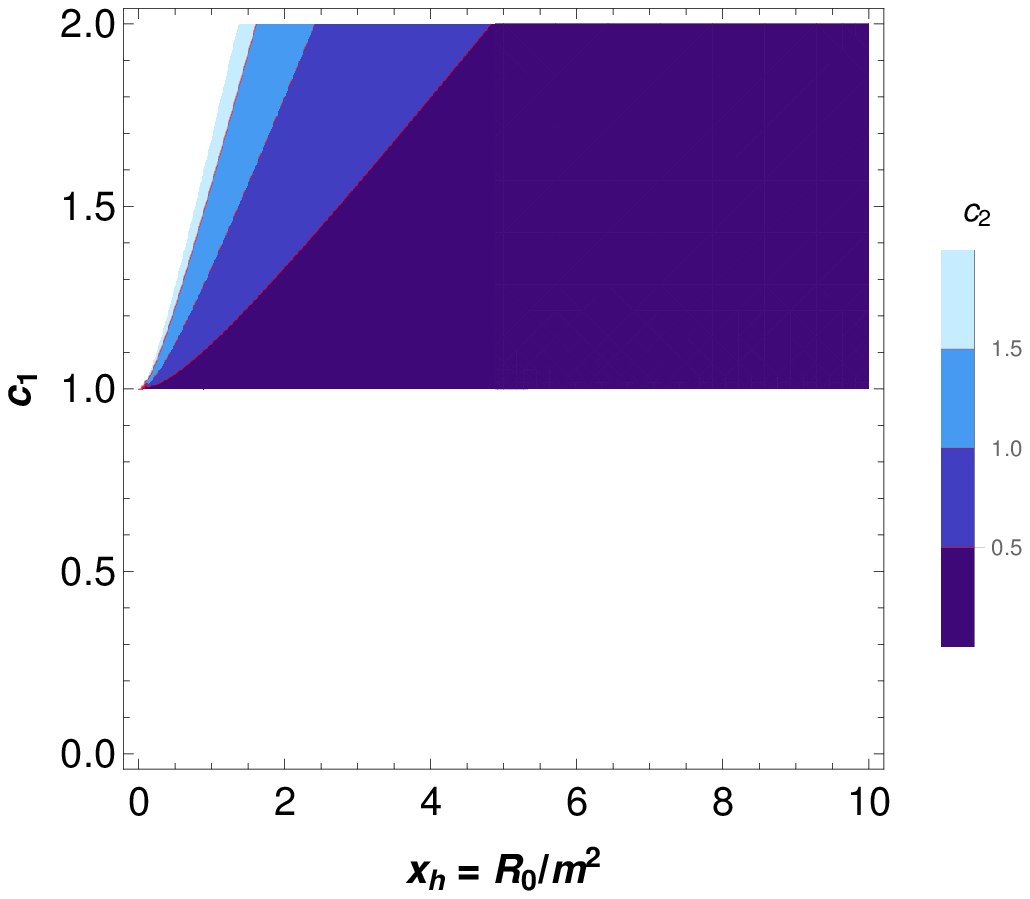}}
\vspace{-0.2cm} 
\caption{Figure on left side shows the de Sitter solutions of the Starobinsky 
model and figure on right side shows the de Sitter solutions for the Hu-Sawicki
 model.}
\label{fig_deSitter_models}
\end{figure}

In this section, we would like to compare the model with two other viable 
models viz., the Starobinsky model and the Hu- Sawicki model in terms of 
stability and nature of the models in local regime. At first, we would like 
study the de Sitter stability of our toy model. For our model the 
Eq.\ \eqref{de_sitter_eq} takes the form:
$$ \frac{2\, \alpha\,  x^2}{x^4+1}+\pi  \left[ \beta  e^{-x} (x+2)+x -2 \beta\right]=2 \alpha  \cot ^{-1}\!\left(x^{-\,2}\right)\!,$$ 
where $x = R_0/R_c$ and $R_c \neq 0$. 
On solving this equation for $\beta$, we get,
\begin{equation} \label{beta_value}
\beta = -\,\frac{e^x \left(-\,\pi  x^5-2 \alpha  x^2+2 \alpha  \cot ^{-1}\left(\frac{1}{x^2}\right)+2 \alpha  x^4 \cot ^{-1}\left(\frac{1}{x^2}\right)-\pi  x\right)}{\pi  \left(-\,x+2 e^x-2\right) \left(x^4+1\right)}.
\end{equation}
The contour plot of $\beta$ as a function of $x$ and $\alpha$ is shown in 
Fig.\ \ref{fig_contour01}.
The Starobinsky model and the Hu-Sawicki model are defined respectively as
\begin{equation}
f_S(R) = R-s_1 m_s^2 \left[1-\left(\frac{R^2}{m_s^4}+1\right)^{-\,n}\right],
\end{equation}
and
\begin{equation}
f_H(R) = R-\frac{m^2 c_1 \left(\frac{R}{m^2}\right)^\mu}{c_2 \left(\frac{R}{m^2}\right)^\mu+1}.
\end{equation}
The de Sitter solution for the Starobinsky model is
\begin{equation} \label{starobinsky_de_sitter}
s_1= \frac{x_s \left(x_s^2+1\right)^{n+1}}{2 \left(x_s^2 \left(x_s^2+1\right)^n+\left(x_s^2+1\right)^n-n x_s^2-x_s^2-1\right)},
\end{equation}
where $x_s = R_0/m_s^2$. Similarly, for the Hu-Sawicki model, using $\mu = 1$ 
the de Sitter solution is found to be,
\begin{equation} \label{hu_sawicki_de_sitter}
c_2 = \frac{c_1+\sqrt{(c_1-1) c_1}-1}{x_h}.
\end{equation}
The contour plots of the parameters $s_1$ and $c_2$ respectively for the 
Starobinsky model and the Hu-Sawicki model are shown in Fig.\ 
\ref{fig_deSitter_models}. From the contours it can be seen that for very 
small values of $n$ and $x_s$ in the Starobinsky model, we do not have de 
Sitter solutions. The parameter $s_1$ rises very rapidly for the increase of 
the parameters $n$ and $x_s$. For the Hu-Sawicki model, we see that, there is 
no de Sitter solutions present for $c_1<1$ and the parameter $c_2$ rises very 
slowly for increments of $c_1$ beyond $1$. On the other hand, for our toy 
model, we see that $\alpha$ and $x$ have a wider region of de Sitter 
solutions. The other parameter $\beta$ has a higher value than $\alpha$ in the 
de Sitter solution space towards higher values of $x$.

\begin{figure}[htb]
\centerline{
   \includegraphics[scale = 0.5]{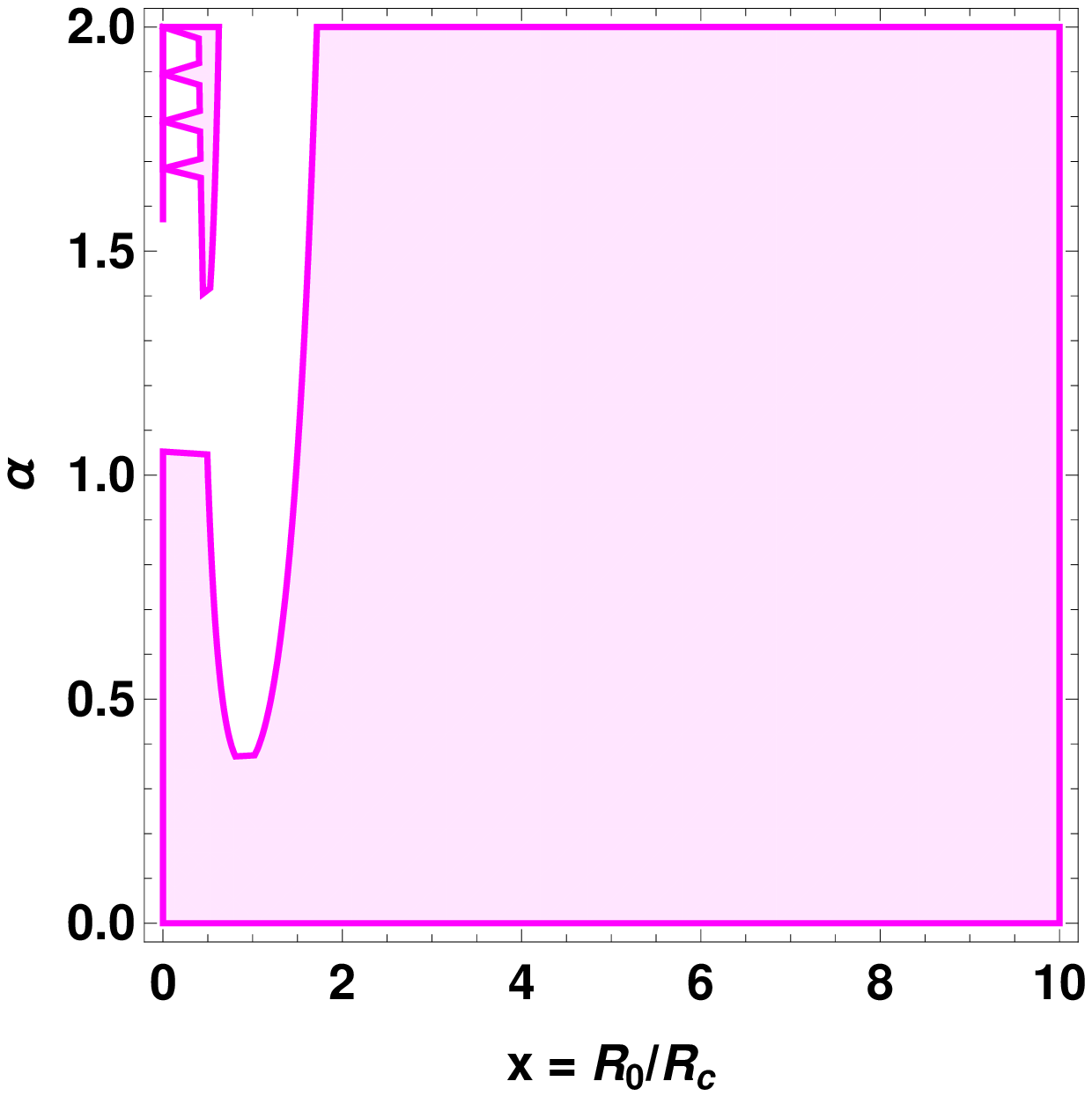}\hspace{1cm}
   \includegraphics[scale = 0.5]{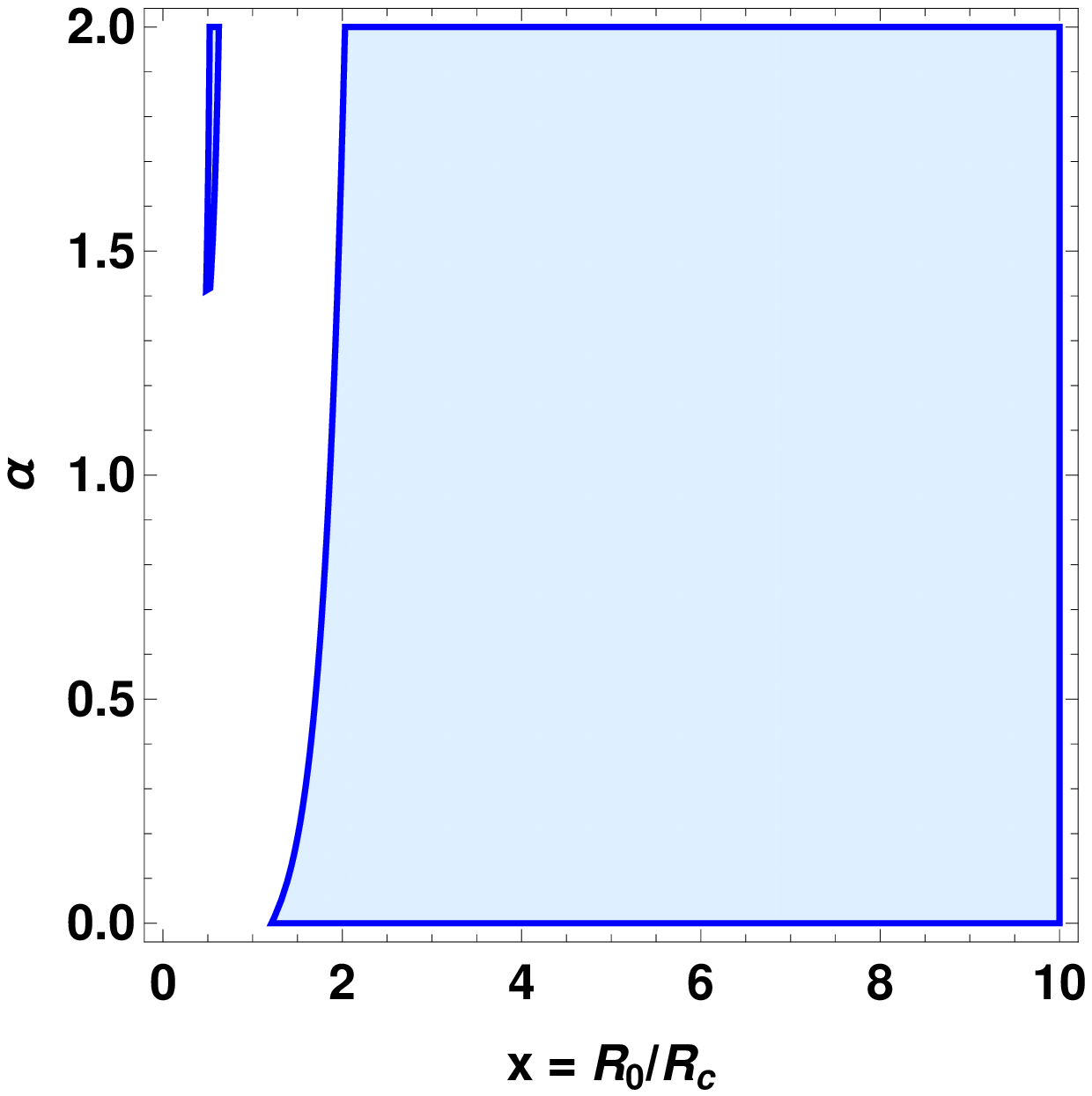}}
\vspace{-0.2cm} 
\caption{Stability region of the model \eqref{model}. Figure on the left side 
shows the stability region allowed by Eq.\ \eqref{stability01} and figure on 
the right side shows the stability region allowed by Eq.\ \eqref{stability02}.}
\label{fig_stability01}
\end{figure}

\begin{figure}[htb]
\centerline{
   \includegraphics[scale = 0.5]{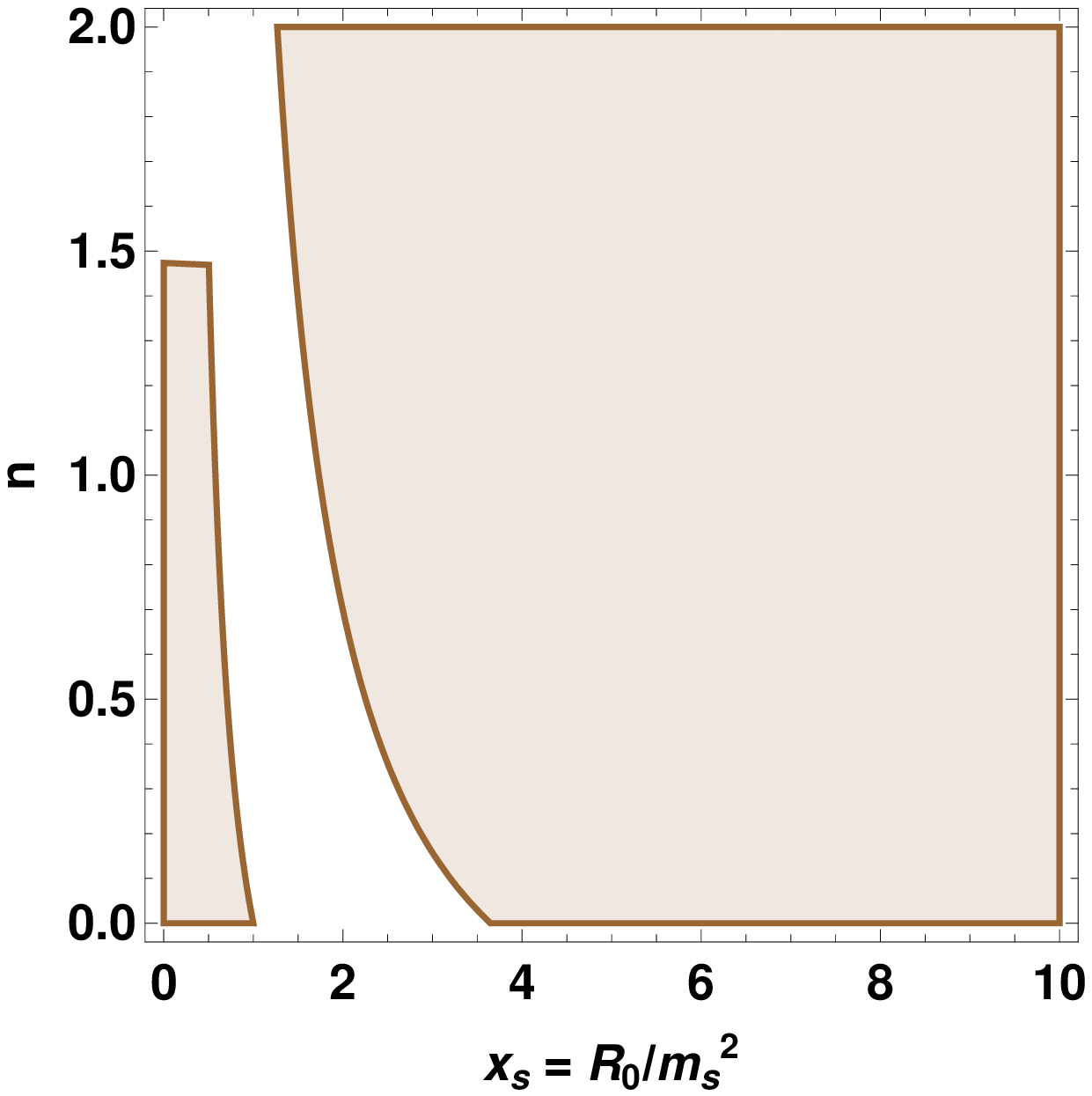}\hspace{1cm}
   \includegraphics[scale = 0.5]{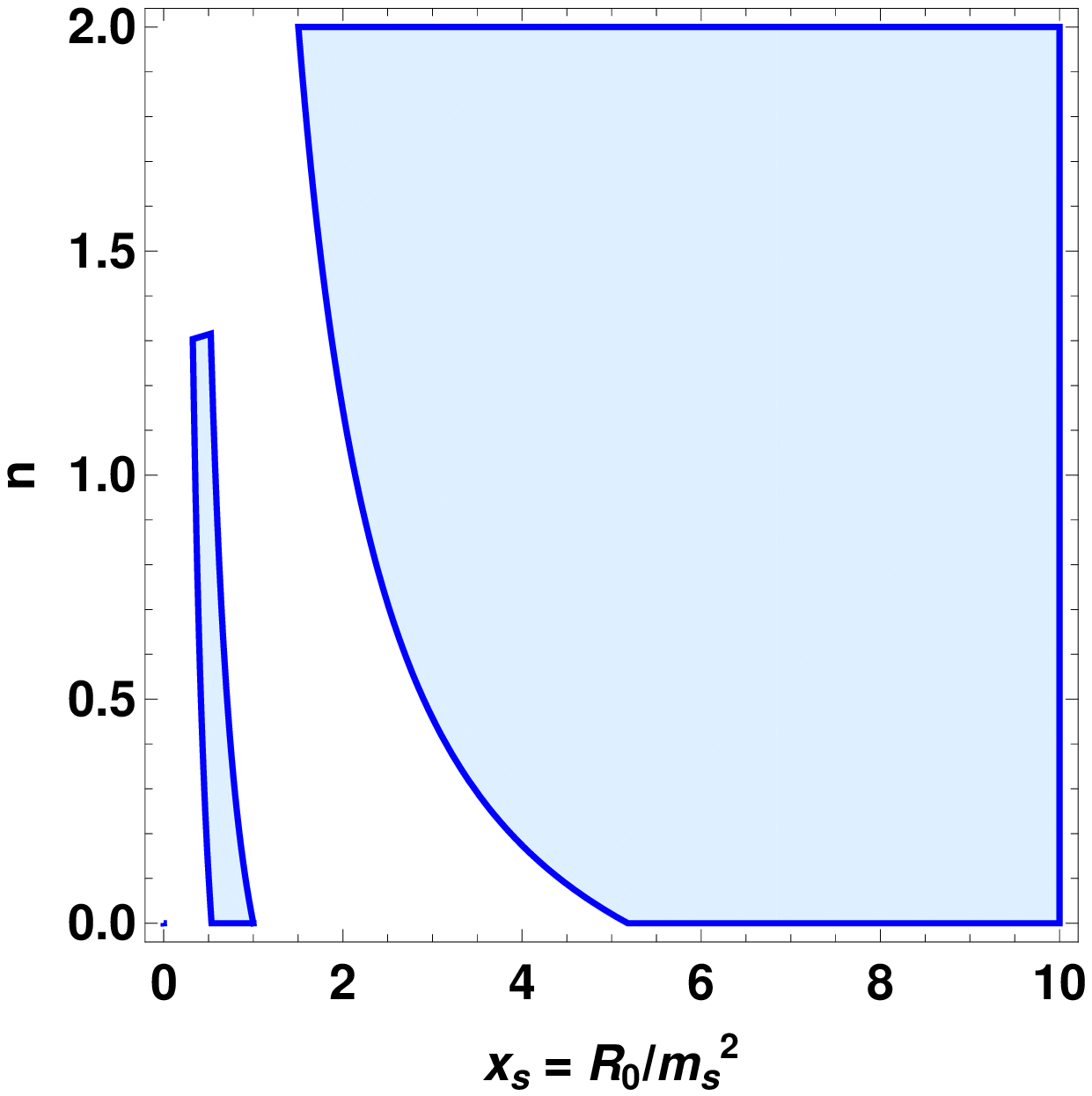}}
\vspace{-0.2cm} 
\caption{Stability region of the Starobinsky model. Figure on the left side 
shows the stability region allowed by Eq.\ \eqref{stability03} and figure on 
the right side shows the stability region allowed by Eq.\ \eqref{stability04}.}
\label{fig_stability02}
\end{figure}

\begin{figure}[htb]
\centerline{
   \includegraphics[scale = 0.5]{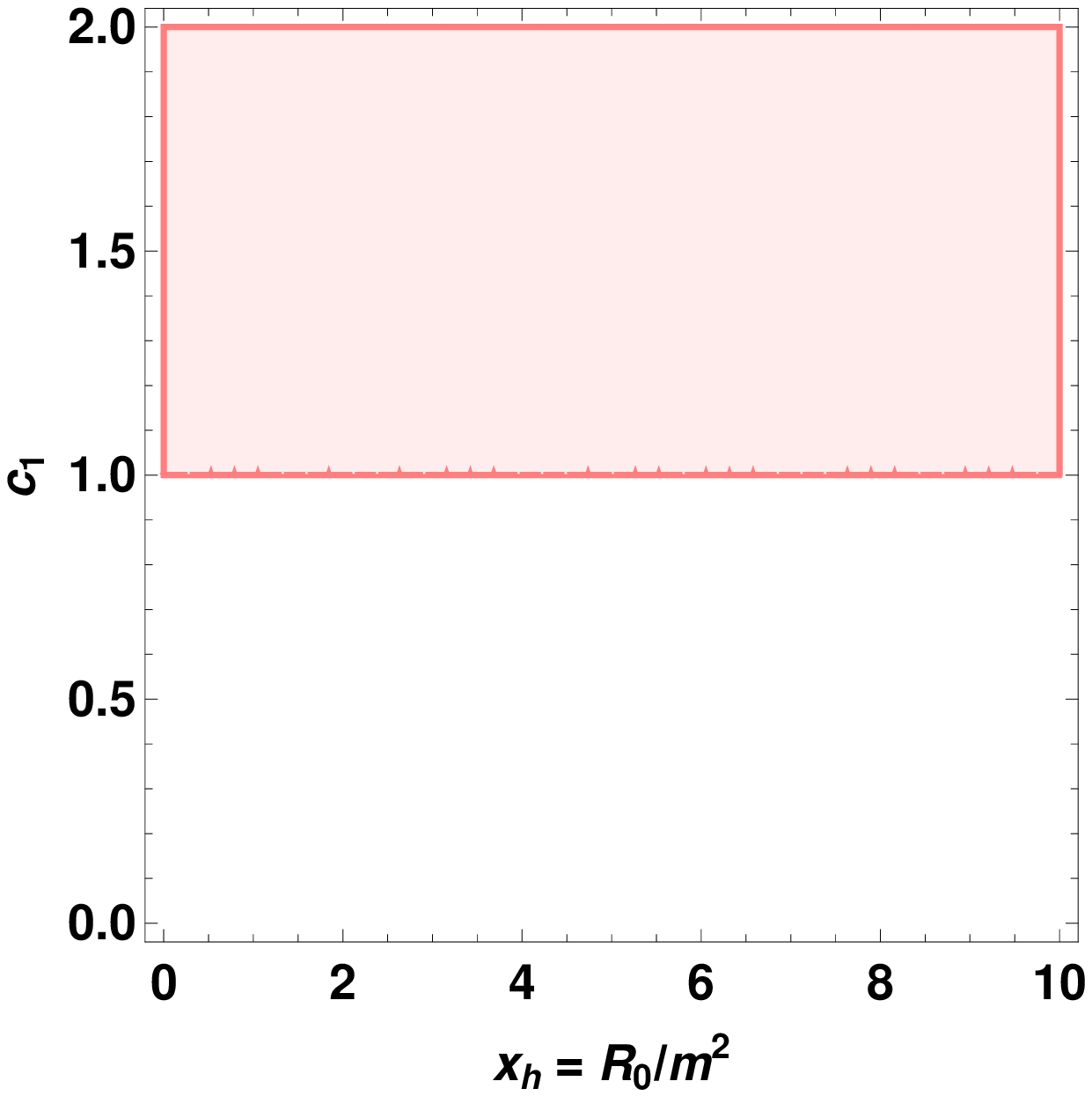}\hspace{1cm}
   \includegraphics[scale = 0.5]{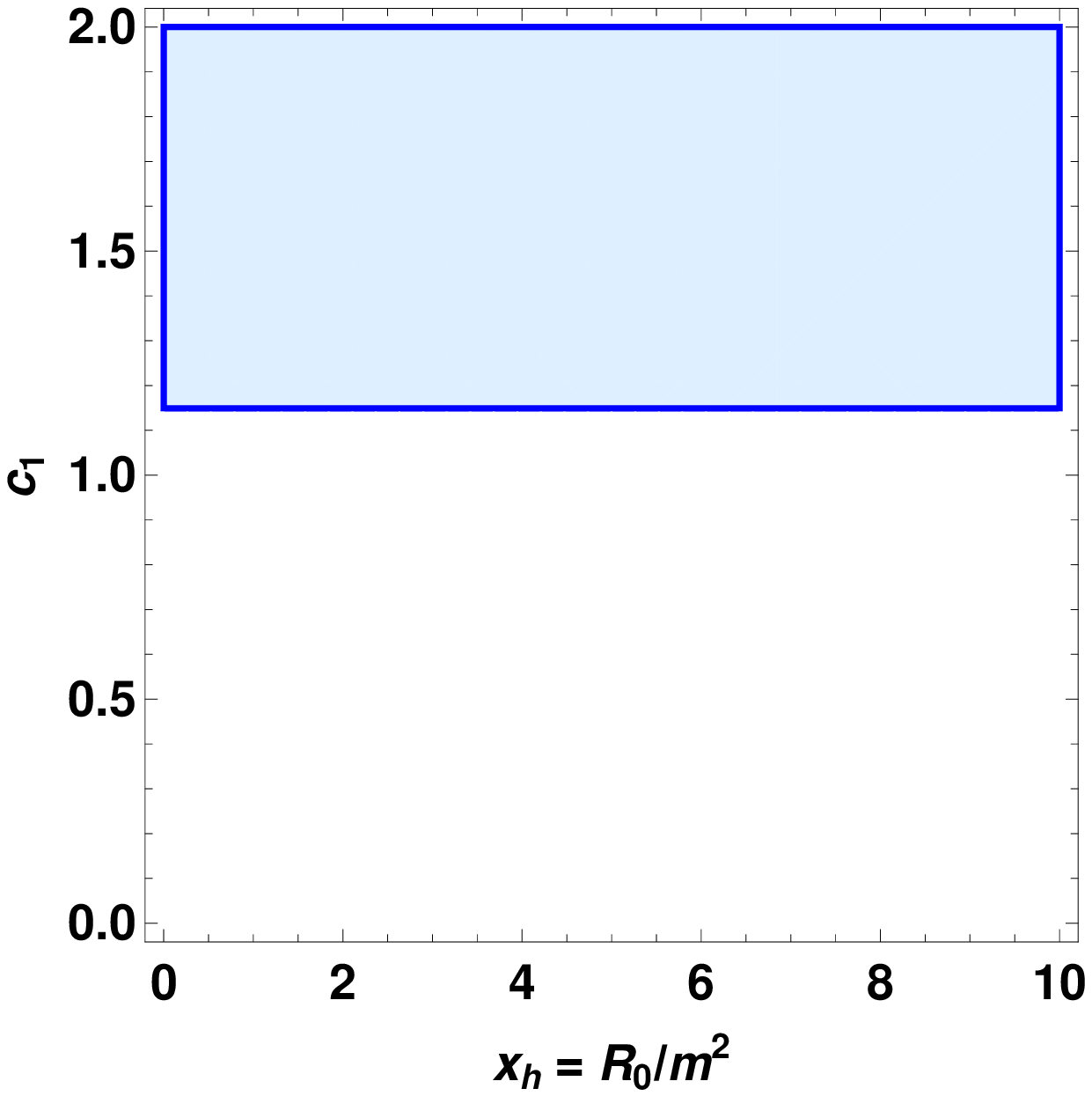}}
\vspace{-0.2cm} 
\caption{Stability region of the Hu-Sawicki model. Figure on the left side shows the stability region allowed by Eq.\ \eqref{stability05} and figure on the 
right side shows the stability region allowed by Eq.\ \eqref{stability06}.}
\label{fig_stability03}
\end{figure}

The de Sitter space defined by the Eq.\ \eqref{beta_value} is stable if it  
satisfies the condition \eqref{stability_primary_eq}. Our model \eqref{model} 
in Eq.\ \eqref{stability_primary_eq} leads to the inequality:
\begin{equation}
\left[2 \alpha  e^x \left(3 x^4-1\right)+\pi  \beta  \left(x^4+1\right)^2\right] \left[\pi  \beta  (x+1) \left(x^4+1\right)^2+e^x \lbrace 8 \alpha  x^5-\pi  \left(x^4+1\right)^2\rbrace \right]<0.
\end{equation}
By using Eq.\ \eqref{beta_value} in the above inequality, we can further 
reduce it to the following form:
\begin{eqnarray}\label{stability01} \notag
\left(x-2 e^x+2\right)^2 \left[\pi  x \left(x^4+1\right)^2-2 \alpha  \left(x^4+1\right)^2 \cot ^{-1}\left(\frac{1}{x^2}\right)+2 \alpha  A \right]\\ \left[2 \alpha  (x+1) \left(x^4+1\right)^2 \cot ^{-1}\left(\frac{1}{x^2}\right)+\pi  \left(-x^2-2 x+2 e^x-2\right) \left(x^4+1\right)^2-2 \alpha  B x^2\right]>0,
\end{eqnarray}
where $A = \left(x^6-3 x^5-6 x^4+e^x \left(6 x^4-2\right)+x^2+x+2\right)$ and $B = \left(x^5-3 x^4+8 \left(e^x-1\right) x^3+x+1\right)$.
When the model satisfies this Eq.\ \eqref{stability01}, it can have stable de Sitter solutions. The stability region of the model in parameter space is shown in Fig.\ref{fig_stability01}. Again, in order to get oscillatory behaviour around the de Sitter space, the following condition needs to be satisfied \cite{motohashi}:
\begin{equation} \label{oscillatory_condition}
\frac{f'(R_0)}{f''(R_0)}>\dfrac{25 R_0}{16}.
\end{equation}
Using our model in this inequality, we get,
\begin{equation}
\frac{\left(x^4+1\right) \lbrace e^x \left(\pi  x^4-2 \alpha  x+\pi \right)-\pi  \beta  \left(x^4+1\right)\rbrace}{2 \alpha  e^x \left(3 x^4-1\right)+\pi  \beta  \left(x^4+1\right)^2}>\frac{25 x}{16}
\end{equation}
Now, eliminating $\beta$ by using Eq.\ \eqref{beta_value}, we may rewrite the 
above expression as
\begin{equation} \label{stability02}
-\,\frac{2 \left(x^4+1\right) \lbrace-\pi  \left(-x+e^x-1\right) \left(x^4+1\right)-\alpha  \left(x^4+1\right) \cot ^{-1}\left(\frac{1}{x^2}\right)+2 \alpha  \left(e^x-1\right) x\rbrace}{\pi  x \left(x^4+1\right)^2-2 \alpha  \left(x^4+1\right)^2 \cot ^{-1}\left(\frac{1}{x^2}\right)+2 \alpha  \lbrace x^6-3 x^5-6 x^4+e^x \left(6 x^4-2\right)+x^2+x+2\rbrace}>\frac{25 x}{16}.
\end{equation}
If the toy model satisfies the above condition, it will have a stable de Sitter solution as well as oscillatory behaviour around the de Sitter solution. Similarly, for the Starobinsky model, the stability condition \eqref{stability_primary_eq} takes the form:
\begin{equation}
s_1 n \lbrace(2 n+1) x_s^2-1\rbrace \lbrace4 s_1 n (n+1) x_s^3-\left(x_s^2+1\right)^{n+2}\rbrace<0.
\end{equation}
Using Eq.\ \eqref{starobinsky_de_sitter} in the above inequality, we obtain,
\begin{equation} \label{stability03}
C \left[x^4 \lbrace-2 n^2+\left(x_s^2+1\right)^n-3 n-1\rbrace+\left(x_s^2+1\right)^n+x_s^2 \lbrace2 \left(x_s^2+1\right)^n-n-2\rbrace-1\right]>0,
\end{equation}
where $C = n x_s \left(x_s^2+1\right)^{2 n+2} \lbrace(2 n+1) x_s^2-1\rbrace \left[\left(x_s^2+1\right)^n+x_s^2 \lbrace\left(x_s^2+1\right)^n-n-1\rbrace-1\right]^2$.
The Starobinsky model gives stable de Sitter solutions for any parameter set 
satisfying the above inequality. For this model, the condition \eqref{oscillatory_condition} gives,
\begin{equation} \label{stability04}
C \left[x_s^4 \lbrace-50 n^2+16 \left(x_s^2+1\right)^n-16-57 n\rbrace+x_s^2 \lbrace 32 \left(x_s^2+1\right)^n-32-7 n\rbrace+16 \left(x_s^2+1\right)^n-16\right]>0.
\end{equation}
Any parameter sets in the Starobinsky model satisfying this condition will 
give oscillations around de Sitter solutions in de Sitter space. Finally for the Hu-Sawicki model, condition \eqref{stability_primary_eq} gives:
\begin{equation}
c_1 \mu  x_h^{\mu +1} \lbrace c_2 (\mu +1) x_h^{\mu }-\mu +1\rbrace \left[ x_h \left(c_2 x_h^{\mu }+1\right){}^3-c_1 \mu  x_h^{\mu } \lbrace c_2 (\mu +2) x_h^{\mu }-\mu +2\rbrace \right]>0.
\end{equation} 
Now using Eq.\ \eqref{hu_sawicki_de_sitter} in the above expression and considering $\mu = 1$, we get,
\begin{equation} \label{stability05}
\left(c_1-1\right) x_h \lbrace \sqrt{\left(c_1-1\right) c_1 x_h^2}+\left(c_1-1\right) x_h\rbrace \lbrace 2 \sqrt{\left(c_1-1\right) c_1 x_h^2}+2 c_1 x_h-x_h\rbrace >0.
\end{equation}
Similarly, for the Hu-Sawicki model, the condition for existence of oscillations around stable de Sitter solutions gives:
\begin{equation} \label{stability06}
c_1^2 x_h \lbrace \sqrt{\left(c_1-1\right) c_1 x_h^2}+\left(c_1-1\right) x_h\rbrace \lbrace \left(32 c_1-41\right) \sqrt{\left(c_1-1\right) c_1 x_h^2}+\left(32 c_1^2-57 c_1+25\right) x_h\rbrace>0.
\end{equation}
We have plotted these inequalities to check the stability region of the models 
in parameter space in in Figs.\ \ref{fig_stability02} and \ref{fig_stability03}.

From the Fig.\ \ref{fig_stability01}, it is seen that the toy model has a 
small region of instability
for lower values of $x$ and comparatively higher values of $\alpha$ (greater than 0.4 approximately). For higher values of $x$, the model has a continuous stability region in the parameter space. In case of the Starobinsky model, from 
Fig.\ \ref{fig_stability02} it is seen that there is a small region of 
instability in the parameter space. But in case of Hu-Sawicki model, the model 
is stable only for the values of $c_1> 1$ (Fig.\ \ref{fig_stability03}). This suggests that, the toy model has a wider range of stability region in the parameter space. Since, the model also has a comparatively larger region admitting Eq.\ \eqref{oscillatory_condition}, it can be easily concluded that the toy model with a selected set of parameters can pass the stability conditions as well as the condition for having oscillatory solutions in the de Sitter space.\\

\begin{figure}[htb]
\centerline{
   \includegraphics[scale = 0.3]{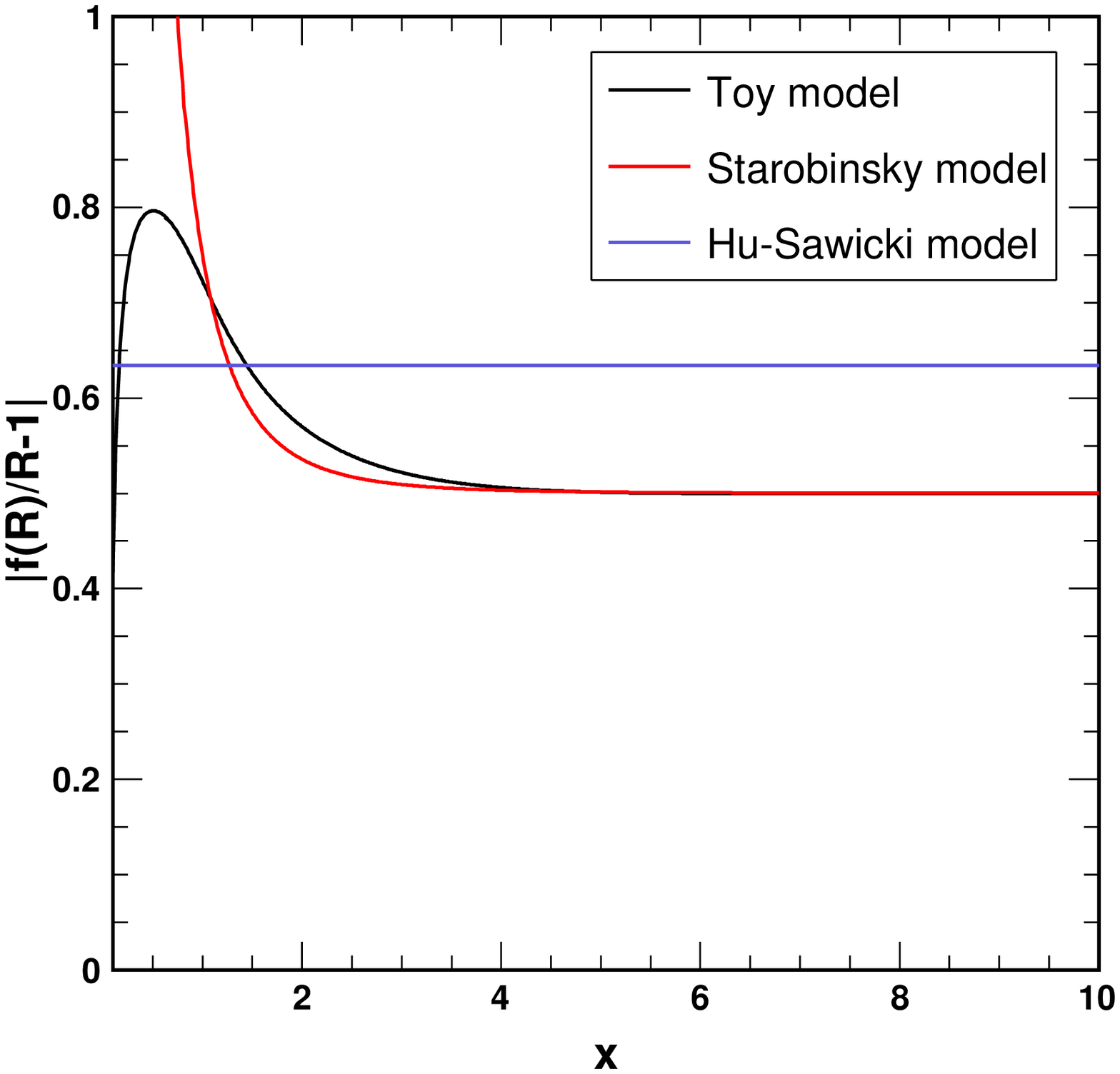}\hspace{0.3cm}
   \includegraphics[scale = 0.3]{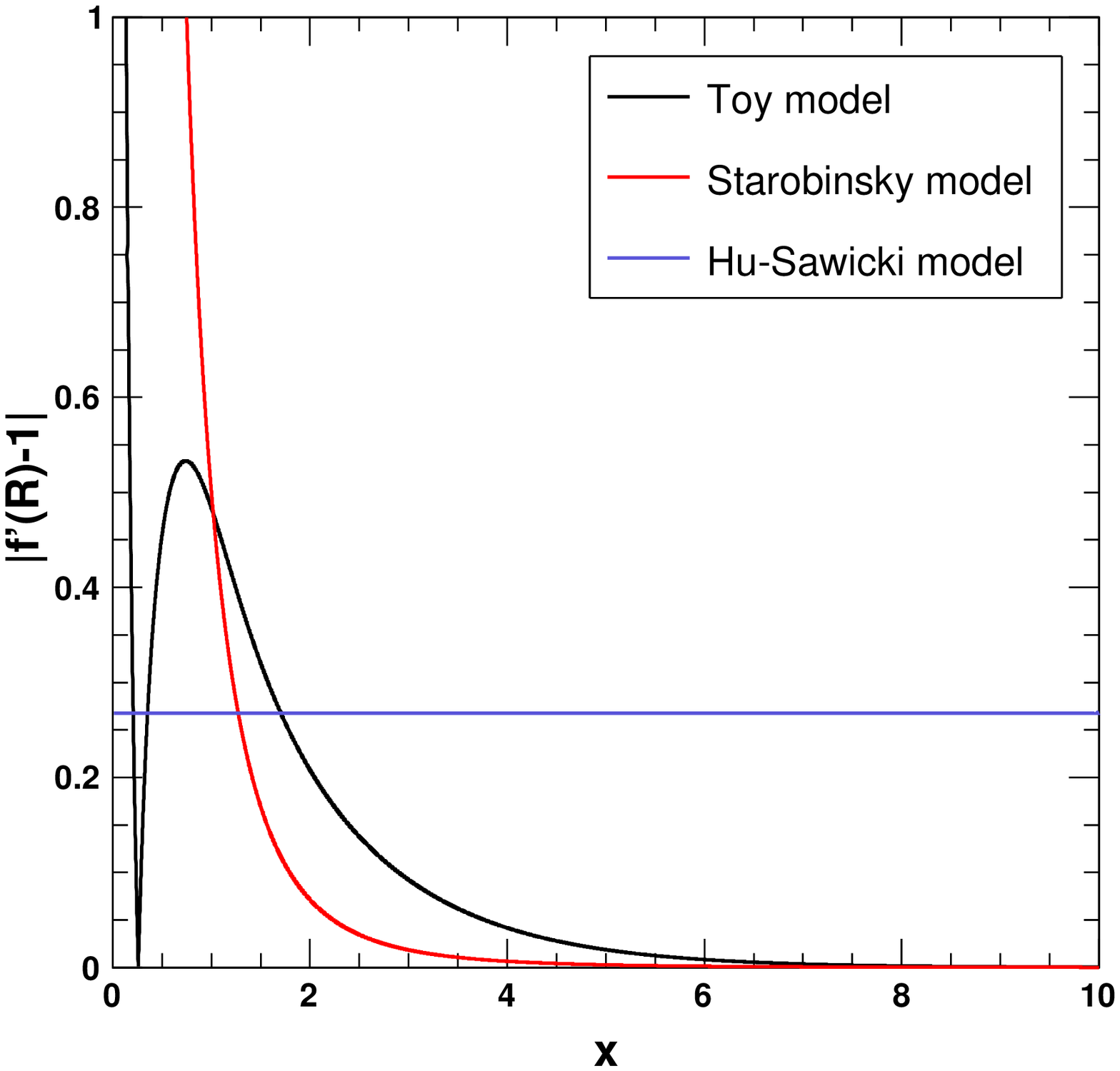}\hspace{0.3cm}
   \includegraphics[scale = 0.3]{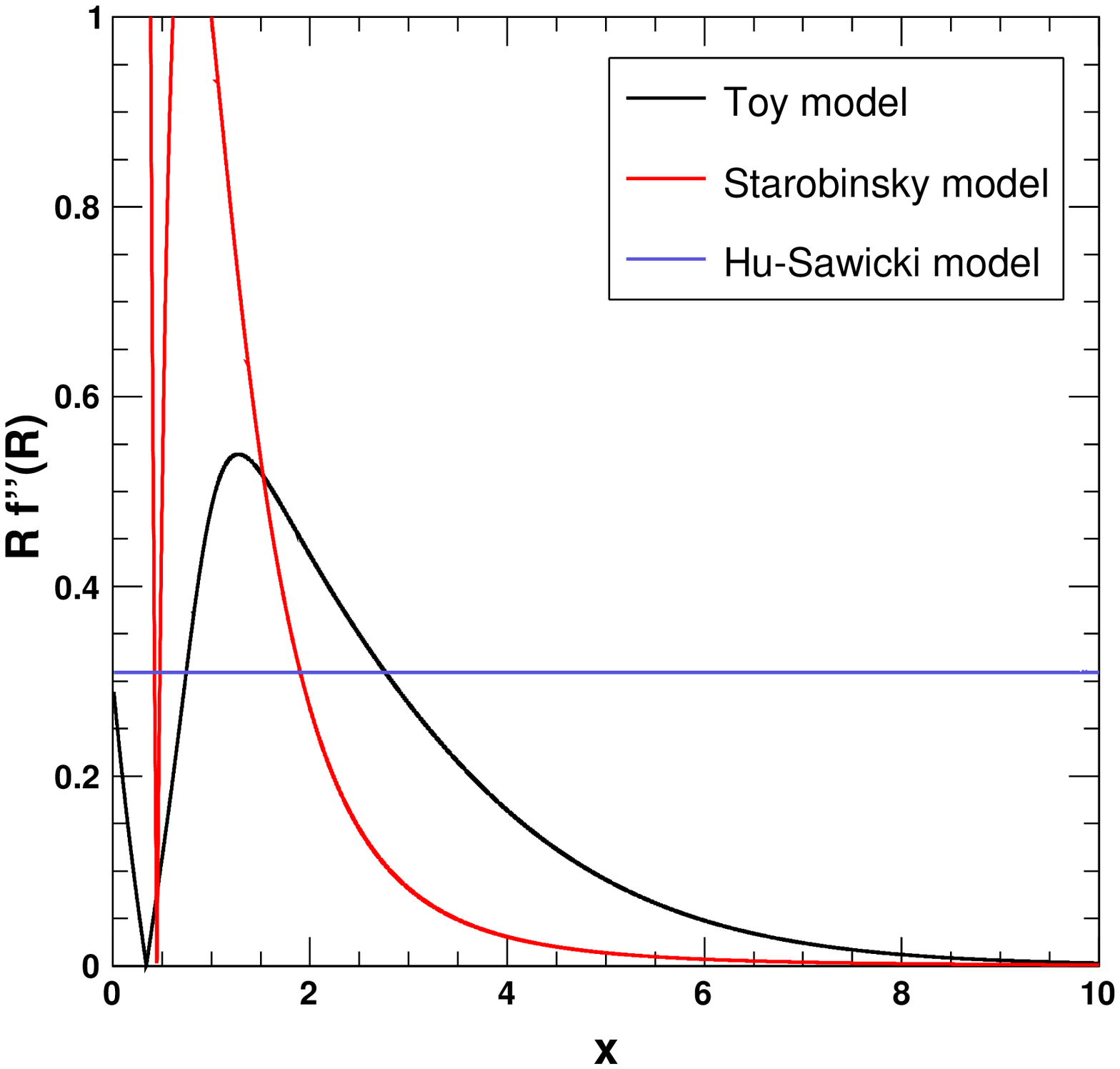}\vspace{-0.1cm}}
\centerline{   
   \includegraphics[scale = 0.3]{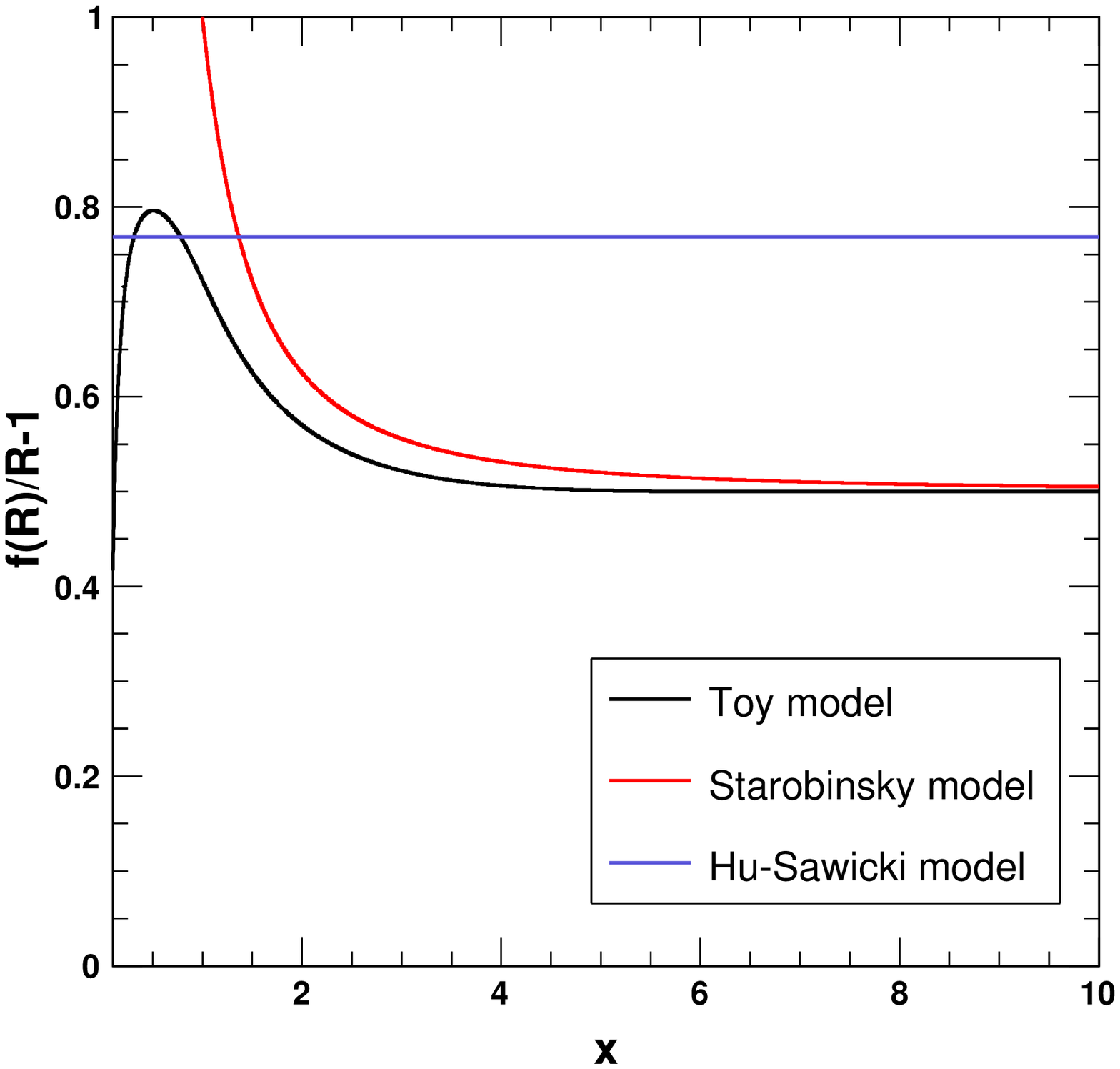}\hspace{0.3cm}
   \includegraphics[scale = 0.3]{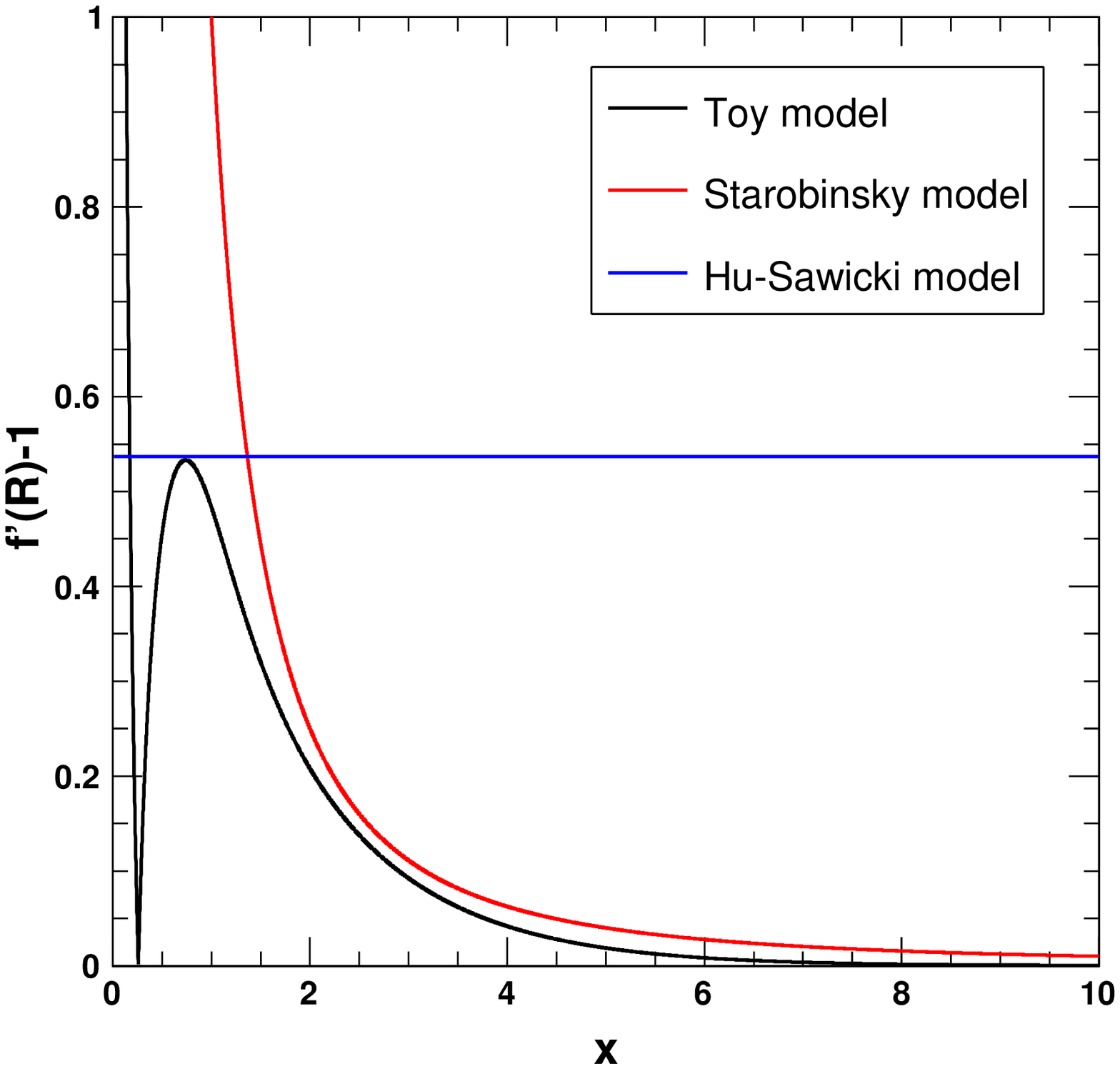}\hspace{0.3cm}
   \includegraphics[scale = 0.3]{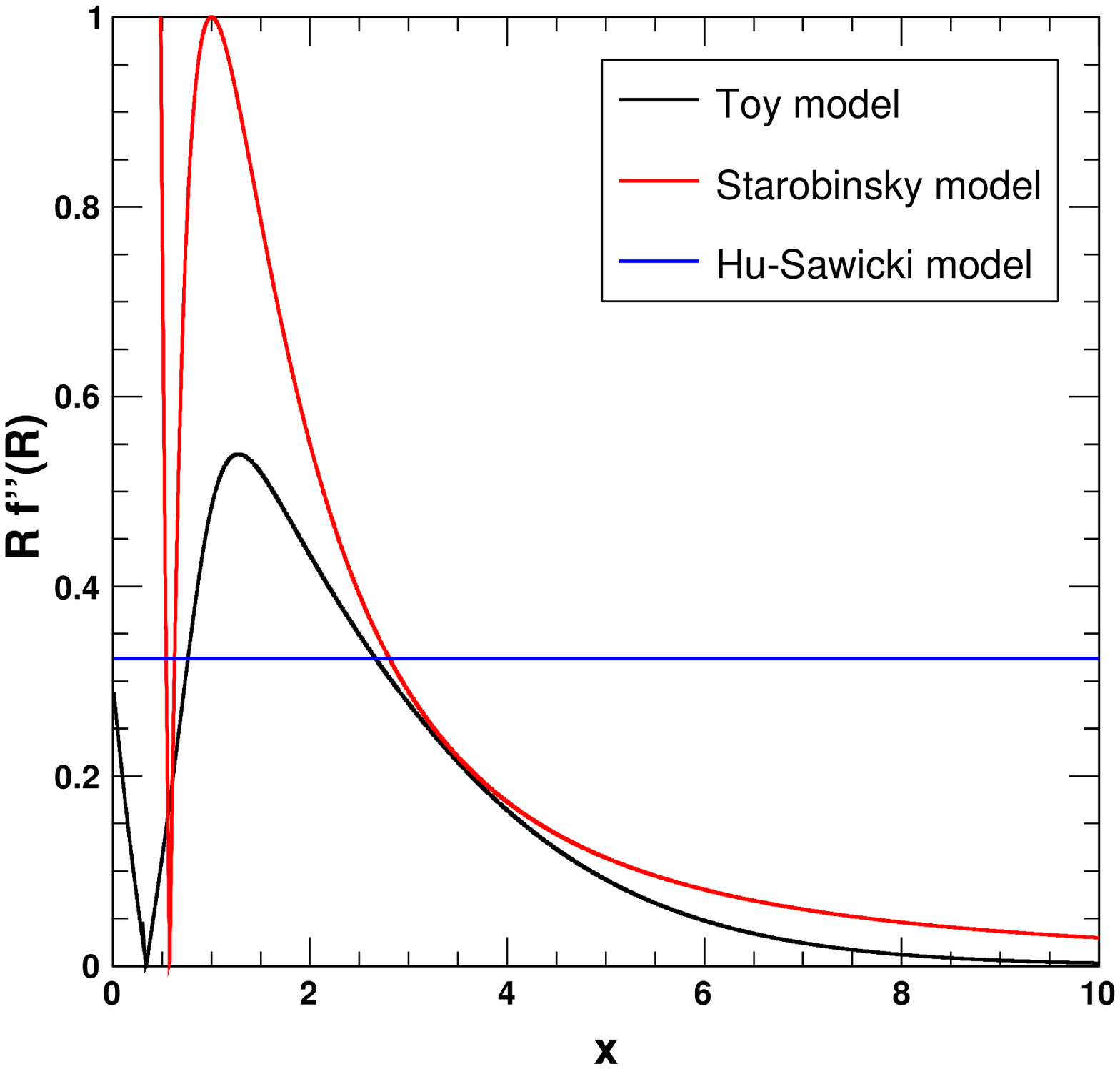}}
\vspace{-0.2cm} 
\caption{Solar system test functions with respect to $x$ for the toy model (with $\alpha = 0.3$), Starobinsky model (with $n$ = 2 at the upper panel and $n=1$ at the lower panel) and Hu-Sawicki model (with $\mu = 1$ and $c_1 = 1.5$ at the upper panel and $c_1 = 1.1$ at the lower panel).}
\label{fig_ss_comparison}
\end{figure}

Now, we would like to compare the bahviour of the toy model with the 
Starobinsky and Hu-Sawicki models using the functions used in the expressions 
\eqref{sstest01}, \eqref{sstest02} and \eqref{sstest03}. These expressions 
ensure the viability of the models in local systems. Although, in the previous 
section, we have chosen the parameters freely to check the solar system 
viabilities, now we shall study the same and compare with the other two models 
in de Sitter sitter stability regions of the parameter spaces. However, 
it should be noted that most of the parameter sets used in that sections are  
found to belong automatically in the de Sitter sitter stability region of the 
model. The function used in relation \eqref{sstest01} takes the following
respective form in the toy model, Starobinsky model and Hu-Sawicki model in 
the de Sitter solution regime,
\begin{equation}
\dfrac{f(R_0) - R_0}{R_0} = \frac{\alpha  \left(x^4+1\right) \cot ^{-1}\left(\frac{1}{x^2}\right)-\left(e^x-1\right) \left(\pi  x^4+2 \alpha  x+\pi \right)}{\pi  \left(-x+2 e^x-2\right) \left(x^4+1\right)},
\end{equation}

\begin{equation}
\dfrac{f_S(R_0) - R_0}{R_0} = -\frac{\left(x_s^2+1\right) \left(\left(x_s^2+1\right){}^n-1\right)}{2 \left(\left(x_s^2+1\right){}^n+x_s^2 \left(\left(x_s^2+1\right){}^n-n-1\right)-1\right)}
\end{equation}
and
\begin{equation}
\dfrac{f_H(R_0) - R_0}{R_0} = -\frac{c_1 \left(x_h\right)_h^{\mu +1}}{\sqrt{\left(c_1-1\right) c_1 \left(x_h\right)_h^2} \left(x_h\right)_h^{\mu }+\left(c_1-1\right) \left(x_h\right)_h^{\mu +1}+\left(x_h\right)_h^2}.
\end{equation}
The second function in \eqref{sstest02} gives,
\begin{equation}
f'(R_0) - 1 = \frac{2 \alpha  \left(x^4+1\right) \cot ^{-1}\left(\frac{1}{x^2}\right)-x \left(\pi  \left(x^4+1\right)+4 \alpha  \left(e^x-1\right)\right)}{\pi  \left(-x+2 e^x-2\right) \left(x^4+1\right)},
\end{equation}

\begin{equation}
f_S'(R_0) - 1 = \frac{n x_s^2}{-\left(x_s^2+1\right){}^n+x_s^2 \left(-\left(x_s^2+1\right){}^n+n+1\right)+1}
\end{equation}
and
\begin{equation}
f_H'(R_0) - 1 = -\frac{c_1 \mu  x_h^{\mu +3}}{\left(\sqrt{\left(c_1-1\right) c_1 x_h^2} x_h^{\mu }+\left(c_1-1\right) x_h^{\mu +1}+x_h^2\right){}^2}.
\end{equation}
The third function for the respective models are calculated as:
\begin{equation}
R_0 f''(R_0) = \frac{2 \alpha  x \left(3 x^4-1\right)}{\pi  \left(x^4+1\right)^2}-\frac{x \left(-\pi  x^5-2 \alpha  x^2+2 \alpha  \cot ^{-1}\left(\frac{1}{x^2}\right)+2 \alpha  x^4 \cot ^{-1}\left(\frac{1}{x^2}\right)-\pi  x\right)}{\pi  \left(-x+2 e^x-2\right) \left(x^4+1\right)},
\end{equation}

\begin{equation}
R_0 f_S''(R_0) = \frac{n x_s^2 \left((2 n+1) x_s^2-1\right)}{\left(x_s^2+1\right) \left(\left(x_s^2+1\right){}^n+x_s^2 \left(\left(x_s^2+1\right){}^n-n-1\right)-1\right)}
\end{equation}
and
\begin{equation}
R_0 f_H''(R_0) = \frac{c_1 \mu  x_h^{\mu +3} \left((\mu +1) \sqrt{\left(c_1-1\right) c_1 x_h^2} x_h^{\mu }+\left(c_1-1\right) (\mu +1) x_h^{\mu +1}-(\mu -1) x_h^2\right)}{\left(\sqrt{\left(c_1-1\right) c_1 x_h^2} x_h^{\mu }+\left(c_1-1\right) x_h^{\mu +1}+x_h^2\right){}^3}.
\end{equation}
These functions for the said models are compared in Fig.\ \ref{fig_ss_comparison}. It is seen that the models are capable of passing the solar system tests in 
the de Sitter stable regime for higher values of $x$. The Hu-Sawicki model in 
this regime shows a constant behaviour. However, the Starobinsky model and the 
toy model give the higher values of the test functions for lower $x$ values. 
In this case, the behaviour of the toy model still indicates the ability to 
pass the solar system tests. One interesting point to note that for the toy 
model and the Starobinsky model, the asymptotic behaviours of the test 
functions are almost same. These results suggest that the behaviour of the toy 
model is closer to the Starobinsky model. However, to make a clear conclusion, 
a detailed study of the models is required and this is beyond the scope of 
this manuscript.  

\section{Constraints on the Model}
For the viability of a model in $f(R)$ gravity it is utmost necessary to
impose constraints on the model on the basis of different available 
observational data. A model which passes such constraints are considered as 
a viable model in $f(R)$ gravity. Starobinsky model and Hu Sawicki model are 
examples of two viable models in this context as mentioned earlier. There are 
several ways to constrain an $f(R)$ gravity model \cite{Boubekeur_2014, 
Desai_2018, Jana_2019, Gupta_2019, Chiba_2007, Cataneo_2015, Nunes_2017}. A 
constrained model is helpful to study different implications of the model. In 
this section, we will try to constrain our toy model using the results 
published in \cite{Boubekeur_2014, Desai_2018} and \cite{Jana_2019}. In Ref.\
\cite{Boubekeur_2014}, authors carried out a Markov chain Monte Carlo (MCMC) 
analysis for GWs from Hu Sawicki model using the data sets of cosmic microwave 
background (CMB) and baryon acoustic oscillations (BAO) together with the 
independent constraints on the relationship between the matter clustering 
amplitude $\sigma_8$ and the matter mass-energy density $\Omega_m$ from 
Planck Sunyaev-Zeldovich (PSZ) cluster number counts and also from the CFHTLens
weak lensing tomography measurements. Combining CMB, BAO and 
$\sigma_8-\Omega_m$ relationship from the PSZ catalog \cite{Ade_2014}, they 
obtained a bound which is still better than the bounds obtained from the GW 
event GW170817 \cite{Jana_2019}. The bound on the parameter $f'(R)$ reported 
by them is
\begin{equation}
-3.7 \times 10^{-6} < f'(R) - 1 < 3.7 \times 10^{-6} ,
\end{equation}
with $95\%$ confidence level at upper bound \cite{Boubekeur_2014}. On the 
other hand in Ref.\ \cite{Desai_2018} a constraint was introduced on the 
Compton wavelength $\lambda_{g}$ of the graviton. From their study, we have a 
constraint on $\lambda_{g}^{-1}$ as given by,
\begin{equation}
0 ~\text{m}^{-1}< \lambda_{g}^{-1}<1.098901099 \times 10^{-23} ~\text{m}^{-1}
\end{equation}
with $90\%$ confidence level on upper bound \cite{Desai_2018}. 

\begin{figure}[!htb]
\centering
\includegraphics[scale = 0.50]{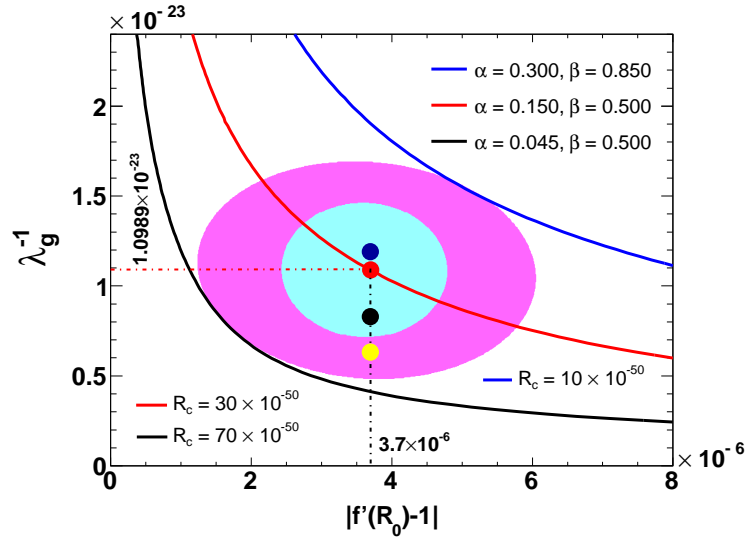}
\vspace{-0.2cm}
\caption{Contours with $95\%$ confidence level on the upper bounds of 
$f'(R) - 1$ \cite{Boubekeur_2014} and $90\%$ confidence level on the upper 
bounds of $\lambda_{g}^{-1}$ \cite{Desai_2018} (the larger contour) and with 
$68\%$ confidence level on the both (the smaller contour). The central red dot 
denotes the $\lambda_g$ corresponding to galaxy cluster Abell 1689 \cite{Desai_2018}, 
blue dot corresponds to Abell 262, black one corresponds to Abell 1991 and 
yellow one corresponds to Abell 383 \cite{Gupta_2019}. data}
\label{fig07}
\end{figure}

Now, we have computed the values of $f'(R) - 1$ and $\lambda_{g}^{-1}$ for 
our model by taking into consideration of above cited respective upper bounds 
with the corresponding confidence levels to constraint our model 
parameters $R_c, \alpha$ and $\beta$. The results of this computation along 
with the contour plots are shown in Fig.~\ref{fig07}. Here we have not considered the constraint on the model parameters coming from the fact that at higher curvatures it goes near to the cosmological constant. However, the parameters are chosen from the stability region of the parameter space. It is seen that the 
model can be a viable one within a proper range of variables. The figure shows 
the contours with $95\%$ confidence level for $f'(R) - 1$ and $90\%$ confidence
level for $\lambda_{g}^{-1}$ (the larger contour) and  with $68\%$ confidence 
level for the both (the smaller contour). The central point denotes the 
boundary value for 
both the parameters $f'(R) - 1$ and $\lambda_{g}^{-1}$ and any value lower 
than the boundary value is viable. This point corresponds to the galaxy
cluster Abell 1689 data \cite{Desai_2018}. In the plots we have considered three sets 
of the parameters and we see that the smaller values of $\alpha$ allow 
the model to pass the constraints easily. In the Fig.~\ref{fig07}, we have 
also shown 3 other points corresponding to galaxy clusters Abell 262, Abell 
1991 and Abell 383 data from the Ref.~\cite{Gupta_2019}. All these points lie
within the confidence level contours. 

The model can be constrained by using the GWs event GW170817 also. In a recent 
study \cite{Jana_2019}, $f(R)$ gravity was constrained by using the GW170817. 
They provided a bound on $f'(R)$, which is
\begin{equation}\label{GW_constraint}
-3 \times 10^{-3} < f'(R)-1 < 3 \times 10^{-3}.
\end{equation}
Using our toy model in this expression, we find,
\begin{equation}
-\,\frac{3}{1000}<\frac{2 \alpha  x}{\pi  x^4+\pi }+\beta  e^{-x}<\frac{3}{1000}.
\end{equation}
Now, using Eq.\ \eqref{beta_value} in the above expression, we can have,
\begin{equation}
-\,\frac{3}{1000}<\frac{\pi  x \left(x^4+1\right)-2 \alpha  \left(x^4+1\right) \cot ^{-1}\left(\frac{1}{x^2}\right)+4 \alpha  \left(e^x-1\right) x}{\pi  \left(-x+2 e^x-2\right) \left(x^4+1\right)}<\frac{3}{1000}.
\end{equation}
Choosing $x = 7.5$ (an arbitrary point in the stability region from Fig. \ref{fig_stability01}) and considering $\alpha$ to be a positive quantity,
the above equation reduces to,
\begin{equation}
0\leq\alpha <0.743783.
\end{equation}


\begin{figure}[!htb]
\centerline{
   \includegraphics[scale = 0.28]{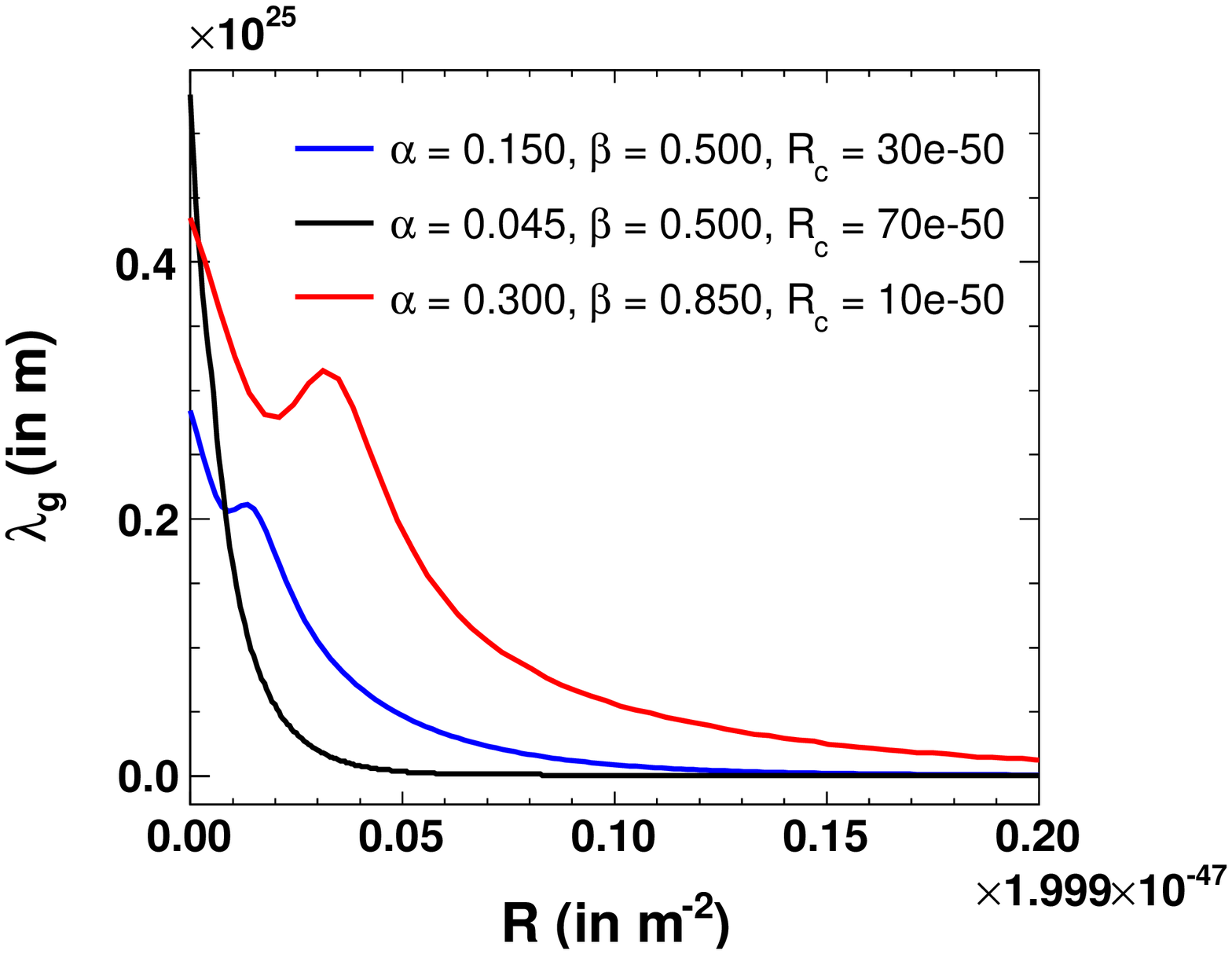}\hspace{0.5cm} 
   \includegraphics[scale = 0.28]{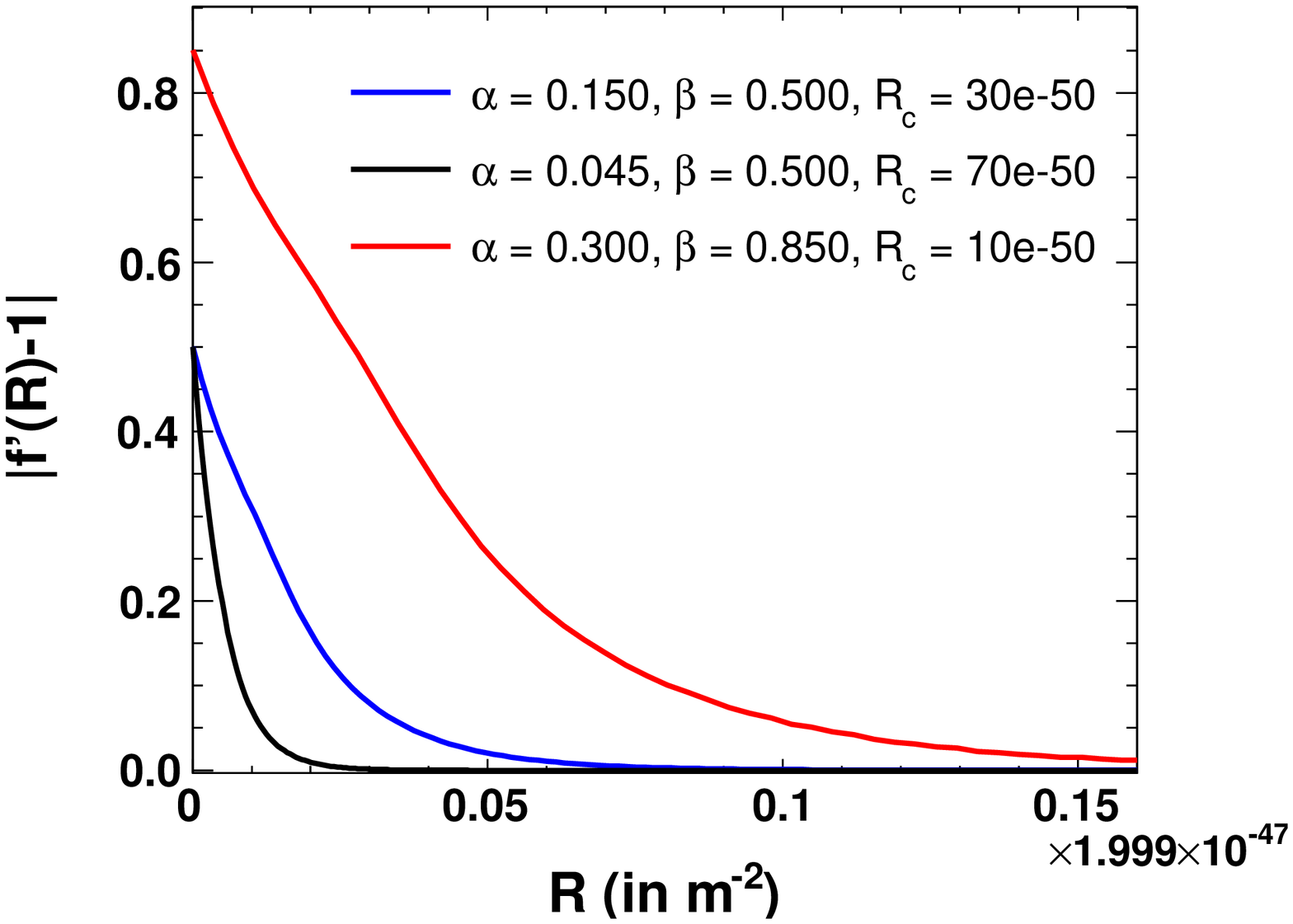} \hspace{0.5cm} 
   \includegraphics[scale = 0.28]{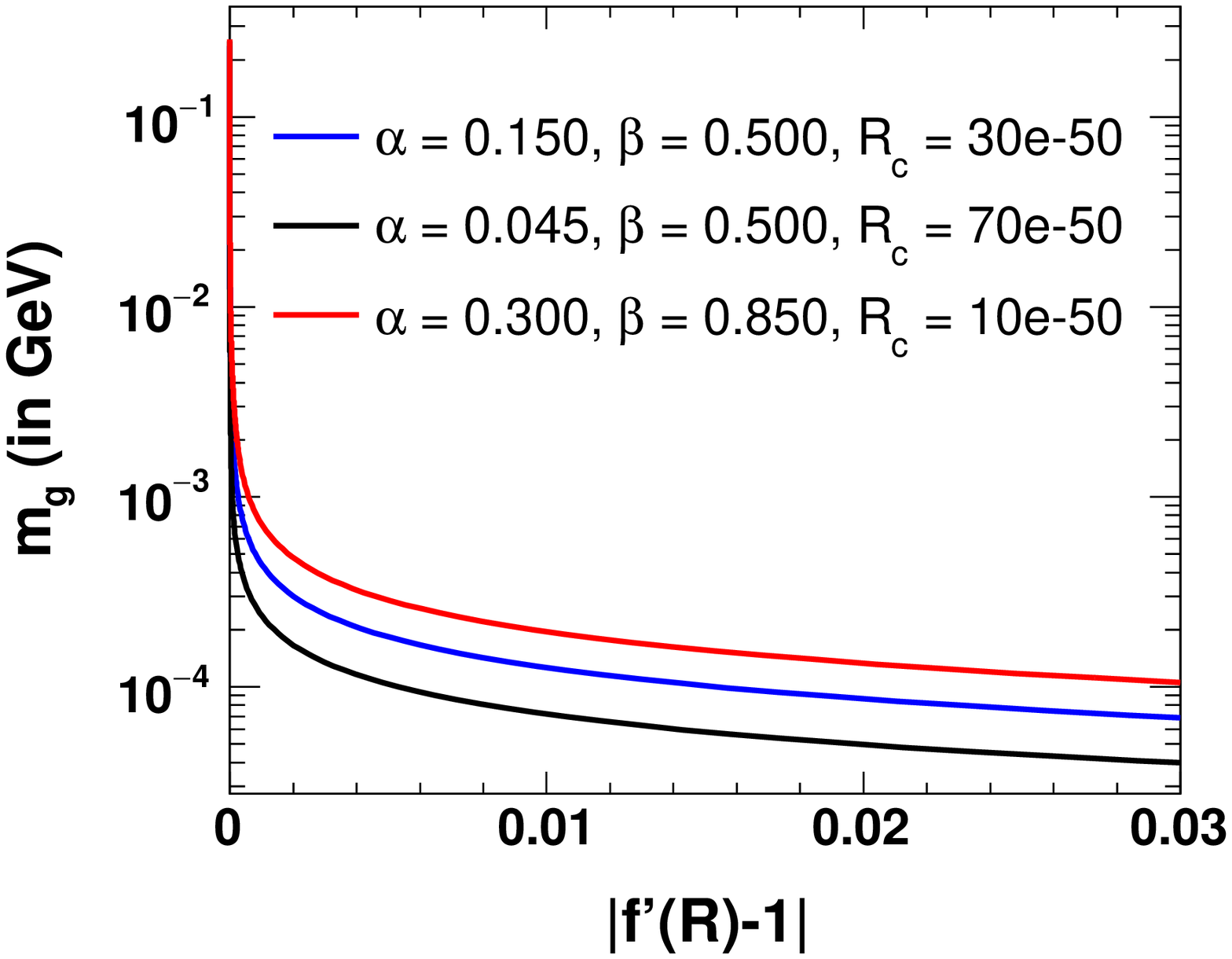}}
\caption{Plot on left shows the variation of $\lambda_{g}$ with respect to 
$R_0$, plot in middle shows the variation of $f'(R)-1$ with respect to $R_0$ 
and plot on right shows variation of $m_{g}$ with respect to $f'(R)-1$ for 
(i) $ \alpha = 0.150, \beta = 0.500, R_c = 30 \times 10^{-50} ~m^{-2}$; 
(ii) $\alpha = 0.045, \beta = 0.500, R_c = 70 \times 10^{-50} ~m^{-2}$; 
(iii) $\alpha = 0.300, \beta = 0.850, R_c = 10 \times 10^{-50} ~m^{-2}.$ These 
are the constrained set of parameters of our model that is done on the basis 
of upper bounds on $f'(R)-1$ \cite{Boubekeur_2014} and 
$\lambda_g$ \cite{Desai_2018}}.
\label{fig08}
\end{figure}

In Fig.~\ref{fig08}, we have shown the variations of $\lambda_{g}$ with respect
to $R_0$, $f'(R)-1$ with respect to $R_0$ and $m_{g}$ with respect to $f'(R)-1$
for the values of the model parameters used in the contour plots in 
Fig.~\ref{fig07}. The hump in the mass of the scalar field encountered in the 
Fig.~\ref{fig01} and in the Fig.~\ref{fig06} are also present in the 
$\lambda_{g}$ vs. $R_0$ curves. However, as mentioned earlier, this hump 
vanishes when $(\beta-\alpha) \gg 0$. For higher curvatures, $\lambda_{g}$ 
rapidly moves towards zero. The function $f'(R)-1$ also has significantly 
higher values near the Minkowski spacetime and as soon as the background Ricci 
curvature increases, this model dependent function decreases rapidly. This 
nature of the model is suitable for overcoming the local system constraints. 
Again as seen from the figure, with the increase of the function $f'(R)-1$ 
mass of the scalar field decreases initially at a faster rate and then becomes 
almost constant at later stage.

\section{Polarization Modes of GWs in the Model}\label{sec4}
In this section, we wish to check the polarization modes of GWs in the model. 
In presence of massive polarization mode, it would be easy to constraint the 
model using the experimental results. To explore the polarization modes of 
GWs in the model, at first we'll introduce the perturbation to the field 
equation.
\subsection{Perturbation to the Field Equation}
If there are propagating GWs in spacetime, then they perturbs the metric 
around its background value. Considering the background metric as 
$\bar{g}_{\mu\nu}$ we may express the spacetime metric to the first order of 
perturbation value $h_{\mu\nu}$, which is usually usually very small, as
\begin{equation}
g_{\mu\nu}=\bar{g}_{\mu\nu} + h_{\mu\nu},\;\; \mbox{where}\; \left|\, h_{\mu\nu} \,\right| \ll \left|\, \bar{g}_{\mu\nu}\, \right|.
\label{eq4}
\end{equation}
Now, expanding the Ricci tensor and the Ricci scalar upto the first order of 
$h_{\mu\nu}$, we may write:
\begin{align}
\notag
 R_{\mu\nu}& \simeq \tilde{R}_{\mu\nu} + \delta R_{\mu\nu} + \mathcal{O}(h^2)\\ 
&= \tilde{R}_{\mu\nu} - \dfrac{1}{2} (\nabla_\mu \nabla_\nu 
h - \nabla_\mu \nabla^{\lambda} h_{\lambda\nu}- \nabla_\nu \nabla^{\lambda} h_{\mu\lambda} + \square h_{\mu\nu}) + \mathcal{O}(h^2)
\label{eq5}
\end{align}
and 
\begin{align}
\notag
 R &\simeq \tilde{R} + \delta R + \mathcal{O}(h^2)\\
 & =\tilde{R} - \square h + \nabla^\mu \nabla^\nu h_{\mu\nu} -\tilde{R}_{\mu\nu} h^{\mu\nu} + \mathcal{O}(h^2).
\label{eq6}
\end{align}
Similarly, we may write for the $f(R)$ and $f'(R)$ as
\begin{align}
 f(R) &\simeq f(\tilde{R}) + f'(\tilde{R})\, \delta R + \mathcal{O}(h^4),\\[5pt]
 f'(R) &\simeq f'(\tilde{R}) + f''(\tilde{R})\, \delta R + \mathcal{O}(h^4),
\label{eq7}
\end{align}
where $\tilde{R}$ is some constant curvature. Thus, due to the perturbation
in spacetime the trace equation (\ref{trace_field_eq}) can be rewritten as
\begin{equation} \label{pert_field_trace}
3 f''(\tilde{R})\, \square \delta R + \left[f''(\tilde{R})\tilde{R} - f'(\tilde{R})\right]\! \delta R = 0,
\end{equation}
where we have used $T_{\mu\nu} = 0$ for the empty space or far away from the 
source. Fixing the gauge to be harmonic gauge with $ \nabla_\mu h^\mu_\nu = \dfrac{1}{2}\,\nabla_\nu h$, which after operating by $\nabla^\nu$ we find,
\begin{equation}\label{gf}
\nabla^\mu \nabla^\nu h_{\mu\nu} = \dfrac{1}{2}\, \square h.
\end{equation}
An important point to be mentioned here is that, the Eq.\ (\ref{field_equation}) is also satisfied by another solution: $R_{\mu\nu} = \Lambda g_{\mu\nu} =\tilde{R}_{\mu\nu} $, giving
\begin{equation}
2f'(\tilde{R}) \tilde{R}_{\mu\nu} -g_{\mu\nu} f(\tilde{R}) = \kappa^2\, T_{\mu\nu}.
\end{equation}
This equation actually corresponds to the Eq.\ (\ref{stationary_condition}), 
which is the stationary condition used earlier in the Sec.\ \ref{sec2}. In 
empty space, this equation has the form:
\begin{equation}
2f'(\tilde{R}) \tilde{R}_{\mu\nu} -g_{\mu\nu} f(\tilde{R}) = 0
\end{equation}
and it leads to have the equation,
\begin{equation}\label{stcn}
2f(\tilde{R}) - \tilde{R} f'(\tilde{R}) = 0.
\end{equation}
This is the stationary condition (\ref{stationary_condition}) of the scalar
field potential in the empty space corresponding to the constant curvature 
$\tilde{R}$ of spacetime. Using the Eq.s (\ref{eq6}), (\ref{gf}) and
(\ref{stcn}) in Eq.~(\ref{pert_field_trace}), we get
\begin{equation}
3 f''(\tilde{R})\, \square^2 h + \left(\frac{5 f(\tilde{R}) f''(\tilde{R})}{f'(\tilde{R})}-f'(\tilde{R})\right)\! \square h+ \left( \frac{2 f(\tilde{R})^2 f''(\tilde{R})}{f'(\tilde{R})^2}-f(\tilde{R}) \right)\!h=0.
\end{equation}
Now, we would like to define $\square h = m^2 h$, where $m$ is the mass of the 
associated scalar field. Using this definition in the above equation we obtain,
\begin{equation}
3 f''(\tilde{R}) \, m^4 + \left(\frac{5 f(\tilde{R}) f''(\tilde{R})}{f'(\tilde{R})}-f'(\tilde{R})\right)\! m^2+ \left( \frac{2 f(\tilde{R})^2 f''(\tilde{R})}{f'(\tilde{R})^2}-f(\tilde{R}) \right)=0.
\end{equation}
This is a quadratic equation in $m^2$ and solution for $m^2$ gives,
\begin{equation}\label{massmod}
m^2 = \frac{f'(\tilde{R})}{3 f''(\tilde{R})}-\frac{2 f(\tilde{R})}{3 f'(\tilde{R})} = \dfrac{1}{3} \left[ \dfrac{f'(\tilde{R})}{f''(\tilde{R})} - \tilde{R} \right]\!,
\end{equation}
and
\begin{equation}
m^2 = -\,\frac{f(\tilde{R})}{f'(\tilde{R})} = -\,\dfrac{\tilde{R}}{2}.
\end{equation}
We see that the second solution corresponds to tachyonic scalar field which 
becomes zero in the Minkowski spacetime or at far distance away from the 
source. The first solution is identical to Eq.\ (\ref{mass_scalar_field}).
Thus the term $m^2$ given by Eq.\ (\ref{massmod}) is exactly same as the 
scalar field mass square term $m_\phi^2$ given in Eq.\ (\ref{mass_model_01}) 
for our model, when $\tilde{R} = R_0$. Therefore this solution suggests that 
there exists a massive scalar mode of polarization of GWs in the theory apart 
from the massless tensor modes. 

At very far distance away from the source, we can consider $\bar{g}_{\mu \nu} 
= \eta_{\mu \nu}$, i.e.\ the Minkowski metric and the background curvature 
$\tilde{R} = 0$. In this case, the Ricci scalar slowly varies near zero, i.e.\ 
$R \simeq 0 + \delta R$. Hence, for the Minkowski space the Eq.~(\ref{eq4}) 
can be written as
\begin{equation}
g_{\mu\nu}=\eta_{\mu\nu} + h_{\mu\nu}.
\end{equation}
And to the first order of $h_{\mu\nu}$, we get
\begin{equation}
\label{pertEq01}
R_{\mu\nu}=\frac{1}{2}\!\left(\partial_\mu\partial_\rho h^\rho_\nu+\partial_\nu\partial_\rho h^\rho_\mu-\partial_\mu\partial_\nu h-\square h_{\mu\nu}\right)\!,
\end{equation}
\begin{equation}
\label{pertEq02}
R=\partial_\mu\partial_\rho h^{\rho\mu}-\Box h,
\end{equation}
where $h=\eta^{\mu\nu}h_{\mu\nu}$.
So for our model, to the first order of perturbation, the 
Eq.~\eqref{field_equation} becomes
\begin{equation}
\label{model_tensor_field_eq}
R_{\mu\nu}-\frac{1}{2}\eta_{\mu\nu}R - \frac{ (2 \alpha - \pi  \beta ) }{ \pi  (\beta -1) R_c}\left(\partial_\mu\partial_\nu R-\eta_{\mu\nu}\,\square R\right)=0.
\end{equation}
Taking the trace of this Eq.\ \eqref{model_tensor_field_eq}, we get
\begin{equation}
\label{kg_eq_48}
(\square-m_0^2)R=0,
\end{equation}
where $$m_0^2=\frac{\pi  (\beta -1) R_c}{6 \alpha -3 \pi  \beta }$$ 
with $\alpha > 0$ and $\beta > 0$. This is also exactly the same mass square
term $m_\phi^2\big|_{R_0\,=\,0}$ in Minkowski space given by the Eq.\ (\ref{mass_model_02}) for our model. 
Next, we introduce a variable
\begin{equation}
\label{gauge_var01}
\bar{h}_{\mu\nu}=h_{\mu\nu}-\frac{1}{2}\eta_{\mu\nu}h- \frac{ (2 \alpha - \pi  \beta ) }{ \pi  (\beta -1) R_c}\, \eta_{\mu\nu} R.
\end{equation}
The trace of this variable is
\begin{equation}
\label{gauge_var02}
\bar{h}=\eta^{\mu\nu}\bar{h}_{\mu\nu}=-\,h-4 \frac{ (2 \alpha - \pi  \beta ) }{ \pi  (\beta -1) R_c} R.
\end{equation}
Using this Eq.\ (\ref{gauge_var02}) in the variable (\ref{gauge_var01}) we find,
\begin{equation}
\label{gauge_var03}
h_{\mu\nu}=\bar{h}_{\mu\nu}-\frac{1}{2}\eta_{\mu\nu}\bar{h}-\frac{ (2 \alpha - \pi  \beta )}{ \pi  (\beta -1) R_c}\, \eta_{\mu\nu} R.
\end{equation}
From Eq.~\eqref{gauge_var01} and Eq.~\eqref{gauge_var03}, one can easily see 
that both $h_{\mu\nu}$ and $\bar{h}_{\mu\nu}$ are interchangeable, i.e.\ 
replacing $\bar{h}_{\mu\nu}$ by $h_{\mu\nu}$ and vice-versa in 
Eq.~\eqref{gauge_var01} gives Eq.~\eqref{gauge_var03}. Again, under an 
infinitesimal coordinate transformation, $x^\mu\rightarrow x^{\mu\prime}=x^\mu+\varsigma^\mu$, we have 
\begin{equation}
\label{gauge_var04}
h_{\mu\nu}'=h_{\mu\nu}-\partial_\mu\varsigma_\nu-\partial_\nu\varsigma_\mu.
\end{equation}
The trace of this equation is
\begin{equation}
\label{gauge_var05}
h'=h-2\,\partial_\mu\varsigma^\mu.
\end{equation}
And
\begin{equation}
\label{gauge_var06}
\bar{h}_{\mu\nu}'=\bar{h}_{\mu\nu}-\partial_\mu\varsigma_\nu-\partial_\nu\varsigma_\mu+\eta_{\mu\nu}\,\partial_\rho\varsigma^\rho.
\end{equation}
The trace trace of this equation gives,
\begin{equation}
\label{gauge_var07}
\bar{h}'=\bar{h}+2\,\partial_\rho\varsigma^\rho.
\end{equation}
Here we raise or lower the indices with the help of $\eta_{\mu\nu}$, i.e.\ with
the Minkowski metric. The Lorentz gauge condition $\partial^\mu \bar{h}_{\mu\nu}'=0$ can be obtained if $\varsigma_\mu$ satisfies $\square \varsigma_\nu=\partial^\mu\bar{h}_{\mu\nu}$.
The Lorentz gauge condition does not constrain the gauge freedom and there is 
always a possibility to choose the transverse and traceless conditions, i.e.\
$\partial^\mu \bar{h}_{\mu\nu}=0$ and $ \bar{h}=\eta^{\mu\nu}\bar{h}_{\mu\nu}=0$ \cite{Liang_2017, Corda_2007, Corda_2008, Capozziello_2008}.
By using the transverse traceless gauge condition and substituting the 
Eq.~\eqref{gauge_var03} into Eq.~\eqref{pertEq01}, we get
\begin{equation}
\label{model_pert_eq56}
R_{\mu\nu}=\frac{1}{2}\!\left[-\,\square\bar{h}_{\mu\nu}+2\, \frac{ (2 \alpha - \pi  \beta )}{ \pi  (\beta -1) R_c}\,\partial_\mu\partial_\nu R+\frac{(2 \alpha - \pi  \beta )}{ \pi  (\beta -1) R_c}\,\eta_{\mu\nu}\,\square R\right]\!.
\end{equation}
Plugging Eq.~\eqref{model_pert_eq56} into Eq.~\eqref{model_tensor_field_eq}, 
we obtain
\begin{equation}
\label{eq57}
3\,\frac{2 \alpha -\pi  \beta }{2 \pi  (\beta -1) R_c}\,\eta_{\mu\nu}(\square-m_0^2)R-\frac{1}{2}\,\square \bar{h}_{\mu\nu}=0.
\end{equation}
Combining Eq.s \eqref{kg_eq_48} and \eqref{eq57}, we get
\begin{equation}
\label{eq58}
\square \bar{h}_{\mu\nu}=0,
\end{equation}
which is the wave equation of the massless tensor field. The solution to this 
Eq.~\eqref{eq58} is \cite{Corda_2007, Corda_2008}
\begin{equation}
\label{eq59}
\bar{h}_{\mu\nu}=e_{\mu\nu}\exp(iq_\mu x^\mu)+\text{c.c.},
\end{equation}
where $\eta_{\mu\nu}q^\mu q^\nu=0$ and $q^\mu e_{\mu\nu}=0$.
Whereas the solution to the massive scalar field Eq.~\eqref{kg_eq_48} is given 
by \cite{Corda_2007, Corda_2008},
\begin{equation}
\label{eq60}
R= \psi =\psi_0 \exp(ip_\mu x^\mu)+\text{c.c.},
\end{equation}
where $\eta_{\mu\nu}p^\mu p^\nu=-\,m^2_0$. 
Assuming the GWs propagation direction along $z$, the general solution can be 
written as \cite{Liang_2017}
\begin{equation} \label{general_solution_01}
h_{\mu\nu} = \bar{h}_{\mu\nu}(t-z) + \frac{2 \alpha -\pi  \beta }{\pi(1-\beta)R_c}\, \eta_{\mu\nu}\, \psi (vt-z),
\end{equation}
where $\bar{h}_{\mu\nu} $ is transverse and traceless
and it represents the standard spin-2 graviton, and $\psi$ represents the 
scalar field which is massive in nature and travels with a speed less than 
$c$. The solution of the scalar part along $z$ axis, i.e.\ $\psi (vt-z)$ can 
be expressed as
\begin{equation}
\psi = \psi_0\, e^{-i\omega t\,+\,ikz}
\end{equation}
and hence the mass of the field in terms of $k$ and $\omega$ is
\begin{equation} \label{eq63}
m_0 = \sqrt{\omega^2 - k^2}.
\end{equation}

\subsection{Calculation of Exact Polarization amplitudes and Newman-Penrose Quantities of the Model}
In 1973, a powerful method was introduced in Ref.\ \cite{Eardley_1973} which 
deals with the study of the properties of GWs in any metric theory of gravity. 
This method involves analysing all the relevant components of Riemann tensor, 
which results relative acceleration between two test particles. They used a 
null-tetrad basis in order to calculate the Newman-Penrose quantities 
\cite{Newman_1962}. In the Newman-Penrose formalism, there are ten $\Psi$'s, 
nine $\Phi$'s, and a $\Lambda$, which are algebraically independent and 
represent the irreducible parts of the Riemann tensor 
$R_{\lambda\mu\kappa\nu}$. They are known as Newman-Penrose quantities. But 
in case of plane waves or nearly plane waves, the differential and symmetry 
properties of $R_{\lambda\mu\kappa\nu}$ reduce the number of independent, 
nonvanishing components, to six. Hence, in this formalism, the set 
$\{\Psi_2,\Psi_3,\Psi_4,\Phi_{22}\}$ is used to describe the six independent 
components of GWs in the metric theory. In the tetrad basis, 
the Newman-Penrose quantities of the Riemann tensor are \cite{Eardley_1973}:
\begin{align}
\Psi_2& = -\,\frac{1}{6}\, R_{lklk},\\[8pt]
\Psi_3& = -\,\frac{1}{2}\, R_{lkl\overline{m}},\\[8pt]
\Psi_4& = -\, R_{l\overline{m}l\overline{m}},\\[8pt]
\Phi_{22}& = -\, R_{lml\overline{m}}.
\end{align}
It should be noted that, $\Psi_3$ and $\Psi_4$ are complex. Therefore, each 
one of them is capable of describing two independent polarizations. One 
polarization mode for the real part and one for the imaginary part. Thus total 
number of polarization modes is 6.

The tetrad components of Ricci tensors can be expressed as \cite{Eardley_1973},
\begin{align}
R_{lk} &= R_{lklk},\\[8pt]
R_{ll}& = 2\, R_{lml\overline{m}},\\[8pt]
R_{lm}& = R_{lklm},\\[8pt]
R_{l\overline{m}}& = R_{lkl\overline{m}},
\end{align}
and the Ricci scalar is
\begin{equation}
R = -\, 2\, R_{lk} = -\, 2\, R_{lklk}.
\end{equation}

In normal coordinate system \cite{Eardley_1973},
\begin{align}
\notag
\Psi_2 &= -\,\dfrac{1}{6}\, R_{ztzt},\\[8pt]\notag
\Psi_3 &= -\,\dfrac{1}{2}\,R_{xtzt}+\dfrac{i}{2}\, R_{ytzt},\\[8pt]\notag
\Psi_4 &= -\,R_{xtxt} + R_{ytyt}+ 2i\, R_{xtyt},\\[8pt]\notag
\Phi_{22} &= -\, R_{xtxt} - R_{ytyt}.
\end{align}

Although the amplitudes $\{\Psi_2,\Psi_3,\Psi_4,\Phi_{22}\}$ of a wave depend 
on the observer \cite{Eardley_1973}, there are certain invariant statements 
about them that hold true for all the standard observers if they hold true for 
any one. These statements characterize the invariant $E(2)$ classes of waves. 
For a standard observer, under the assumptions that (a) the wave travels in 
the $+z$ direction, and (b) the same frequency for a monochromatic wave is 
observed, the $E(2)$ classes are:
\begin{itemize}
 \item \textbf{Class} $II_6$: $\Psi_2 \neq 0$. Standard
     observers measure the same non-vanishing amplitude in the
     $\Psi_2$ mode. Presence or absence of all other
     modes is observer-dependent;
 \item \textbf{Class} $III_5$: $\Psi_2 = 0,~\Psi_3 \neq 0$.
     Standard observers measure the absence of $\Psi_2$
     and the presence of $\Psi_3$. Presence or absence
     of $\Psi_4$ and $\Phi_{22}$ is observer-dependent;
 \item \textbf{Class} $N_3$: $\Psi_2 = \Psi_3 = 0,~\Psi_4 \neq
     0, \Phi_{22} \neq 0$. Presence or absence of all modes is
     observer-independent;
 \item \textbf{Class} $N_2$: $\Psi_2 = \Psi_3 = \Phi_{22} =
     0;~\Psi_4 \neq 0$. Observer-independent;
 \item \textbf{Class} $O_1$: $\Psi_2 = \Psi_3 = \Psi_4 =
     0;~\Phi_{22} \neq 0$. Observer-independent;
 \item \textbf{Class} $O_0$: $\Psi_2 = \Psi_3 = \Psi_4 =
     \Phi_{22} = 0$. Observer-independent. All standard
     observers measure no wave.
\end{itemize}

In $f(R)$ gravity, the field equation derived from the Lagrangian in metric 
formalism results dynamical expressions for Ricci tensor and Ricci scalar,
as we have already seen. 
Using this method, the expressions for Ricci tensor and scalar are calculated 
in weak field limit, i.e.\ far from the GWs source considering that the GW is 
propagating along $z$ axis. Thus, the Ricci tensor components corresponding to 
directions other than $z$ and $t$ will vanish. But from earlier sections, we 
see that a massive scalar mode of polarization is present in this model and so 
it is not possible to use this Newman-Penrose formalism formalism which is 
developed for null waves 
\cite{Liang_2017}. For waves with massive propagation mode, modified 
Newman-Penrose formalism has to be applied \cite{Hyun_2019}. In modified 
Newman-Penrose formalism, the polarization amplitudes as well as 
Newman-Penrose quantities are calculated for a massive wave subject to proper 
gauge condition. Considering the monochromatic wave solution of the form: 
\begin{align}
h_{\mu\nu}=C_{\mu\nu}e^{-i\omega t+ikz}
\, ,
\label{mwave}
\end{align}
where $\omega$ is the frequency and $k$ is the wave number.
However, we have noticed that, the modified Newman-Penrose scalars and polarization amplitudes introduced in \cite{Hyun_2019} using Lorentz gauge condition can not distinguish breathing mode when the scalar field becomes massless. It is due to the fact that, they have used the transverse traceless condition to make the non-tensor modes vanish when  $\omega = k$. But $\omega = k$ can't demand $p_{6}^{(b)}=0$ because, the breathing mode is 
massless in nature \cite{Gogoi_2019}.  Breathing modes satisfy the transverse 
condition but not the traceless condition. It implies that a model having 
$\omega = k$ can have massless breathing mode of polarization 
\cite{Gogoi_2019}. Keeping this fact in mind, we have modified the polarization amplitudes and 
Newman-Penrose quantities. According to our calculations, 
modified polarization amplitudes are expressed as:
\begin{align}
p_{1}^{(l)}\,\,&=  \frac{1}{2}\! \left( \frac{ \omega^{2}-k^{2} }{\omega^{2}+k^{2}} \right)\omega^{2}(h_{tt}+h_{zz})-\frac{1}{2}\left( \omega^{2}-k^{2} \right) h_{tt}\,,\nonumber\\[5pt]
p_{2}^{(x)}\,&=  \frac{1}{2}\left( \omega^{2}-k^{2}\right) h_{xz}\,,\nonumber\\[5pt]
p_{3}^{(y)}\,&=  \frac{1}{2}\left( \omega^{2}-k^{2}\right) h_{yz}\, , \nonumber\\[5pt]
p_{4}^{(+)}&= \frac{1}{2}\,\omega^{2}\left( h_{xx}-h_{yy} \right)\,,\nonumber\\[5pt]
p_{5}^{(\times )}&=  \frac{1}{2}\,\omega^{2}h_{xy},\nonumber\\[5pt]
p_{6}^{(b)}\,&= \frac{1}{2}\,\omega^{2}(h_{xx}+h_{yy})\, .
\label{polarization_amplitudes}
\end{align}
Here, we have not applied the traceless condition to the breathing mode. These are the exact polarization amplitudes of the wave. These expressions are valid for any metric theory. The modified Newman-Penrose quantities now can be 
expressed as:

\begin{align}
\Psi_{2}&=-\,\frac{1}{24}\left( \frac{\omega^{2}-k^{2}}{\omega^{2}+k^{2}} \right)\left[(3k^{2}-\omega^{2})h_{tt}+(k^{2}-3\omega^{2})h_{zz}\right]\,,\nonumber\\[5pt]
\Psi_{3}&=\frac{1}{8}\,\frac{(\omega-k)(\omega+k)^{2}}{\omega}\,(h_{xz}-ih_{yz})\,,\nonumber\\[5pt]
\Psi_{4}&=\frac{1}{8}\,(\omega+k)^2(h_{xx}+h_{yy})-\frac{1}{4 }(\omega+k)^{2}(h_{yy}+ih_{xy})\,,\nonumber\\[5pt]
\Phi_{22}&=\frac{1}{8}\,(\omega+k)^2(h_{xx}+h_{yy})\,.
\label{NP_quantities}
\end{align}
These equations differ from those in Ref.~\cite{Hyun_2019}.

Now, for our model the general wave solution is given by Eq.~\eqref{general_solution_01}. Using this Eq.~\eqref{general_solution_01} in the above set of 
Eq.s~\eqref{polarization_amplitudes}, we found the polarization amplitudes for 
our model as
\begin{align}
p_{1}^{(l)}\,\,&= \dfrac{1}{2}\,m^2_0\, C_1 \psi\,,\nonumber\\[5pt]
p_{2}^{(x)}\,&=  0\,,\nonumber\\[5pt]
p_{3}^{(y)}\,&=  0\, , \nonumber\\[5pt]
p_{4}^{(+)}&= -\, \dfrac{1}{2}\,(\ddot{\bar{h}}_{xx} - \ddot{\bar{h}}_{yy})\, , \nonumber\\[5pt]
p_{5}^{(\times )}&= -\, \frac{1}{2}\, \ddot{\bar{h}}_{xy}\,,\nonumber\\[5pt]
p_{6}^{(b)}\,&= \omega^2 C_1 \psi \,, 
\end{align}
where $m^2_0$ is given by Eq.~\eqref{eq63} and $C_1 = \frac{2 \alpha -\pi  \beta }{\pi(1-\beta)R_c}$ (see Eq.~\eqref{general_solution_01}). From the above 
expressions we can calculate the Newman-Penrose quantities. Note 
that, above results suggest, there are $4$ non-zero polarization amplitudes in 
the theory. Using Eq.s~\eqref{NP_quantities}, we've calculated the 
Newman-Penrose quantities for the model as

\begin{align}
\Psi_{2}&=\dfrac{1}{12}\, m^2_0\, C_1 \psi\,,\nonumber\\[5pt]
\Psi_{3}&=0\,,\nonumber\\[5pt]
\Psi_{4}&=(\ddot{\bar{h}}_{yy}+i\ddot{\bar{h}}_{xy})\,,\nonumber\\[5pt]
\Phi_{22}&=\dfrac{1}{4}\,(\omega +k)^2\, C_1 \psi\,. \label{model_NP}
\end{align}
Thus the $E(2)$ classification of the model is $II_6$. The model exhibits non-zero Newman-Penrose quantities for $4$ 
polarization modes viz., tensor plus, tensor cross, scalar transverse massless 
breathing mode and scalar longitudinal massive mode of polarization. However, the degrees of freedom associated with the theory is $3$. This suggests that the 
breathing mode and the longitudinal mode exist in a mixed state to give rise to a single polarization mode.
If $m_0 =0$, the massive longitudinal mode will vanish, giving  
$\Psi_{2}=0$. Note that $m_0 =0$ is not a sufficient condition to imply the 
absence of scalar degrees of freedom in the theory. It is because, in $f(R)$ 
theory there exists massless breathing mode which is transverse but not 
traceless. Absence of scalar degrees of freedom requires both $\Psi_{2}=0$ and 
$\Phi_{22} = 0$. When both $m_0$ and $\psi(vt-z)$ vanish, the theory reduces 
to GR giving only tensor modes of polarizations.

\section{Detection of Polarization Modes of GWs - A Review}
Experimental detection of polarization modes of GWs is very important to know 
the exact nature of GWs and hence in checking the viabilities of modified 
gravity theories. In this section we discuss the Pulsar Timing Arrays (PTAs) 
as a tool to distinguish between different polarization modes. Moreover, we 
include a discussion on the results based on our model.

PTAs play a significant role in the indirect detection of GWs. They are 
also used for numerous astrophysical applications. In 1968, Counselman \& 
Shapiro explained that the observations of pulsars could be used to test 
GR \cite{Counselman_1968}. Later in 1982, the first millisecond pulsar was 
discovered \cite{Backer_1982}. Till now, a pretty good number of millisecond 
pulsars has been discovered. The advantage of these pulsars over the normal 
pulsars is that they are very stable. Their arrival times can be measured and 
predicted with a good accuracy. This allows to use these pulsars as a probe to 
search for GWs. In 2004, the Parkes Pulsar Timing Array (PPTA) project began 
with the Parkes 64 m telescope \cite{Manchester2013, Oslowski2019}. After three years, the North 
American Nanohertz Observatory for Gravitational Waves (NANOGrav) in North 
America was founded \cite{Brazier2019, Arzoumanian2016}. NANOGrav uses the Arecibo and Green Bank 
telescopes to observe around 36 pulsars. In the same year, the European Pulsar 
Timing Array (EPTA) project was also founded \cite{Desvignes2016}. Using the Sardinian, 
Effelsberg, Nancay, Westerbork and Jodrell Bank telescopes, EPTA observes 
around 42 pulsars. Later, by combining these three projects, the International 
Pulsar Timing Array (IPTA) was formed \cite{Verbiest2016, Perera2019}. In this study, we have used 
some selected data from PPTA \cite{Manchester2013} and IPTA \cite{Verbiest2016, Perera2019}.

\begin{figure}[!htb]
\centerline{
   \includegraphics[scale = 0.4]{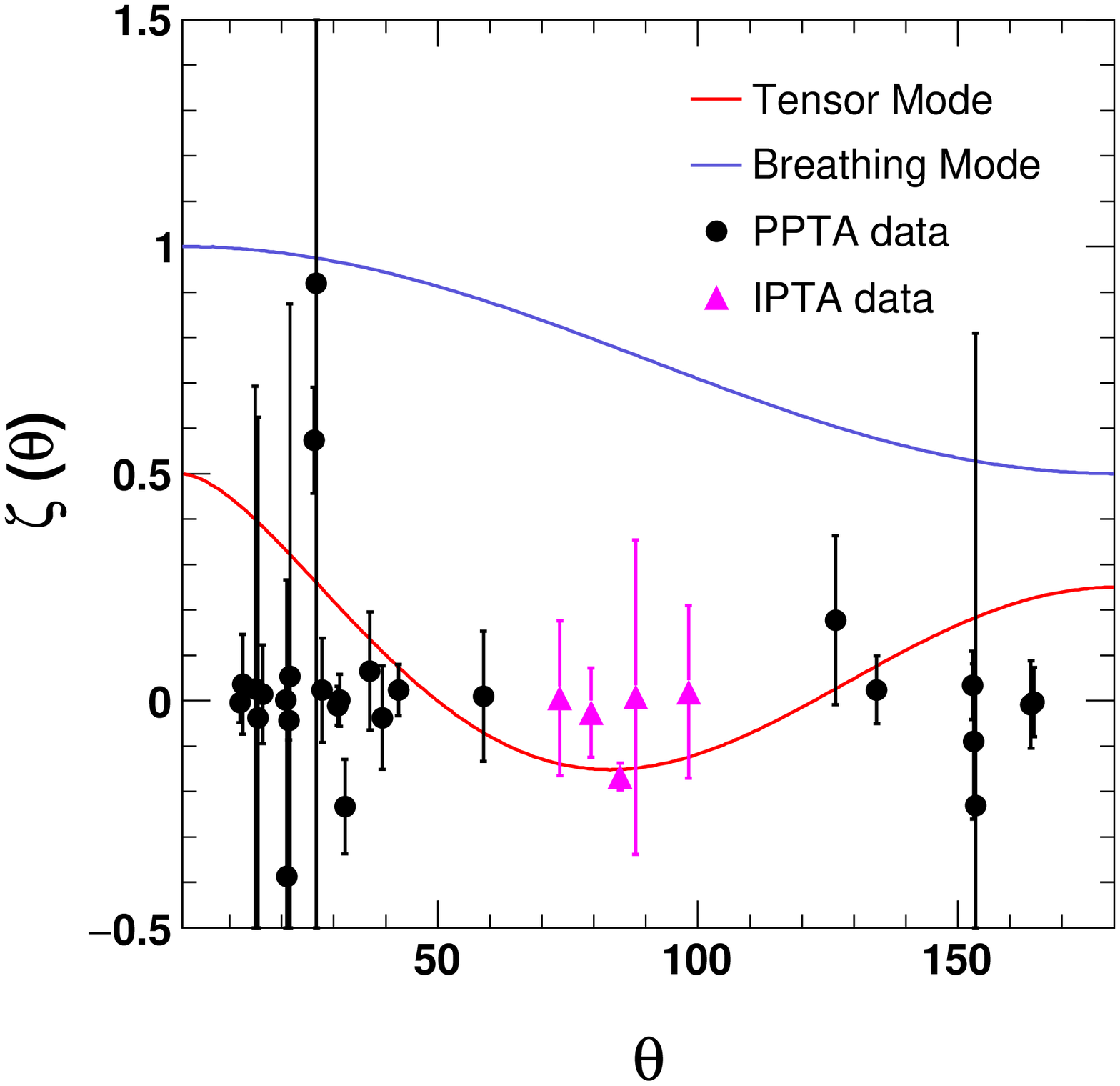}\hspace{1cm} 
  \includegraphics[scale = 0.4]{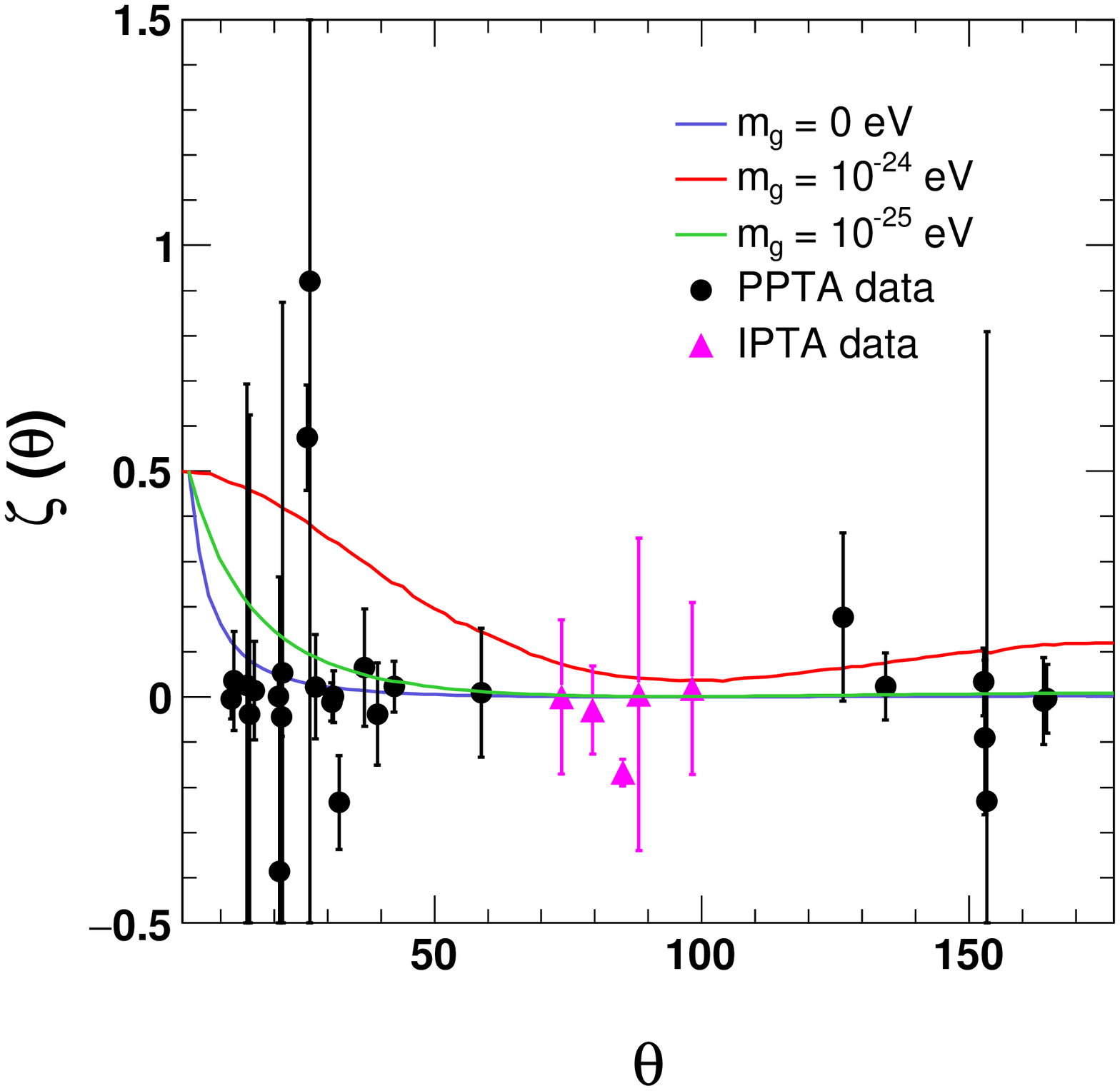}}
\caption{Variation of correlation functions $(\zeta(\theta))$ with respect to 
$\theta$. Plot on the left shows the correlation function for tensor modes and 
massless breathing mode of polarization of GWs. Plot on the right shows the 
correlation function for the longitudinal mode of polarization for different
values of the mass of the mode as predicted by the model (\ref{model}) along 
with correlation functions for some selected pulsars obtained from PPTA 
\cite{Manchester2013} and IPTA \cite{Verbiest2016, Perera2019} data.}
\label{pta_fig01}
\end{figure}

From the Ref.~\cite{Lee_2008}, we can have the correlation functions for 
different polarization modes. The calculation of correlation functions for the 
tensor and breathing modes are model independent, but in the case of massive 
longitudinal mode, the correlation function is model dependent as it depends 
on the mass of the scaler graviton. 
The correlation function for tensor modes is \cite{Lee_2008} 
\begin{equation}
C^{+,\times}(\theta) = \xi^{GR}(\theta)\int_0^\infty\dfrac{|h_c^{+,\times}|^2}{24\pi^2 f^3}\,df,
\end{equation}
where $$\xi^{GR}(\theta)=\dfrac{3\,(1-\cos\theta)}{4}\log\left(\dfrac{1-\cos\theta}{2}\right) + \dfrac{1}{2} - \dfrac{1-\cos\theta}{8} + \dfrac{\delta(\theta)}{2},$$
and $\theta$ is the angular separation between two pulsars. For the scalar 
modes it is \cite{Lee_2008}
\begin{equation}
C^{b}(\theta) = \xi^{b}(\theta)\int_0^\infty\dfrac{|h_c^{b}|^2}{12\pi^2 f^3}\,df,
\end{equation}
where $$\xi^{b}(\theta) = \dfrac{1}{8}\big[\cos\theta + 3 + 4\,\delta(\theta)\big].$$ 
The normalized correlation function in general is given by:
$$ \zeta (\theta) = \dfrac{C(\theta)}{C(0)}. $$
These are the correlation functions for tensor modes and massless breathing 
mode of GW polarization. But in our model, there exists a massive longitudinal 
mode. Thus to see the effect of massive mode, we follow the Refs.~\cite{Jenet_2005,Lee_2013, Lee_2010} in which the timing residual induced by GWs is 
expressed as
\begin{equation}
	R=-\,\frac{1}{\cal S}\, A^{ij} H_{ij},\label{res}
\end{equation}
here ${\cal S}=2\left (1+({c}/{{\omega_{ g}}}){{\mathbf{k}_{ g}}} \cdot \hat{\rm \bf n}\right )$ gives the dispersion relation of the GWs, the terms $\omega_{ g}$ and $k_{ g}$ connect the mass of the longitudinal mode of polarization $m_g$ by the relation $m_g^2 = \omega_{g}^2 - k_{ g}^2$, $A^{ij} \equiv \hat{\rm \bf n}^{i}\hat{\rm \bf n}^{j}$ and $H_{ij}=  \int_{0}^{\tau}h_{ij}(\tau,0)-h_{ij}(\tau-|{\rm \bf D}|/c,{\rm \bf D} )\,d\tau$. $\rm \bf D$ is the displacement 
vector from the observer to the pulsar and $\hat{\rm \bf n}^{i}$ and $\hat{\rm \bf n}^{j}$ are two unit vectors pointing to two pulsars. The correlation 
coefficient $C$ between two different pulsars is given by $$C_{1,2}(\theta)=\langle R_{1} R_{2}\rangle ={A}_1{A}_2 \langle {\cal S}_1 {\cal S}_2 {H}_1 {H}_2\rangle ,$$ where the sub-scripts are indices for the pulsars. With these 
assumptions and following Ref.~\cite{Lee_2008}, the correlation functions are 
calculated numerically for different values of $m_g$ 
(see Fig. \ref{pta_fig01}). In terms of our model, $$ m_g^2 \equiv m^2_{\phi} =\left[\frac{R_c e^{R/R_c} \left(\pi  \left(R^4+R_c^4\right)^2-8 \alpha  R^5 R_c^3\right)-\pi  \beta  (R+R_c) \left(R^4+R_c^4\right)^2}{3 \pi  \beta  \left(R^4+R_c^4\right)^2-6 \alpha  R_c^4 e^{R/R_c} \left(R_c^4-3 R^4\right)}\right]_{R\, =\, R_0}\!\!\!\!\!. $$
All these numerically calculated correlation functions are plotted with respect
to $\theta$ in Fig.\ \ref{pta_fig01}. In this figure we have used three values 
of $m_g$ for the following sets of parameters:\\

(i) $m_g = 0$ for $R = 0 ~m^{-2}, R_c = 0 ~m^{-2}, ~\alpha, \beta \in [0, 1]$,\\

(ii) $m_g = 10^{-24}$ eV for $ R=0 ~m^{-2}, R_c= 10^{-50} ~m^{-2}, \alpha = 0.05, \beta = 0.031831$ and\\

(iii) $m_g = 10^{-25}$ eV for $R= 4.44\times10^{-52} ~m^{-2}, R_c = 5\times 10^{-51} ~m^{-2}, \alpha = 0.0314565, \beta = 0.17268558.$\\

To check the experimental viability of our model, we have calculated the 
correlation functions for GWs using some selected data from IPTA and PPTA data 
set \cite{Manchester2013, Verbiest2016, Perera2019, Hobbs2006, Edwards2006} as mentioned above, which are also plotted in Fig.\ \ref{pta_fig01}. 
Although we do not have 
a clear conclusion, which requires more observation period as well as data, we 
can still see that these experimental data could not directly rule out the 
existence of extra polarization modes of GWs. However, to distinguish between 
polarization modes, we need to wait for more PTA data with GW events.


\section{Summary and Conclusion}
In this work, we have introduced a new $f(R)$ gravity toy model and studied 
the polarization modes of GWs in it. The study shows that, in metric formalism,
there exists $3$ polarization modes of GWs viz., tensor plus mode, tensor cross 
mode and scalar mode. The scalar mode is a mixed state of massless breathing mode and massive longitudinal mode. 
The tensor modes of 
polarization are transverse, traceless and massless in nature. The scalar 
breathing mode is transverse but exists with non vanishing trace and massless 
in nature. On the other hand, the scalar longitudinal mode is massive in nature
and hence propagates with speed less than that of tensor modes. When the scalar field becomes massless, the longitudinal 
mode vanishes and only the massless scalar breathing mode exists in the scalar 
degrees of freedom. As an experimental correspondence of our model prediction
on the modes of GWs, we have compared the correlation function of massive 
longitudinal mode as predicted by the model with that of the some selected 
PTAs data of PPTA and IPTA. The result is found to be quite encouraging.   

We have also shown that a wisely selected set of the 
parameters easily allows the model to pass the solar system tests, which is a 
very important requirement for the viability of a model. Further, the model 
has been constrained using combined CMB, BAO and $\sigma_8-\Omega_m$ 
relationship from the PSZ catalog and Abell 1689 galaxy cluster data. The 
model can be easily constrained with the help of the parameters $R_c$, 
$\alpha$ and $\beta$. Further, we have also constrained the model using the GW 
event GW170817. It is seen that, the model 
can withstand the constraints put by GW170817 and hence can be included as a 
post GW170817 viable model.

In the present work, we have studied only few properties of the model. 
However, for a proper understanding of the model characteristics, a detailed 
study is required. As such, in future the model can be checked also for 
various stabilities and constraints as well as in different cosmological and 
astrophysical contexts, which will give us more information about the 
viability of the model. Moreover, it is to be noted that in the Palatini 
formalism the polarization modes of GWs in $f(R)$ gravity is model 
independent. So, this formalism can not distinguish our model from 
other $f(R)$ gravity models in this context. Nevertheless, there may be some 
cosmological variations and stellar structure differences of the model in 
Palatini formalism, which might be useful to study the generation of GWs in 
such situations.

\section*{Acknowledgement}
A part of this work was done during a visit of authors to IUCAA, Pune. Authors 
are grateful to IUCAA for the hospitality during their stay.

\bibliographystyle{apsrev}
\end{document}